\documentclass[12pt]{article}
\usepackage{a4wide}
\usepackage{amstext,amsmath,amssymb,amsfonts}
\usepackage{graphicx}
\usepackage{color}
\usepackage{hyperref}

\usepackage{rotating}
\usepackage{ytableau}

\usepackage{tikz,pbox}
\usetikzlibrary{shapes,arrows}
\usetikzlibrary{calc,decorations.markings}

\def\Mult{ \hbox{Mult} }

\newcommand{\mytikz}[1]{
	\pbox{\textwidth}{\begin{tikzpicture}[>=stealth,decoration={
    markings,
    mark=at position 0.5 with {\arrow{>}}}]	
		#1
	\end{tikzpicture}}
}

\usepackage{array}
\usepackage{fancybox}

\newtheorem{lemma}{Lemma}

\newtheorem{proposition}{Proposition}

\newcommand{\be}{\begin{equation}}
\newcommand{\ee}{\end{equation}}
\newcommand{\beq}{\begin{equation}}
\newcommand{\eeq}{\end{equation}}
\newcommand{\bea}{\begin{eqnarray}\displaystyle}
\newcommand{\eea}{\end{eqnarray}}
\newcommand{\nnm}{\nonumber}

\newcommand{\ba}{\begin{array}}
\newcommand{\ea}{\end{array}}
\newcommand{\ben}{\begin{enumerate}}
\newcommand{\een}{\end{enumerate}}
\newcommand{\bi}{\begin{itemize}}
\newcommand{\ei}{\end{itemize}}
\newcommand{\bc}{\begin{center}}
\newcommand{\ec}{\end{center}}
\newcommand{\bfig}{\begin{figure}}
\newcommand{\efig}{\end{figure}}
\newcommand{\bq}{\begin{quotation}}
\newcommand{\eq}{\end{quotation}}
\newcommand{\bt}{\begin{table}}
\newcommand{\et}{\end{table}}
\newcommand{\btab}{\begin{tabular}}
\newcommand{\etab}{\end{tabular}}
\newcommand{\bmi}{\begin{minipage}}
\newcommand{\emi}{\end{minipage}}
\newcommand{\bs}{\begin{slide}}
\newcommand{\es}{\end{slide}}

\newcommand{\cN}{ {\mathcal N} }

\newcommand{\cO}{ { \mathcal{ O } } }

\newcommand{\IC}{{\mathbb C}}
\newcommand{\sC}{{\sf{C}}}

\newcommand{\la}{\langle}
\newcommand{\ra}{\rangle}

\newcommand{\cZ}{ \mathcal{Z} }

\newcommand{\bdel}{ {\boldsymbol{\delta}} }

\newcommand{\ung}{ {\rm{un}} }

\newcommand{\mC}{\mathbb{C}} 

\newcommand{\diag}{ {\rm Diag} } 
\newcommand{\Sym}{ {\rm Sym} } 
 
\newcommand{\Aut}{ {\rm Aut} }

\newcommand{\col}{{\rm{cs}}}
\newcommand{\cy}{{\bf{c}}}

\newcommand{\Tr}{{\rm Tr}}
\newcommand{\tr}{{\rm tr}}

\newcommand{\Dim}{ {\rm Dim } } 

\newcommand{\qed}{\hfill$\square$} 
\newcommand{\proof}{{\bf Proof.}\;}

\def\cA{{\cal A}}  
  
\def\cG{{\cal G}}  
 \def\cK{{\cal K}} 
 \def\cN{{\cal N}} \def\cO{{\cal O}}

 \def\cZ{{\cal Z}}

\def\wTp{ { \widehat{T_p} } }
 
\newcommand{\idtr}{ {
\scalebox{0.3}{
\begin{tikzpicture}[xscale=1,yscale=1.7]
\draw[solid] (1.5,1.5) circle (0.1);
\fill[fill=black] (2.25,1.5) circle (0.1);
\draw[solid] (1.5,0.5) circle (0.1);
\fill[fill=black] (2.25,0.5) circle (0.1);
\fill[fill=black] (1.5,1) circle (0.1);
\draw[solid] (2.25,1) circle (0.1);
\draw[shorten <= 0.1cm, shorten >= 0.1cm] (1.52,1.52) to[out=50, in=130] (2.23,1.52);
\draw[shorten <= 0.1cm, shorten >= 0.1cm] (1.52,1.48) to[out=-50, in=-130] (2.23,1.48);
\draw[shorten <= 0.1cm, shorten >= 0.1cm] (1.52,0.52) to[out=50, in=130] (2.23,0.52);
\draw[shorten <= 0.1cm, shorten >= 0.1cm] (1.52,0.48) to[out=-50, in=-130] (2.23,0.48);
\draw[shorten <= 0.1cm, shorten >= 0.1cm] (1.52,1.02) to[out=50, in=130] (2.23,1.02);
\draw[shorten <= 0.1cm, shorten >= 0.1cm] (1.5,0.98) to[out=-50, in=-130] (2.23,0.98);
\draw (1.6,1.5) -- (2.15,1.5) ;
\draw (1.6,0.5) -- (2.15,0.5) ;
\draw (1.6,1) -- (2.15,1) ;
\end{tikzpicture}} 
}}

\newcommand{\cc}[1]{  
\scalebox{0.3}{ 
\begin{tikzpicture}[xscale=1.3,yscale=1.5]
\node (r2) at (0.13,1.4)  {{\large{#1}}};  
\fill[fill=black] (1.25,1.5) circle (0.1);
\draw[solid] (0.5,1.5) circle (0.1);
\draw[solid] (0.5,0.5) circle (0.1);
\fill[fill=black] (0,1) circle (0.1);
\fill[fill=black] (1.25,0.5) circle (0.1);
\draw[solid] (1.7,1) circle (0.1);
\draw (1.31,1.43) -- (1.66,1.09); 
\draw (0.6,0.5) -- (1.15,0.5);
\draw (0.07,1.06) -- (0.43,1.43); 
 \draw[shorten <= 0.1cm, shorten >= 0.1cm] (1.25,0.53) to[out=90, in=180] (1.67,1.005);
\draw[shorten <= 0.1cm, shorten >= 0.1cm] (1.25,0.48) to[out=0, in=-90] (1.71,0.97);
 \draw[shorten <= 0.1cm, shorten >= 0.1cm] (0.5,0.53) to[out=90, in=0] (0.026,1.005); 
 \draw[shorten <= 0.1cm, shorten >= 0.1cm] (0.48,0.48) to[out=180, in=-90] (0,0.98); 
\draw[shorten <= 0.1cm, shorten >= 0.1cm] (0.52,1.52) to[out=50, in=130] (1.23,1.52);
\draw[shorten <= 0.1cm, shorten >= 0.1cm] (0.52,1.48) to[out=-50, in=-130] (1.23,1.48);
\end{tikzpicture}
}
}

\newcommand{\ac}[1]{ 
\scalebox{0.3}{  
\begin{tikzpicture}[xscale=1.1,yscale=1.7]
\node (r2) at (0.9,1.15)  {{\large{#1}} };  
\draw[solid] (0.5,1.5) circle (0.1);
\draw[solid] (0.5,0.5) circle (0.1);
\fill[fill=black] (0.5,1) circle (0.1);
\fill[fill=black] (1.25,0.5) circle (0.1);
\draw[solid] (1.25,1) circle (0.1);
\fill[fill=black] (1.25,1.5) circle (0.1);
\draw (0.5,0.6) -- (0.5,1.4); 
\draw (1.25,0.6) -- (1.25,0.9); 
\draw (1.25,1.1) -- (1.25,1.4); 
\draw (0.6,1) -- (1.15,1);
\draw[shorten <= 0.1cm, shorten >= 0.1cm] (0.52,0.52) to[out=50, in=130] (1.25,0.5);
\draw[shorten <= 0.1cm, shorten >= 0.1cm] (0.51,0.47) to[out=-50, in=-130] (1.25,0.5);
\draw[shorten <= 0.1cm, shorten >= 0.1cm] (0.52,1.52) to[out=50, in=130] (1.25,1.5);
\draw[shorten <= 0.1cm, shorten >= 0.1cm] (0.52,1.48) to[out=-50, in=-130] (1.25,1.5);
\end{tikzpicture}
} 
}

\newcommand{\dtwo}[1]{ 
\scalebox{0.3}{ 
\begin{tikzpicture}[xscale=1.35,yscale=1.7]
\draw[solid] (0.5,1.5) circle (0.1);
\fill[fill=black] (1.25,1.5) circle (0.1);
\draw[shorten <= 0.1cm, shorten >= 0.1cm] (0.52,1.53) to[out=50, in=130] (1.25,1.52);
\draw[shorten <= 0.1cm, shorten >= 0.1cm] (0.52,1.47) to[out=-50, in=-130] (1.25,1.47);
\draw (0.6,1.5) -- (1.15,1.5) ;
\node (r1) at (0.3,0.75)  {{\large{#1}}};   
\draw[solid] (0.5,0.5) circle (0.1);
\fill[fill=black] (1.25,0.5) circle (0.1);
\fill[fill=black] (0.5,1) circle (0.1);
\draw[solid] (1.25,1) circle (0.1);
\draw (0.5,0.6) -- (0.5,0.9); 
\draw (1.25,0.6) -- (1.25,0.9); 
 \draw[shorten <= 0.1cm, shorten >= 0.1cm] (0.525,0.53) to[out=50, in=130] (1.25,0.5);
\draw[shorten <= 0.1cm, shorten >= 0.1cm] (0.525,0.47) to[out=-50, in=-130] (1.25,0.5);
\draw[shorten <= 0.1cm, shorten >= 0.1cm] (0.5,1) to[out=50, in=130] (1.23,1.02);
\draw[shorten <= 0.1cm, shorten >= 0.1cm] (0.5,1) to[out=-50, in=-130] (1.21,0.99);
\end{tikzpicture}
}
}

\newcommand{\ktr}{   
\scalebox{0.3}{
\begin{tikzpicture}[xscale=0.85,yscale=1.7]
\draw[solid] (0.5,1.5) circle (0.1);
\draw[solid] (0.5,0.5) circle (0.1);
\fill[fill=black] (0.05,1) circle (0.1);
\fill[fill=black] (1.25,0.5) circle (0.1);
\draw[solid] (1.7,1) circle (0.1);
\fill[fill=black] (1.25,1.5) circle (0.1);
\draw (1.3,1.43) -- (1.66,1.09); 
\draw (0.6,0.5) -- (1.15,0.5);
\draw (0.05,1.05) -- (0.43,1.43); 
\draw (0.6,1.5) -- (1.2,1.5);  
\draw (1.67,0.91) -- (1.32,0.55); 
\draw (0.41,0.54) -- (0.05,0.95); 
\draw (0.565,0.565) -- (1.25,1.5); 
\draw (0.56,1.42) -- (1.27,0.48); 
\draw (0.05,1) -- (1.6,1); 
\end{tikzpicture}
} 
}

\newcommand{\ses}{ {\setminus} }

\def\s{ \sigma } 
\def\g { \gamma }

\begin{document}

\begin{flushright}
QMUL-PH-17-13
\end{flushright}

\bigskip

\begin{center}

{\Large \bf Tensor Models, Kronecker coefficients and 

 }
 \medskip 

{ \Large \bf   Permutation Centralizer Algebras}

\bigskip

{
Joseph Ben Geloun$^{a,c,*}$
 and Sanjaye Ramgoolam$^{b , d ,\dag}  $}

\bigskip
$^a${\em Laboratoire d'Informatique de Paris Nord UMR CNRS 7030} \\
{\em Universit\'e Paris 13, 99, avenue J.-B. Clement,
93430 Villetaneuse, France} \\
\medskip
$^{b}${\em School of Physics and Astronomy} , {\em  Centre for Research in String Theory}\\
{\em Queen Mary University of London, London E1 4NS, United Kingdom }\\
\medskip
$^{c}${\em International Chair in Mathematical Physics
and Applications}\\
{\em ICMPA--UNESCO Chair, 072 B.P. 50  Cotonou, Benin} \\
\medskip
$^{d}${\em  School of Physics and Mandelstam Institute for Theoretical Physics,} \\   
{\em University of Witwatersrand, Wits, 2050, South Africa} \\
\medskip
E-mails:  $^{*}$bengeloun@lipn.univ-paris13.fr,
\quad $^{\dag}$s.ramgoolam@qmul.ac.uk

\begin{abstract}
 We show that the counting  of observables and correlators for a 3-index  tensor model 
  are organized by the structure of a family of permutation centralizer algebras. 
  These algebras are shown to be semi-simple and  their Wedderburn-Artin decompositions  into matrix blocks are given in terms of Clebsch-Gordan coefficients of 
  symmetric groups.  The matrix basis for the algebras also gives an   orthogonal basis for the tensor observables which diagonalizes the Gaussian two-point functions. The centres of the algebras are associated with correlators which are expressible in terms of Kronecker coefficients (Clebsch-Gordan multiplicities of symmetric groups). The color-exchange symmetry present in the Gaussian model, as well as a large class of interacting models, is used to refine the description of  the permutation centralizer algebras. This discussion is extended to a general number of colors $d$: it  is used to  prove  the  integrality of an infinite family of number sequences related to color-symmetrizations of  colored graphs, and expressible in terms of symmetric group representation theory data.  Generalizing a connection between matrix models and Belyi maps, correlators in Gaussian tensor models are interpreted in terms of covers of singular 2-complexes. There is an intriguing difference, between matrix and higher rank tensor models, in the computational complexity of superficially comparable correlators of observables parametrized by Young diagrams. 

\end{abstract}

\end{center}


\newpage 

\tableofcontents

\section{Introduction}

Introduced as generalizations of  matrix models \cite{Di Francesco:1993nw,GM9304} to study the discrete-to-continuum transition 
for discretized path integrals in quantum gravity, tensor models \cite{ambj3dqg,sasa1,mmgravity} 
and their further generalizations \cite{Oriti:2006se}   were found to be 
tremendously more difficult to handle than the theory of matrices. 
One the main sources of difficulties in the study of tensor models at that time was the absence of an organizing principle for their partition function. Matrix models
are organized by the  large $N$ expansion \cite{'tHooft:1973jz} which sorts maps by their genus, and typically a world-sheet `t Hooft coupling constant at fixed genus.  
After approximately two decades, significant progress on tensor models emerged in a series of papers \cite{LargeN,GurRiv,Gur4,Gurau:2009tw,Gurau:2010nd,colorN}. The large $N$ expansion for colored tensors was characterized in terms of sets of ribbon graphs known as ``jackets'' and new double scaling limits involving ``melons'' were found.  Since then, many results on 
random tensors \cite{tensors} have been achieved from  statistical 
mechanics, to quantum field theory but as well in combinatorics and
probability theory (see \cite{Bonzom:2011zz,BGR,TFT}
and the reviews \cite{tensortrack} and \cite{Gurau:2016cjo}).  
Recently,  the large $N$ expansion for tensors added another twist in this already-remarkable story:
the large $N$ limit of the famous Sachdev-Ye-Kitaev  (SYK) condensed matter model 
\cite{SYK,malstan,JSY1603,polros,Maldacena:2015waa} 
matches with the same limit of a quantum mechanical model built with colored tensors without disorder \cite{wit}. The  SYK model is an active topic of research, its   connections being explored with  black hole physics, AdS/CFT correspondence, quantum gravity and condensed matter physics. The new connection between tensor  
and SYK models  has thus come to be of relevance to  
several areas of  theoretical physics (see for instance \cite{Klebanov:2016xxf,Carrozza:2015adg,Gurau:2016lzk,Gurau:2017qna,Giombi:2017dtl,
Ferrari:2017ryl, DGT1707} and references therein).

A better understanding of the combinatorics of tensor models will be crucial in identifying and characterizing  their holographic duals. There are two closely related aspects to the combinatorics, in the first instance the enumeration of observables, in the second, the 
understanding of the correlators. The former has immediate implications for 
thermodynamic questions related to these models (for recent investigations focused on these aspects see \cite{BT1705}),  the spectrum of physical excitations in the holographic duals
and since observables can be used to parametrize deformations of a given model, for the space of possible holographic duals.  The implications of the counting for the intricacy of the holographic dual has been discussed in \cite{Klebanov:2016xxf}, see also \cite{Bulycheva:2017ilt, Choudhury:2017tax}.
In the context of the AdS/CFT correspondence, correlators in the gauge theory or matrix models provide refined information about the holographic dual: interactions of gravitons, strings and branes. From a mathematical point of view, they are related to a host of interesting algebraic structures, notably integrability and hidden symmetries
        \cite{BalIntegRev}.

In the paper \cite{Sanjo}, we  developed a variety of 
 counting formulae starting from  the complete set of invariant 
observables in tensor models  with a complex tensor 
field having $d$ indices transforming in the fundamental of $ U(N)^{ \times d }$. 
The invariant observables are constructed from $n$ copies of the complex field 
$ \Phi $ as well as $n$ copies of the conjugate field $ \bar \Phi$. They are in 1-1 correspondence with graphs with two types of vertices, one for the $\Phi$'s and one for the 
$\bar \Phi$'s, and colored edges, one for each type of index. 
The invariants are constructed by contracting the indices in  the fundamental of 
each $ U(N)$ with the indices in the anti-fundamental. They are parametrized 
by a sequence of $d$  permutations, each in $S_n$,  one for each factor in the unitary symmetry group. These permutations are subject to equivalence relations, which characterize  permutation sequences corresponding to  the same invariant operator. It was observed that the counting formulae can be expressed in terms of topological field theory based on symmetric groups.
 The permutation description was used to give formulae for ``normal-ordered correlators'' in  Gaussian tensor models.

An important additional symmetry exists in these Gaussian models, and indeed in a large class of interacting models. 
It is the group of all permutations of the $d$ types of indices, which form the 
symmetric group $S_d$. It had already been understood  in the tensor model literature that 
the counting of the above  tensor invariants can be expressed as a counting of colored graphs. It had also been understood that by taking advantage of the $S_d$ symmetry, it is useful to consider ``color-symmetrized graphs'' which are defined by additional equivalences generated by the $S_d$ action of color-exchanges on the graphs. 
In \cite{Sanjo} we found counting formulae for the ``color-symmetrized graphs'' in terms of the permutation tuples. These generated sequences of positive integers.  Intriguingly the formulae we obtained for the color symmetrized graphs were expressed as fractional sums of expressions, which turned out to themselves be integers. These sequences were denoted by $S^{(d)}_{ p} (n) $  : specifying  an integer $d$, along with a partition of $d$, we have a sequence of integers as $n$ ranges over positive integers. One of the  results  which follows from the detailed treatment of color exchange symmetry 
  in this paper is to explain the integrality of these additional sequences, and give general expressions for them for any $d$ and $p \vdash d$. 

The permutation approach to counting and correlators has been used in a 
number of papers in the context of AdS/CFT. It was used in the half-BPS 
sector in \cite{cjr,cr} to find orthogonal bases for operators, which are useful in identifying CFT duals of giant gravitons. Following investigations of strings attached to giant gravitons in \cite{BBFH}, orthogonal bases for multi-matrix operators in CFT were found in \cite{BHR1,KR1,BCD1,BDS,KR2,BHR2}. The key idea is to parametrize gauge invariants 
using permutations subject to equivalences, and to understand these equivalence classes 
using Fourier transformation on symmetric groups based on representation theory, to go  from 
the equivalence classes of permutations to representation theoretic bases. A short review is in \cite{PermsGI}. It has been realized that an important role in understanding these orthogonal bases is played by permutation centralizer algebras (PCAs) \cite{PCA1601}. The basic observation is that the once we have found a formulation of 
a counting of invariants in terms of permutations subject to equivalences, it is useful to go to the group  algebras of permutations and consider their sub-algebras associated to the equivalence classes. These algebras  are semi-simple,  i.e. they are associative 
and have a  non-degenerate bilinear pairing. These properties are inherited from 
the underlying permutation group algebras. As a result, from the Wedderburn-Artin (WA)  theorem \cite{wallach,ram} 
on the structure of  these algebras, we have a matrix decomposition of these algebras. The construction of orthogonal bases is, in many cases studied so far, closely connected to the workings of the WA theorem. This was studied in depth for the 2-matrix problem in \cite{PCA1601} (closely related developments from the perspective of open-closed topological field theory are in \cite{Kimura1403,Kimura1701}). In particular, the important role of the centre of the PCA  was noted,   in identifying a sector of correlators which  are computable using characters, without requiring more refined representation theoretic quantities. The appropriate PCA for tensor model counting was identified and a basis in terms of Clebsch-Gordan coefficients of $S_n$ was described. It was called $ \cK ( n)$ and its dimension was shown to be the sum of squares of Kronecker coefficients. 

In this paper, we present a  systematic study of $ \cK (n )$ and highlight the role of its structure, particularly in connection with the WA theorem, in correlators and orthogonal bases for tensor models. The role of color-exchange symmetry in the structure of $ \cK ( n)$ is another important  theme, which leads to new results on the  integer sequences, denoted $ S^{(d)}_p(n ) $,  which arose among the counting of color-symmetrized invariants in \cite{Sanjo}. 
A  key result of this paper is the formula for the dimension of the color-symmetrized sub-algebra of $ \cK ( n ) $ as a sum of squares of representation-theoretic  quantities.

The paper is organized as follows. In section 2, we introduce the  tensor model we will be discussing, based on complex tensors with $d$ indices transforming under $ U(N)^{ \times d } $. We review the permutation approach to tensor models from \cite{Sanjo}.

The algebras $ \cK ( n )$ are defined in section \ref{sec:PCA}. There are two equivalent descriptions of the algebra. In one description, they are  sub-algebras of 
the tensor products $ \mC ( S_n) \otimes \mC ( S_n)$ which commute with the diagonally embedded $ \mC ( S_n)$. In equivalent terminology, $ \cK (n)$ is the centralizer of 
the diagonal $ \mC ( S_n)$ in the tensor product $ \mC ( S_n ) \otimes \mC ( S_n)$, hence the name ``permutation centralizer algebra'' (PCA). Another type of PCA has been found to 
underlie a variety of results on correlators for the 2-matrix problem \cite{PCA1601}:  they are thus emerging as fundamental to the application of permutation group and representation theory techniques to matrix/tensor correlators.  In the other description, they are sub-algebras of $ \mC ( S_n ) \otimes \mC ( S_n) \otimes \mC ( S_n)$ which are invariant under 
left and right diagonal actions. Both descriptions are based on the fact that  tensor 
invariants can be described by sequences of permutations, subject to equivalences defined 
in terms of group multiplications in $S_n$. The second description is an algebra structure on a space of double cosets, so that $ \cK (n)$ is a double coset algebra. 
By partially gauge fixing the equivalences in the double coset description, we arrive at the 
PCA description. $ \cK (n)$ is a semi-simple associative algebra.  As a result of the WA theorem, such algebras are isomorphic to direct sums of matrix algebras. 
We describe this direct sum decomposition of $ \cK ( n )$. We can think of $ \cK (n)$ as made of matrix blocks. The terms in the sum 
are labelled by triples of Young diagrams with $n$ boxes,  with non-vanishing Kronecker coefficient. These are triples $ R_1 , R_2 , R_3$ of representations of 
$S_n$ such that the tensor product $ R_1 \otimes R_2 \otimes R_3$ contains the trivial under the action of the diagonal $S_n$. The algebra elements belonging to the matrix block labelled by the ordered triple $ [ R_1 , R_2 , R_3 ] $, along with more refined data 
associated with the Kronecker multiplicities, are constructed using Clebsch-Gordan coefficients of the symmetric group. These are denoted $ Q^{ R_1 , R_2 , R_3}_{ \tau_1 , \tau_2 } $. $ \cK (n)$ is a non-commutative algebra for generic $n$, so that the centre $ \cZ ( \cK ( n)) $ of $\cK(n)$ is a  proper subspace. Triples of Young diagrams label an overcomplete basis for the centre and triples with non-vanishing Kronecker coefficient label a basis. 

$ \cK (n) $ also has an interpretation as a graph algebra. We explain this in section \ref{cKgraph}. 
The identity of $\cK(n)$ is the maximally disconnected melonic graph. 
For the lower orders $n\leq  4$, this algebra turns out to be commutative. We give illustrations of this 
algebra at $n=2$ and $n=3$ (multiplication tables). At $n=3$, there are elements which could
factorize in different ways and this might lead to interesting properties.  

Section \ref{sect:PCAandCorr} shows how the structure of $ \cK (n)$ described in section \ref{sec:PCA} organizes the properties of correlators in the Gaussian model. We consider two types of correlators: two-point functions of 
normal-ordered invariants, and one-point functions without normal ordering. 
We develop explicit formulae using known results on Kronecker coefficients 
for specific Young diagrams. The discussion of correlators makes it natural to consider the PCAs $ \cK ( n)$ for all $n$ at once, where $n$ labels the number of $ \Phi $ and $ \bar \Phi$ in the invariant.  Hence we consider and discuss the algebra
\be
\cK^{\infty} = \bigoplus_{ n =0}^{ \infty} \cK ( n ) 
\ee
using the convention that at $n=0$, $\cK(0)=\mC$.

In section \ref{sec:tft2}, we describe how tensor model correlators 
can be described by two-dimensional topological field theory of permutations
on  2-complexes. The permutation-TFT2 description of counting and correlators for matrix theories is reviewed and applied to general quiver gauge theories in \cite{quivcalc}. 
Some results on the connection between counting of tensor invariants and permutation-TFT2
were given in \cite{Sanjo}. Here we consider correlators of these invariants as amplitudes in permutation TFT2. 

In section \ref{sect:colsym}, we use the color-exchange symmetry $S_3$ 
of the rank-$3$ tensor model in order to give a refined description of 
$ \cK ( n )$ in terms of irreducible representations (irreps) of $S_3$.  
We prove integrality of some sequences of numbers, which were observed in \cite{Sanjo} 
but not proved. The subspace invariant under color-exchange is a closed sub-algebra. 
We give a formula for  the dimension of this sub-algebra  as a sum of squares which leads to the understanding of the WA-decomposition of  the algebra.

Section \ref{concl} gives a summary of our results and outlines interesting future directions  
for research. Among those directions,  we mention a new type of statistical models based on  Young diagrams,  the quest for holographic duals of tensor models and an intriguing connection between Computational 
Complexity Theory and correlators in matrix and tensor models.

 In the last part of the paper, we have four appendices: appendix \ref{app:symrep} 
gathers basics of representation theory of the symmetric group $S_n$ which 
is used thoroughly in the text.  Appendix \ref{app:algebra} consists in proofs of
statements about PCAs, properties of their bases and their centre.  
Appendix \ref{app:11m} provides an illustration of $\cK(n=3)$ as a graph algebra
and gives its multiplication table. Finally, in 
appendix \ref{app:correlators}, we give a summary of the calculation
of Gaussian correlators (one-point and two-point functions).

\

{\bf Note Added:} While this paper was being completed, a few papers with some overlap \cite{DGT1707,MM1706,DR1706} appeared. We will be pointing out the specific overlaps 
in key points as they arise, particularly in section \ref{sect:PCAandCorr}. Representation theory and Young diagram combinatorics  have also been employed in an SYK context in \cite{KK1706}.

\section{Observables in tensor models using permutations}
\label{sect:Rev} 

We start by giving  a summary of the description of tensor model observables  in terms
of permutations which was introduced in \cite{Sanjo}.

Consider $\{V_i\}_{i=1,\cdots,d}$, a family of complex vector spaces of
respective dimensions $N_1$, $N_2$, \dots, $N_d$. 
Let  $\Phi$ be a rank $d\geq 2$ covariant tensor
with components $\Phi_{i_1 , \cdots , i_d  }$, with $i_a  \in \{1,\dots, N_a\}$, $a=1,2,\dots, d$, transforming as  $\otimes_{a=1}^d V_a$. No symmetry 
is assumed under permutation of the indices of $\Phi_{i_1 , \cdots , i_d  }$. 
The tensor $\Phi$ transforms under the action of the  tensor product of fundamental representations of unitary groups $\otimes_{a=1}^{d} U(N_a)$
where each $U(N_a)$ independently acts on a tensor index $i_a $. The complex conjugate $\bar{\Phi }_{i_1 i_2 \dots i_d}$ of $\Phi_{ i_1 i_2  \dots i_d }$ is a contravariant tensor of the same rank $d$. 
The following transformation rule holds:
\bea
\Phi_{i_1i_2 \dots i_d} &=& \sum_{j_1, \dots, j_d} 
U^{(1)}_{i_1 j_1} U^{(2)}_{i_2 j_2} \dots U^{(d)}_{i_d j_d}  \Phi_{j_1 j_2 \dots j_d } \cr
\bar{\Phi}_{i_1i_2 \dots i_d} &=& \sum_{j_1, \dots, j_d} 
\bar{U}^{(1)}_{i_1 j_1} \bar{U}^{(2)}_{i_2j_2} \dots \bar{U}^{(d)}_{i_dj_d}
  \bar{\Phi }_{j_1j_2 \dots j_d } 
\label{tut}
\eea
where $U^{(a)}$ are unitary belonging to $U(N_a)$, $a=1,2,\dots, d$ and
may be all distinct. 
The rank $d=2$ will be referred to as matrix case and will be useful to make contact with known results in matrix models. We will however focus in the rank $d\geq 3$. 

Unitary invariants with respect to the action 
\eqref{tut} are built by contracting pairs of indices of (covariant and contravariant) tensors. 
 These contractions are in bijection with regular bipartite $d$--colored graphs (see section 2.1 in \cite{Sanjo}, for illustrations). 

\begin{figure}[h]\begin{center}
     \begin{minipage}[t]{.8\textwidth}\centering
\includegraphics[angle=0, width=8cm, height=3cm]{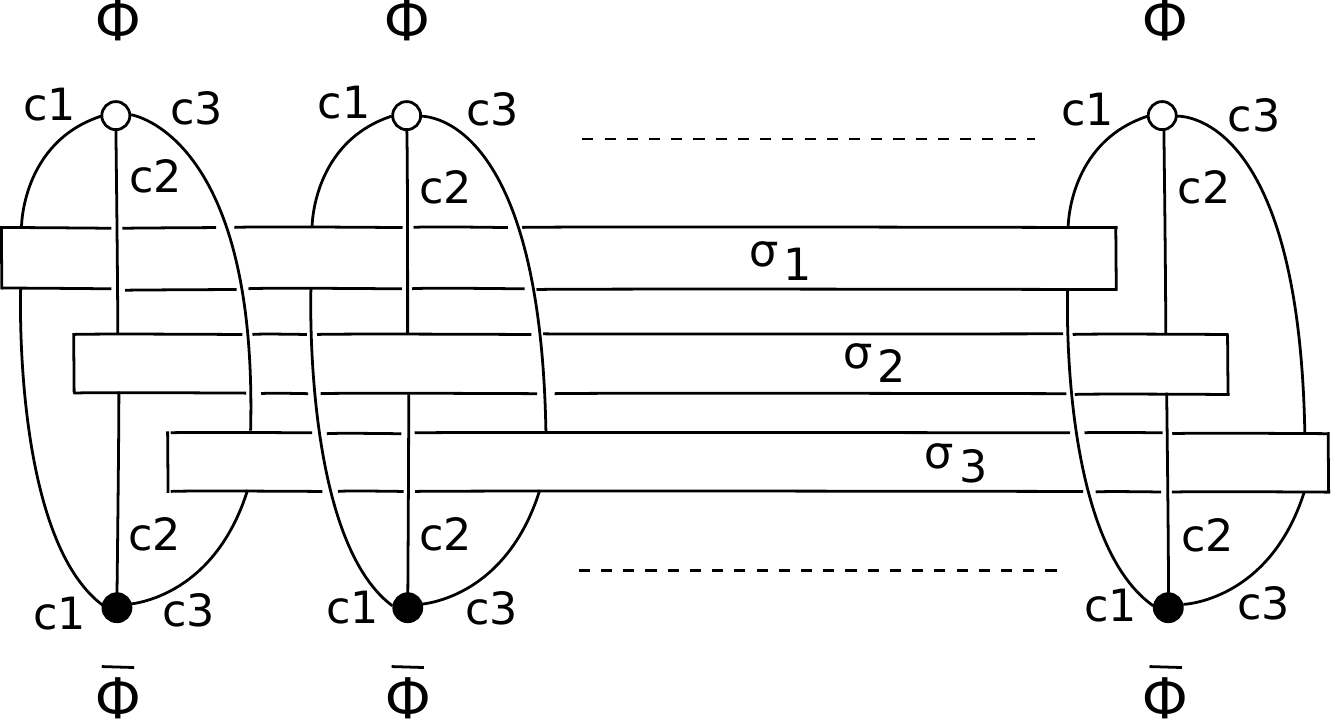}
\vspace{0.3cm}
\caption{ {\small Diagrammatic rank-3 tensor contraction defining $(\s_1,\s_2,\s_3)$}} 
\label{fig:sss0}
\end{minipage}
\end{center}
\end{figure}

The unitary invariants are called observables of  tensor models.
Take $n$ covariant tensor fields $\Phi$ and $n$ contravariant tensor fields $\bar\Phi$. 
Invariants of the unitary group action built from these are polynomial functions 
of the tensor variables which we will refer to as tensor invariants of degree $n$. 
The observables are constructed by contracting indices from the $n$ copies of $\Phi$ 
and  the $n$ copies of $ \bar \Phi$. The different contractions 
are labelled  by $d$ permutations, $ \sigma_1 , \sigma_2 , \dots , \sigma_d \in S_n $ 
and the corresponding observables are denoted $ \cO_{ \sigma_1 , \sigma_2 , \cdots , \sigma_d } (\Phi,\bar\Phi)$ (see Figure \ref{fig:sss0}).  There is an equivalence under right and left diagonal action of $S_n$ on $S_n^{\times d}$ as \cite{Sanjo}:
\bea \label{gau}
( \sigma_1 , \sigma_2 , \cdots , \sigma_d ) \sim ( \mu_1 \sigma_1 \mu_2 ,\mu_1 \sigma_2 \mu_2 , \cdots , \mu_1 \sigma_d \mu_2 ) 
\eea
where $\mu_{1,2}\in S_n$.
Equivalent permutation tuples give rise to the same observable. 
 Thus, counting observables is counting points in the double coset 
\be
 \diag ( S_n ) \ses  ( S_n \times S_n\times \cdots \times S_n ) / \diag ( S_n ) 
\ee
We denote the number of points in this double coset as $ Z_d(n)$. 
 Using the Burnside lemma, we obtain the counting \cite{Sanjo}: 
\be\label{counttf}
Z_{d}(n) = \frac{1}{(n!)^{2}}  \sum_{\mu_i \in S_n}\sum_{\s_i \in S_n} 
\prod_{i=1}^d\delta( \mu_1 \sigma_i \mu_2 \s_i^{-1}) 
\ee
which can be simplified to 
\be\label{sol3} 
Z_d(n) = \sum_{ p\, \vdash n }  ( \Sym ( p ) )^{d-2} \,,
\qquad 
\Sym(p):= \prod_{i=1}^n (i^{p_i})(p_i!)
\ee
where the sum is performed over partitions of $n$ specified by $p= \{p_1 , p_2 , \cdots , p_n  \} $  where the partition has $p_1$ copies of $1$, $p_2 $ copies of $2$ etc. so that   $n=\sum_i i p_i$.  A partition of $p$ of $n$, denoted $ p \vdash n $, specifies a cycle structure of permutations $ \sigma \in S_n$. A cycle structure corresponds to a conjugacy class in $S_n$. The conjugacy class corresponding to $p$ will be denoted $T_p$. 
 A permutation in $T_p$ has a symmetry factor, which is the number of $S_n$ permutations 
 leaving it unchanged under conjugation. This is denoted as $ \Sym ( \sigma )$ and is the same for any permutation $ \sigma \in T_p$, so we also denote this number as $ \Sym ( p)$. 
 
An important feature of  this paper is that we will extend the  correspondence between permutations in $ S_n^{ \times d }$ and observables 
\be
( \sigma_1 , \cdots , \sigma_d ) \rightarrow \cO_{ \sigma_1 , \cdots , \sigma_d } ( \Phi , \bar \Phi )  
\ee
to the group algebra $ (\mC ( S_n))^{ \otimes d }$ by linearity 
\be\label{mapPermsObs}  
\sum_{ \sigma_1 , \cdots , \sigma_d } 
\lambda^{ \sigma_1, \cdots , \sigma_d } 
( \sigma_1 \otimes \cdots \otimes \sigma_d )  \rightarrow 
\sum_{ \sigma_1 , \cdots , \sigma_d } 
\lambda^{ \sigma_1, \cdots , \sigma_d } \cO_{ \sigma_1 , \cdots , \sigma_d } ( \Phi , \bar \Phi ) 
\ee
We will consider sub-algebras  of $ (\mC ( S_n))^{ \otimes d } $  
defined by these equivalences (\ref{gau}). We  will describe special bases
in these sub-algebras constructed using representation theory, which reveal the matrix structure of these sub-algebras expected from the WA theorem.

We now introduce the second fundamental ingredient for our analysis, the measure over the complex tensors  
\be\label{GM}
d \mu( \Phi , \bar \Phi ) \equiv \prod_{i_k}
d\Phi_{i_1i_2 \dots i_d}d\bar \Phi_{i_1i_2 \dots i_d}\, e^{- \sum_{i_k}\Phi_{i_1i_2 \dots i_d}
\bar \Phi_{i_1i_2 \dots i_d}}
\ee
which defines the Gaussian tensor models of interest. 
Expectation values of observables $\cO_{ \sigma_1 , \sigma_2 , \cdots , \sigma_d}(\Phi,\bar\Phi) $ (which we will denote as  $\cO_{ \sigma_1 , \sigma_2 , \cdots , \sigma_d}$ for 
brevity) are defined as 
\bea 
\Big\langle \cO_{ \sigma_1 , \sigma_2 , \cdots , \sigma_d}\Big\rangle 
= { \int d \mu ( \Phi , \bar \Phi )  \cO_{ \sigma_1 , \sigma_2 , \cdots , \sigma_d} 
 \over \int d \mu ( \Phi , \bar \Phi )  } 
\eea
These correlators can be evaluated by summing over Wick contractions, which can be parametrized by permutations $ \gamma \in S_n $. 
Thus a graph configuration in the expansion of
$\la \cO_{ \sigma_1 , \sigma_2 , \cdots , \sigma_d} \ra$, is determined
by $(\gamma, \s_1,\s_2, \cdots, \s_d)$.  Summing over all $\gamma$'s
gives the full correlator. 
Fixing $N_a = N$, for all $a=1,\dots,d$, after some algebra one gets (see appendix \ref{app:correlators} for a derivation of the following equality)
\be\label{corrd} 
\Big\langle  \cO_{ \sigma_1 , \sigma_2 , \cdots , \sigma_d} \Big\rangle  = 
\sum_{ \gamma \in S_n } N^{ \cy ( \gamma \sigma_1 ) + \cy ( \gamma \sigma_2 ) + \dots + \cy (\gamma \sigma_d )   }
\ee
where $ \cy ( \alpha ) $  is the number of cycles of $\alpha$. 
This formula reflects the fact  that correlators depend on the total number of cycles of
compositions of the permutations  $\g \s_i$, $i=1,\dots, d$. 
The formula (\ref{corrd}) has also been obtained in \cite{MM1706}.
Another natural type of correlator considered in \cite{Sanjo} is the insertion of 
a product of observables in the integral, with the prescription that we do not allow 
Wick contractions within the observable. These are referred to as ``normal ordered'' correlators. They are discussed further in the appendix \ref{app:correlators} 
and section \ref{sect:PCAandCorr}. 

\section{The permutation centralizer  algebra $ \cK ( n ) $}
\label{sec:PCA}

In  this section we will show that the equivalence classes of permutations which  define tensor invariants of degree $n$ in the rank $d=3$ case form an associative algebra $ \cK ( n )$. The structure of the  algebra  $ \cK ( n )$ is intimately related to Kronecker coefficients. Its dimension  is equal to the sum of squares of Kronecker coefficients.  
For higher $d$, we have analogous algebras $ \cK ( n , d ) $ with dimensions equal to 
a sum of squares which can be expressed in terms of  higher order products of Kronecker coefficients. This is  shown below  in section \ref{sec:countKron}. 
In subsequent subsections, we will primarily focus the analysis to the rank $d=3$ 
case and the algebra $ \cK (n)$.    

\subsection{Counting observables and Kronecker coefficients}\label{sec:countKron} 
 
Consider the counting of tensor invariants of degree $n$.
 We rewrite the above counting 
\eqref{sol3} as 
\bea 
Z_3(n) = \sum_{ p \vdash n } \Sym ( p )   = 
 \sum_{ p \vdash n } { \Sym   ( p )  \over n!  }
 \sum_{ \sigma \in T_p  } \Sym  ( p )  
\eea
For each partition $p$ of $n$, we are summing over all the permutations in that conjugacy class $\sigma \in T_p$, and dividing 
by the size of the conjugacy class $T_p$, denoted $ |T_p|$. 
 We use the fact that 
\bea 
|T_p| = {  n! \over \Sym (p) }  
\eea
Using  identities in appendix \ref{sapp:syrep}, 
we write 
\bea 
 Z_3(n)  &=& \sum_{ p \vdash n } { \Sym  ( p )  \over n!  } \sum_{ \sigma \in T_p  } \Sym ( p )  
= { 1 \over n! }  \sum_{ \sigma \in S_n }  \Sym ( \sigma ) \Sym ( \sigma )  \cr 
&=&  { 1 \over n! } \sum_{ \sigma \in S_n}  \sum_{ \gamma_1, \g_2 \in S_n }   \delta ( \gamma_1 \sigma \gamma_1^{-1 } \sigma^{-1} ) 
\delta ( \gamma_2 \sigma \gamma_2^{-1} \sigma^{-1} ) 
  \cr 
& =&  { 1 \over n! } \sum_{ \sigma  \in S_n} \sum_{ R_1 , R_2\vdash n  } \chi^{ R_1  } ( \sigma )  \chi^{ R_1} ( \sigma ) \chi^{ R_2  } ( \sigma )  \chi^{ R_2} ( \sigma )  \cr 
&=&  { 1 \over (  n! )^2 }  \sum_{\gamma \in S_n}  \sum_{ \sigma_l  \in S_n } \sum_{ R_1 , R_2 \vdash n} \chi^{ R_1  } ( \sigma_1 )  \chi^{ R_2} ( \sigma_1 ) \delta ( \sigma_1 \gamma \sigma_2 \gamma^{-1} ) 
\chi^{ R_1  } ( \sigma_2  )  \chi^{ R_2} ( \sigma_2  )  \cr 
&=&  { 1 \over ( n! )^2 } \sum_{ \sigma_l  \in S_n } \sum_{ R_1 , R_2 \vdash n} \chi^{ R_1  } ( \sigma_1 )  \chi^{ R_2} ( \sigma_1 )
 \left ( \sum_{ S\vdash n } \chi^S ( \sigma_1 ) \chi^S ( \sigma_2 )  \right ) 
\chi^{ R_1  } ( \sigma_2  )  \chi^{ R_2} ( \sigma_2  )  \cr 
&=&  \sum_{ R_1 , R_2 , S \vdash n} (  \sC ( R_1 , R_2 , S ) )^2  
\label{coefKon3}
\eea
where the symbol 
\be 
\sC ( R_1 , R_2 , R_3 ) = \frac{1}{n!}
\sum_{\s \in S_n}  \chi^{ R_1  } ( \sigma )  \chi^{ R_2} ( \sigma) 
\chi^{ R_3} ( \sigma) 
\ee 
 is the Kronecker coefficient or multiplicity 
of the irreducible representation (irrep)  $R_3$ in the tensor product of the
 irreps  $R_1$ and  $R_2$. Equivalently it is the multiplicity of the one-dimensional 
 representation in the tensor product $ R_1 \otimes R_2 \otimes R_3$. 
Similar manipulations show that the same counting is also equal to 
\be
Z_3(n)  =
\sum_{ R_1 , R_2 , S \vdash n} \sC ( R_1 , R_1, S ) \sC ( R_2  , R_2 , S )  
\ee
Hence,  counting observables of  tensor model of rank 3 coincides
with a sum of squares (or product) of Kronecker coefficients. That sum is also 
the dimension of an algebra $\cK ( n )$ that we will discuss in the next section.  
The connection between the counting of  tensors and Kronecker coefficients 
has also been discussed in the physics literature in \cite{PCA1601,DR1706,DGT1707} 
and in the mathematics literature in \cite{HeWi09}. For future reference in this paper, a key point from the above discussion is 
\bea\label{dimKKron} 
 \dim ( \cK ( n ) ) =  \sum_{ R_1 , R_2 , S \vdash n} (  \sC ( R_1 , R_2 , S ) )^2
\eea

\noindent
{\bf Counting in rank-$d$ tensors -}
The above counting generalizes quite naturally at any rank as 
\bea 
 Z_d(n)  &=& \sum_{ p \vdash n } { \Sym  ( p )^{d-2}  \over n!  } \sum_{ \sigma \in T_p  } \Sym ( p )  
= { 1 \over n! }  \sum_{ \sigma \in S_n }  \Sym ( \sigma )^{d-1} \cr 
&=&  { 1 \over n! } \sum_{ \sigma \in S_n}  \sum_{ \gamma_1, \g_2, \dots,\g_d  \in S_n }   \delta ( \gamma_1 \sigma \gamma_1^{-1 } \sigma^{-1} ) 
\delta ( \gamma_2 \sigma \gamma_2^{-1} \sigma^{-1} ) 
\dots \delta ( \gamma_{d-1} \sigma \gamma_{d-1}^{-1} \sigma^{-1} )
  \cr 
&=&  { 1 \over n! } \sum_{ \sigma \in S_n} \prod_{i=1}^{d-1}\Big[\sum_{ R_i \vdash n  } \chi^{ R_i} ( \sigma ) \chi^{ R_i} ( \sigma )\Big]
\eea
where we used \eqref{delgsg} of appendix \ref{sapp:syrep}.
Then we re-introduce delta-functions which couple different permutations 
as: 
\bea
 Z_d(n) &=&  { 1 \over (  n! )^2 }  \sum_{\gamma \in S_n}  \sum_{ \sigma_l  \in S_n } \sum_{ R_1 , R_2 \vdash n} \chi^{ R_1  } ( \sigma_1 )  \chi^{ R_1} ( \sigma_1 ) \delta ( \sigma_1 \gamma \sigma_2 \gamma^{-1} ) 
\chi^{ R_2 } ( \sigma_2 )  \chi^{ R_2} ( \sigma_2  ) \crcr
&& \times 
\prod_{i=3}^{d-1}\Big[\sum_{ R_i \vdash n  } \chi^{ R_i} ( \sigma_2 ) \chi^{ R_i} ( \sigma_2 )\Big]
  \cr 
&=&  { 1 \over ( n! )^3 }
 \sum_{ \sigma_l  \in S_n } \sum_{ R_1 , R_2 \vdash n} \chi^{ R_1  } ( \sigma_1 )  \chi^{ R_1} ( \sigma_1 )
 \Big( \sum_{ S_1 \vdash n } \chi^{S_1} ( \sigma_1 ) \chi^{S_1}  ( \sigma_2 )  \Big) 
\chi^{ R_2  } ( \sigma_2  )  \chi^{ R_2} ( \sigma_2  )  \cr 
&&\times 
\Big[ \sum_{ S_2\vdash n } \chi^{S_2} ( \sigma_2 ) \chi^{S_2} ( \sigma_3 )  \Big] 
\chi^{ R_3} ( \sigma_3 ) \chi^{ R_3} ( \sigma_3 )
\prod_{i=4}^{d-1}\Big[\sum_{ R_i \vdash n  } \chi^{ R_i} ( \sigma_3 ) \chi^{ R_i} ( \sigma_3 )\Big] \crcr
&=&\sum_{ R_i , S_i \vdash n} \sC ( R_1 , R_1 , S_1 ) 
\sC_4 ( R_2 , R_2 , S_1,S_2 )  
\sC_4 ( R_3 , R_3 , S_2,S_3 )  
\dots  \crcr
&& 
\times 
\sC_4 ( R_{d-2} , R_{d-2} , S_{d-3},S_{d-2} )  
\sC ( R_{d-1} , R_{d-1} ,S_{d-2} )  \crcr
&=&\sum_{ R_i , S_i \vdash n} \sC ( R_1 , R_1 , S_1 ) 
\Big[\prod_{i=2}^{d-2}
\sC_4 ( R_i , R_i , S_{i-1},S_{i} )   
\Big] \sC ( R_{d-1} , R_{d-1} ,S_{d-2} )  
\eea
where the symbol $\sC_4(\cdot)$ stands for
\bea
\sC_4 ( R_1 , R_2 , R_3,R_4 ) &=& \frac{1}{n!}\sum_{\s\in S_n}
\chi^{R_1}(\s)\chi^{R_2}(\s)\chi^{R_3}(\s)\chi^{R_4}(\s)
 \cr
&=& { 1 \over  (n!)^2 }  \sum_{ \gamma , \sigma_1 , \sigma_2 \in S_n} 
\chi^{ R_1} ( \sigma_1  ) \chi^{R_2} ( \sigma_1  ) \left (  \delta ( \sigma_1 \gamma \sigma_2 \gamma^{-1} ) \right )  
\chi^{ R_3} ( \sigma_2  ) \chi^{R_4} ( \sigma_2  )   \cr 
&=& { 1 \over (n!)^2 }   \sum_{  \sigma_1 , \sigma_2\in S_n } 
\chi^{ R_1} ( \sigma_1  ) \chi^{R_2} ( \sigma_1  ) 
\left (  \sum_{ S\vdash n } \chi^{ S } ( \sigma_1 ) \chi^S ( \sigma_2 )  \right ) 
\chi^{ R_3 } ( \sigma_2  ) \chi^{R_4} ( \sigma_2  ) \cr 
&=& \sum_{S\vdash n}\sC(R_1,R_2,S)\sC(S,R_3,R_4) 
\label{c4=c32}
\eea 
Thus at any rank the counting of observables of tensor
models maps to a sum of products of Kronecker coefficients. 
For example, at rank $d=4$, we obtain
\be
 Z_d(4)  = \sum_{ R_i , S_i \vdash n} 
\sC(S_2, S_{1},S_{3} )  
\sC ( R_1 , R_1 , S_1 ) 
\sC ( R_2 , R_2,S_2)  
\sC ( R_{3} , R_{3} ,S_{3} ) 
\ee
To write a compact formula as a sum of squares, we introduce 
\bea 
\sC_{ k } ( R_1 , R_2 , \cdots , R_k  ) = 
{ 1 \over n! } \sum_{ \sigma \in S_n } \chi^{ R_1} ( \sigma )  \chi^{ R_2} ( \sigma )
\cdots \chi^{ R_k} ( \sigma )  
\eea
This counts the multiplicity of the one-dimensional  $S_n$ irrep in the tensor product of irreps $ R_1 \otimes \cdots \otimes R_k$. It can be expressed in terms of products of Kronecker coefficients. The dimension of $ \cK ( n , d )$ is 
\bea\label{Kndsumsqrs} 
Z_d ( n ) = \dim ( \cK ( n , d )) = \sum_{ R_1 , \cdots , R_{d-1} , S \vdash n } 
 (\sC_d  ( R_1 , R_2 , \cdots , R_{d-1} , S )  )^2 
\eea

\subsection{$\cK(n)$ as a centralizer algebra in  $ \mC ( S_n ) \otimes \mC ( S_n )$} 
\label{sec:algKn} 

The permutation equivalence classes in $ S_n \times S_n \times S_n $ described earlier (\ref{gau}) have a gauge-fixed formulation involving pairs of permutations. One way to see this \cite{Sanjo} is by manipulating the symmetric group delta functions which implement the Burnside lemma counting. For example, we can choose $ \mu_1 = \sigma_1^{-1} $ which maps  the triple 
\bea 
( \sigma_1 , \sigma_2 , \sigma_3 ) \rightarrow ( 1 , \sigma_1^{-1} \sigma_2 , \sigma_1^{-1} \sigma_3 ) \equiv ( 1 , \tau_1 , \tau_2 ) 
\eea
The $ \mu_2 = \mu  $  equivalence now acts on  $ ( \tau_1 , \tau_2 )  $ as 
\bea 
( \tau_1 , \tau_2  ) \sim  ( \mu \tau_1 \mu^{-1} , \mu \tau_2 \mu^{-1} ) 
\eea
We will therefore define  $\cK(n)$ as the 
 sub-algebra of group algebra $ \mC ( S_n ) \otimes \mC ( S_n )$ 
 which is invariant under conjugation by  the diagonally embedded   
$  S_n    $. In this section, we detail the structure
of this sub-algebra.

Consider the elements of $ \mC ( S_n ) \otimes \mC ( S_n )$  
obtained by starting with a tensor product  $ \sigma_1 \otimes \sigma_2 $ 
and summing all their conjugates by $ \gamma$ acting diagonally 
\be \label{invgam}
\sigma_1 \otimes \sigma_2 \rightarrow \sum_{ \gamma \in S_n } \gamma \sigma_1 \gamma^{-1} \otimes 
\gamma \sigma_2 \gamma^{-1} 
\ee
Now, $\cK(n) \subset  \mC ( S_n ) \otimes \mC ( S_n )$ 
 is the  vector space over $\mC$ spanned by all $\sum_{ \gamma \in S_n } \gamma \sigma_1 \gamma^{-1} \otimes  \gamma \sigma_2 \gamma^{-1} $, 
$\s_1$ and $\s_2 \in S_n$: 
\be
\cK(n) = {\rm Span}_{\mC}\Big\{    
\sum_{ \gamma \in S_n } \gamma \sigma_1 \gamma^{-1} \otimes  \gamma \sigma_2 \gamma^{-1} , \; \s_1, \s_2 \in S_n
\Big\}
\label{graphbasis}
\ee
By construction, the elements of $\cK(n)$ are invariants under the action 
of $\diag(\mC(S_n))$. To verify this, we evaluate any element $A\in \cK(n)$
\be
(\tau \otimes \tau ) \cdot A \cdot (\tau^{-1} \otimes \tau^{-1} ) = 
\sum_{\s_1,\s_2,\g \in S_n} c_{\s_1,\s_2} \; 
\tau \gamma \sigma_1 \gamma^{-1} \tau^{-1}\otimes  
\tau \gamma \sigma_2 \gamma^{-1} \tau^{-1} = A 
\ee
where we redefine $\g \to \tau\gamma$.

\begin{proposition}
$\cK(n)$ is an associative unital sub-algebra of $ \mC ( S_n ) \otimes \mC ( S_n )$. 
\end{proposition}
\proof 
We verify that $\cK(n)$ is a closed under multiplication. 
Take two elements of $\cK(n)$, $A = \sum_{\s_1,\s_2,\g_1 \in S_n} c_{\s_1,\s_2}
\gamma_1 \sigma_1 \gamma_1^{-1} \otimes  \gamma_1 \sigma_2 \gamma_1^{-1}
$ and $B = \sum_{\tau_1,\tau_2,\g_2 \in S_n} c'_{\tau_1,\tau_2}
\gamma_2 \tau_1 \gamma_2^{-1} \otimes  \gamma_2 \tau_2 \gamma_2^{-1}
$, with  coefficients $c_{\s_1,\s_2}$ and $c'_{\tau_1,\tau_2}$, 
\bea 
A B
& =& \sum_{\s_i,\tau_i \in S_n} c_{\s_1,\s_2} c'_{\tau_1,\tau_2} \sum_{ \gamma_1,  \gamma_2 } \gamma_1 \sigma_1 \gamma_1^{-1}\gamma_2 \tau_1 \gamma_2^{-1}
\otimes \gamma_1 \sigma_2 \gamma_1^{-1} \gamma_2 \tau_2 \gamma_2^{ -1} \cr 
& =&
 \sum_{ \gamma } \sum_{\s_i,\tau_i \in S_n} c_{\s_1,\s_2} c'_{\tau_1,\tau_2}   \sum_{ \gamma_1 }  \gamma_1 ( \sigma_1 \gamma \tau_1 \gamma^{-1} )\gamma_1^{-1} 
\otimes \gamma_1 ( \sigma_2 \gamma \tau_2 \gamma^{-1})  \gamma_1^{-1}  
\eea
where we redefined $ \gamma = \gamma_1^{-1} \gamma_2 , \gamma_2^{-1} = \gamma^{-1} \gamma_1^{-1} $. Clearly, the last line shows that
$AB$ belongs to $\cK(n)$ as a linear combination of
basis elements. 

The unit of $\mC(S_n)^{\otimes 2}$ is $id \otimes id$ which also belongs to $\cK(n)$. One can also check that $\cK(n)$ is an associative algebra because $\mC(S_n)$ is associative. 
Hence $\cK(n)$ is a sub-algebra of $ \mC ( S_n ) \otimes \mC ( S_n )$. 

\qed

The dimension of $\cK(n)$  is associated with the number of observables of  tensor models. 
Indeed, each colored tensor graph is associated with an equivalence relation \eqref{invgam}
also associated with one basis element of $\cK(n)$.
Thus $\dim_{\mC}\cK(n)  =  Z_3(n)$.  In the following,
 the elements of $\cK(n)$ are then called as and identified with ``graphs'' 
and the basis \eqref{graphbasis} will be called graph-basis.

\ 

\noindent{\bf A Fourier basis of invariants -}
The Fourier transform of the
basis \eqref{invgam} of $\cK(n)$ determines another basis of invariants for $\cK(n)$. The elements of this basis are labelled by $ ( R , S , T , \tau_1 , \tau_2 ) $ and are of the form
\be\label{qbasis}
Q^{R,S,T}_{\tau_1,\tau_2} = 
\kappa_{R,S}
\sum_{\s_1, \s_2 \in S_n}
\sum_{i_1,i_2,i_3, j_1,j_2}
C^{R , S ; T , \tau_1  }_{ i_1 , i_2 ; i_3 } C^{R , S ; T , \tau_2  }_{ j_1 , j_2 ; i_3 } 
 D^{ R }_{ i_1 j_1} ( \sigma_1  ) D^S_{ i_2 j_2 } ( \sigma_2 )  \,  \sigma_1 \otimes \sigma_2 
\ee
where $\kappa_{R,S} = \frac{d(R)d(S)}{(n!)^2}$, $i_1$ and $j_1$ (resp. $i_2$ and $j_2$) are positive integers bounded by 
the dimension $d(R)$ (resp. $d(S)$) of the representation of $S_n$,
and $i_3$ by $d(T)$. Meanwhile, $C^{R , S ; T , \tau_1  }_{ i_1 , i_2 ;i_3 } $ are Clebsch-Gordan coefficients involved in the tensor products
of representations of $S_n$, see appendix \ref{sapp:cgc} for 
a brief definition and properties that we will use hereafter; 
the multiplicities $\tau_1$ and  $\tau_2\in  [\![1, \sC(R,S,T) ]\!]$. 
We can check that, by acting by the diagonal action, the 
 basis elements are invariant (for the proof see \eqref{inq1}
in appendix \ref{app:baskn}): 
\be\label{qinv}
(\gamma\otimes \gamma)\cdot Q^{R,S,T}_{\tau_1,\tau_2} \cdot
(\gamma^{-1}\otimes \gamma^{-1}) = Q^{R,S,T}_{\tau_1,\tau_2} 
\ee

The basis $\{Q^{R,S,T}_{\tau_1,\tau_2}\}$, shortly called $Q$-basis in the following,  
 makes explicit that the dimension of the algebra  $\cK(n)$  is given by 
\be
\sum_{ R , S , T } \sC ( R , S , T )^2 
\ee
An important property of the $Q^{R,S,T}_{\tau_1,\tau_2}$'s is that
they are matrix bases of $\cK(n)$. We have (the proof of the
following is detailed in \eqref{qqmatrix} of appendix \ref{app:baskn}): 
\be
Q^{R,S,T}_{\tau_1,\tau_2}Q^{R',S',T'}_{\tau_2',\tau_3} 
=  \delta_{RR'} \delta_{SS'}  \delta_{TT'}\delta_{\tau_2 \tau_2'}  Q^{R,S,T}_{\tau_1,\tau_3}  
\label{qqmatrix0}
\ee
Finally, noting that $\sC(R,S,T)$ is at most 1 for $n\leq 4$, 
then the matrices $Q^{R,S,T}_{\tau_1,\tau_2}$ are $1\times 1$
hence are commuting. Consequently, at lower order
in $n\leq 4$, $\cK(n)$ is commutative. 

\

\noindent{\bf Orthogonality of the $Q$-basis -} 
Consider the bilinear pairing $\bdel_2 : \mC(S_n)^{\otimes d}\times 
\mC(S_n)^{\otimes d} \to \mC$, $d\geq 0$,  
\be
\bdel_d (  \otimes_{i=1}^d \s_i ; \otimes_{i=1}^d \s'_i ) = 
\prod_{i=1}^d \delta (\s_i\s'^{-1}_i) 
\ee
which extends to linear combination with complex coefficients 
naturally: 
\be
\bdel_d \Big(\sum_{\s_l}c_{\{\s_l\}}\otimes_{l=1}^d \s_l ;
\sum_{\s'_l}c'_{\{\s'_l\}} \otimes_{l=1}^d \s'_{l} \Big)  = 
\sum_{\s_l} c_{\{\s_l\}} c'_{\{\s_l\}}  
\ee

We can also consider an inner product, i.e. a sesquilinear pairing, where we would have 
$ \bar c_{\{\s_l\}}$ on the r.h.s. above. The inner product will have the same non-degeneracy property we discuss below for the bilinear form. 

The following proposition can be easily checked. 
\begin{proposition}
$ \bdel_{ d } $ is a non-degenerate pairing on $ \mC(S_n)^{\otimes d} $
 $\forall d\geq 1$. 
\end{proposition}

 Inspecting the pairing of basis elements of $\{ \sum_{\g}\gamma \sigma_1 \gamma^{-1}  \otimes \gamma \sigma_2 \gamma^{-1}\}$ in $\cK(n)$, we have
\begin{align}
&
\bdel_{2}( \sum_{\g_1}\gamma_1 \sigma_1 \gamma_1^{-1}  \otimes \gamma_1 \sigma_2 \gamma_1^{-1} ;   \sum_{\g_2}\gamma_2 \tau_1 \gamma_2^{-1} \otimes \gamma_2 \tau_2 \gamma_2^{ -1} )= \crcr
 &
\sum_{\g_1,\g_2}
\delta( \gamma_1 \sigma_1 \gamma_1^{-1}  (\gamma_2 \tau_1 \gamma_2^{-1})^{-1} )
\delta( \gamma_1 \sigma_2  \gamma_1^{-1} (\g_2 \tau_2 \gamma_2^{-1})^{ -1} )\crcr
&
= \sum_{\g_1,\g_2}
\delta( \gamma_1 \sigma_1 \gamma_1^{-1}  \gamma_2 \tau_1^{-1} \gamma_2^{-1} )
\delta( \gamma_1 \sigma_2  \gamma_1^{-1} \g_2 \tau_2^{-1} \gamma_2^{-1}  )\crcr
\crcr
& = 
\sum_{\g_1,\g_2}
\delta( \gamma_1 \sigma_1  \gamma_2 \tau_1^{-1}(\g_1\gamma_2)^{-1} )
\delta( \gamma_1 \sigma_2  \gamma_2 \tau_2^{-1} (\g_1\gamma_2)^{ -1} ) \crcr
&
=n!\sum_{\g}
\delta(  \sigma_1  \gamma \tau_1^{-1}\gamma^{-1} )
\delta(  \sigma_2  \gamma \tau_2^{-1} \gamma^{ -1} )
\end{align}
which is not vanishing whenever the $\s_i$'s are conjugate to $\tau_i$'s, $i=1,2$. 
This is precisely saying that the two basis elements and the corresponding
graphs are in the same class. The above  sum over $\g$ computes to the 
order of the automorphism group of the graph associated with 
any of the basis element. 
Consider two colored tensor graphs $G_{\s_1,\s_2}$ 
and $G_{\s_1',\s_2'}$ associated with 
the basis elements 
$\sum_{\g}\gamma \sigma_1 \gamma^{-1}  \otimes \gamma \sigma_2 \gamma^{-1}$
and 
$\sum_{\g}\gamma \sigma_1' \gamma^{-1}  \otimes \gamma \sigma_2' \gamma^{-1}$, 
respectively, then we write 
\be\label{delgg=aut}
\bdel_{2}( \sum_{\g_1}\gamma_1 \sigma_1 \gamma_1^{-1}  \otimes \gamma_1 \sigma_2 \gamma_1^{-1} ;   \sum_{\g_2}\gamma_2 \tau_1 \gamma_2^{-1} \otimes \gamma_2 \tau_2 \gamma_2^{ -1} )=n!\delta(G_{\s_1,\s_2},G_{\s'_1,\s'_2}) |\Aut(G_{\s_1,\s_2})| 
\ee
where $\delta(G_{\s_1,\s_2},G_{\s'_1,\s'_2})=1$ if the graphs
are equivalent and 0 otherwise, and $|\Aut(G_{\s_1,\s_2})|$ is the order the automorphism 
group $\Aut(G_{\s_1,\s_2})$ of the graph $G_{\s_1,\s_2}$. 
In the end, the restriction of $\bdel_2 $ to  $\cK(n)$ is non degenerate and
the basis of invariants is orthogonal with respect to that product. 
The following statement is therefore obvious 
\begin{proposition}
$\cK(n)$ is an associative unital  semi-simple algebra. 
\end{proposition}
Semi-simple algebras and their isomorphism  to 
a direct sum of matrix algebras (WA theorem) are explained in 
 \cite{wallach,ram}. 

The $Q$-basis proves to be orthogonal with respect to the bilinear pairing 
 $\bdel_2$ (see \eqref{orhoqq} of appendix \ref{app:baskn})
\be
\bdel_2 (Q_{\tau_1,\tau'_1}^{R,S,T}; Q_{\tau_2,\tau'_2}^{R',S',T'})
 =  \kappa_{R,S} d(T)\,  \delta_{RR'} \delta_{SS'}
\delta_{TT'} \delta_{\tau_1\tau_2} \delta_{\tau'_1\tau'_2}    
\label{orhoqq0}
\ee
Note that we could have changed the normalization $\kappa_{R,S}$
of $Q^{R,S,T}_{\tau_1,\tau_2}$ to make that
basis orthonormal with respect to $\bdel_2$. However, the previous choice
of making as simple as possible the matrix multiplication of the $Q$'s
has fixed the normalization $\kappa_{RS}$. Another
option to make the bilinear pairing  of $Q$'s  normalized 
is to change the definition of the pairing  itself,
but we will keep the present definition of $\bdel_d$ for simplicity. 
The orthogonality relation \eqref{orhoqq0} reveals 
that the basis $\{Q_{\tau,\tau'}^{R,S,T}\}$ decomposes $\cK(n)$
in orthogonal blocs labelled by $(R,S,T)$ and for each such triple
an orthogonal square bloc labelled by $(\tau,\tau')$.

We can address the expansion of the graph-basis 
in terms of the $Q$-basis (inverse transform):  
\be
\sum_{\g} \g \s_1 \g^{-1} \otimes  \g \s_2 \g^{-1} 
 =
 \sum_{R,S,T,\tau_1,\tau_2}
\Big[\sum_{\g} 
\bdel_2(Q^{R,S,T}_{\tau_1,\tau_2}; 
\g \s_1 \g^{-1} \otimes  \g \s_2 \g^{-1} )\Big] 
Q^{R,S,T}_{\tau_1,\tau_2}
\ee
where the coefficients calculate as 
\bea
&& 
\sum_{\g} 
\bdel_2(Q^{R,S,T}_{\tau_1,\tau_2}; 
\g \s_1 \g^{-1} \otimes  \g \s_2 \g^{-1} ) 
= \sum_{\g} 
\bdel_2( (\g \otimes \g )\cdot Q^{R,S,T}_{\tau_1,\tau_2}\cdot (\g^{-1}\otimes \g^{-1}); \s_1  \otimes  \s_2  )\cr\cr
&& 
=\sum_{\g}\bdel_2(  Q^{R,S,T}_{\tau_1,\tau_2}; \s_1  \otimes  \s_2  ) = 
n! \kappa_{R,S}
\sum_{i_l, j_l}
C^{R , S ; T , \tau_1  }_{ i_1 , i_2 ; i_3 } C^{R , S ; T , \tau_2  }_{ j_1 , j_2 ; i_3 } 
 D^{ R }_{ i_1 j_1} ( \sigma_1  ) D^S_{ i_2 j_2 } ( \sigma_2 )  
\label{projGQ}
\eea
where used has been made of \eqref{qinv}, namely the invariance of the
$Q$-basis. 
The coefficient \eqref{projGQ}  can be interpreted  as the projection of a  graph onto the $Q$-basis. 

\subsection{The centre $\cZ(\cK(n))$ of $\cK(n)$}
\label{sect:ZKn}
 
Using the basis elements $Q^{R,S,T}_{\tau_1,\tau_2}$, we build elements of 
the centre $\cZ(\cK(n))$ of $\cK(n)$ by taking their trace at fixed
$(R,S,T)$:
\be
P^{R,S,T} = \sum_{\tau} Q^{R,S,T}_{\tau,\tau} 
\ee
To prove that $P^{R,S,T}$ is in the centre $\cZ(\cK(n))$, it is sufficient
to show that it is commuting with the basis elements $Q^{R,S,T}_{\tau_1,\tau_2}$  of $\cK(n)$:  
\begin{align}
&
 Q^{R',S',T'}_{\tau_1,\tau_2 } \cdot P^{R,S,T} 
= \sum_{\tau}  Q^{R',S',T'}_{\tau_1,\tau_2} Q^{R,S,T}_{\tau,\tau} 
 = 
\sum_{\tau} 
 \delta_{RR'} \delta_{SS'}  \delta_{TT'}\delta_{\tau_2 \tau}  Q^{R,S,T}_{\tau_1,\tau}  
=
\delta_{RR'} \delta_{SS'}  \delta_{TT'} Q^{R,S,T}_{\tau_1,\tau_2} 
 \crcr
& 
 =  \sum_{\tau} \delta_{RR'} \delta_{SS'}  \delta_{TT'} \delta_{\tau\tau_1}
Q^{R',S',T'}_{\tau,\tau_2}
=P^{R,S,T}  \cdot Q^{R',S',T'}_{\tau_1,\tau_2 }     
\end{align}
The orthogonality  of the $P$'s follows from the orthogonality
of the $Q$-basis \eqref{orhoqq0}:  
\bea
&&
\bdel_{2}(P^{R,S,T}; P^{R',S',T'})
=\sum_{\tau,\tau'} \bdel_{2}(Q^{R,S,T}_{\tau,\tau};Q^{R',S',T'}_{\tau',\tau'})  = \cr\cr
&& 
 \kappa_{R,S}\, d(T)\,  \delta_{RR'} \delta_{SS'}
\delta_{TT'} \sum_{\tau,\tau'}  \delta_{\tau\tau'}
 = \kappa_{R,S} \,d(T)\,\sC(R,S,T)  \delta_{RR'} \delta_{SS'}
\delta_{TT'}  
\eea

\begin{proposition}\label{prop:PgenZ}
The set $\{P^{R,S,T}\}$ is a basis of  $\cZ(\cK(n))$
and 
\be
\dim \cZ(\cK(n)) = \text{ number of non vanishing Kronecker coefficients }
\ee
\end{proposition}
 \proof  $\cK(n)$ decomposes in irreducible blocs 
labelled by $(R,S,T)$ and, associated with each of the triples, 
a matrix $Q^{R,S,T}_{\tau,\tau'}$. In that vector 
space, for a given $(R,S,T)$, 
 $P^{R,S,T}$ is the sum of diagonal elements of 
 $Q^{R,S,T}_{\tau,\tau'}$. Collecting all possible
diagonals hence $P^{R,S,T}$ spans the centre $\cZ(\cK(n))$. 

The dimension of $\cZ(\cK(n))$ is given by the number
of non vanishing Kronecker coefficients: a triple $(R,S,T)$,
such that $\sC(R,S,T)\ne 0$ yields a non vanishing $Q^{R,S,T}_{\tau,\tau'}$ and contributes to a single $P^{R,S,T}$. The result on the dimension
of $\cZ(\cK(n))$ follows. 

\qed

\noindent{\bf An overcomplete basis of central elements -}
Here we will show how we can start with a triple of irreps of  $ \mC ( S_n)$ 
and construct central elements of $ \cK ( n)$ from them. These will form an overcomplete basis of central elements, which will be demonstated by taking the pairing 
with the basis $P^{R,S,T}$ described above. 

First consider 
a partition $R$ of $n$ and  the element  
$z_{R} = \sum_{\s} \chi^{R}(\s)\s$  that is 
central in $\mC(S_n)$. Indeed for any $R \vdash n$, 
choose $\g \in S_n$ any arbitrary basis element of 
$\mC(S_n)$, and calculate: 
\be\label{gzg}
\g \,z_{R} \,\g^{-1}  = \sum_{\s} \chi^{R} (\s)\g \s \g^{-1}  =  \sum_{\s} \chi^{R} (\g \s \g^{-1}) \s =  z_{R}  
\ee
Let $R_1$ and $R_2$ be  two partitions of $n$ from which
we introduce  the central elements $z_{R_i} = \sum_{\s} \chi^{R_i}(\s)\s$, $i=1,2$, then build
\be
z_{R_1,R_2} = z_{R_1}\otimes z_{R_2} = \sum_{\s_1, \s_2 \in S_n}
\chi^{R_1} (\s_1)\chi^{R_2} (\s_2) \,
\s_1 \otimes \s_2  
\ee
that one can show to be central because is a tensor product
of central elements (use \eqref{gzg} twice on each sector). 

Another possible element of the centre obtained from a
single partition $R \vdash n$ is
\be
z_R = \sum_{\s \in S_n} \chi^{R}(\s)\, \s \otimes \s 
\ee
One can quickly verify that $z_{R_1,R_2;R_3} = z_{R_1,R_2} \cdot z_{R_3} \in \cK(n)$: 
\bea
&& 
(\g \otimes \g)\cdot 
z_{R_1,R_2;R_3}  \cdot (\g^{-1} \otimes \g^{-1})
 = 
 \sum_{\s_i \in S_n} \chi^{R_1}(\s_1) \chi^{R_2}(\s_2)
 \chi^{R_3}(\s_3)\, \g\s_1\s_3 \g^{-1}\otimes \g \s_2\s_3 \g^{-1} \cr\cr
&&
= \sum_{\s_i \in S_n} \chi^{R_1}(\s_1) \chi^{R_2}(\s_2)
 \chi^{R_3}(\s_3)\, \g\s_1\s_3 \g^{-1}\otimes \g \s_2\s_3 \g^{-1}\cr\cr
&& 
 = \sum_{\s_i \in S_n} \chi^{R_1}(\s_1) \chi^{R_2}(\s_2)
 \chi^{R_3}(\s_3)\, \s_1\s_3 \otimes \s_2\s_3  
 = z_{R_1,R_2;R_3} 
\eea
where we change variables $\g \s_{i=1,2} \to \tilde\s_{i=1,2}\g$,
 $\g \s_3 \g^{-1}\to \tilde\s_3$, and rename 
$\tilde \s_{i=1,2,3} $ as $\s_{i=1,2,3} $. 

We arrive at the following statement:  
\begin{proposition}
The set $\{z_{R_1,R_2;R_3}\}$, with $R_i\vdash n$, is an overcomplete basis of the centre $\cZ(\cK(n))$. 
\end{proposition}
\proof
We project $P^{R_1,R_2,R_3}$ onto $z_{R'_1,R'_2;R'_3}$ and
check that the coefficients are not vanishing (see details in \eqref{eq:Pz} 
in appendix \ref{app:baskn}): 
\be
\bdel_2(P^{R_1,R_2,R_3};z_{R'_1,R'_2;R'_3})  
=
 n!\, \delta_{R_1R_1'}\delta_{R_2R_2'}\delta_{R_3R_3'}
\sC(R_1,R_2,R_3)  
\ee
Hence $P^{R_1,R_2,R_3}$ admits a decomposition in terms
of $z_{R'_1,R'_2;R'_3}$. The overcompleteness follows from the number of elements of 
$\{z_{R'_1,R'_2;R'_3}\}$ is $p(n)^3$ which 
is larger than the number of non vanishing Kroneckers
the dimension of $\cZ(\cK(n))$. 

\qed

A general study of central elements in algebras constructed as subgroup-centralizers in a group algebra is given in \cite{DEM2013}. 

\subsection{Double coset algebra }
\label{sect:ungaug}

The algebra $ \cK ( n)$ introduced in the previous sections as a 
sub-algebra of $ \mC ( S_n) \otimes \mC ( S_n)$ has another description 
as a sub-algebra of $ \mC ( S_n) \otimes \mC ( S_n) \otimes \mC ( S_n) $. 
As we will see shortly,  in this latter description, we have an algebra of double cosets. 
The former description as a centralizer algebra  is a gauge-fixed version. 
Hence the double coset description is an un-gauge-fixed version. 
 For this reason, we will refer to the double coset algebra in this section as 
 $\cK_{\ung}(n)$ and establish its isomorphism with $ \cK ( n)$. 
In the rest of the paper, we will use $ \cK ( n)$ for either description of the algebra, 
and it will be clear from the context whether we are working with the gauge-fixed 
(centralizer algebra)  or un-gauge-fixed (double coset) description.
 While the centralizer algebra  is a more economical description, being embedded in a 
smaller algebra, $ \cK_{ \ung} ( n)$ arises more immediately from inspection of the permutation equivalences relevant to  tensor models, as reviewed in section \ref{sect:Rev}.

\

\noindent{\bf  $\cK_{\ung}(n)$  as a double coset algebra in  $ \mC (S_n)^{ \otimes 3} $ -}
Consider elements $\sigma_1 \otimes \sigma_2  \otimes \sigma_3 \in \mC(S_n)^{\otimes 3}$ and the left and right actions of $\diag(\mC (S_n))$ on these triples as: 
\be
\sigma_1 \otimes \sigma_2  \otimes \sigma_3 \to 
\sum_{\g_1,\g_2 \in S_n} \g_1 \sigma_1 \gamma_2  \otimes \gamma_1 \sigma_2 \gamma_2 \otimes \gamma_1  \sigma_3 \gamma_2   
\ee
 $\cK_{\ung}( n ) $ is the vector space and sub-algebra of 
$ \mC ( S_n ) \otimes \mC ( S_n ) \otimes \mC ( S_n ) $ which 
is invariant under left and right actions by the diagonal $\diag( \mC ( S_n))$: 
\be
\cK_{\ung}(n) = {\rm Span}_{\mC}\Big\{    
\sum_{ \g_1, \g_2 \in S_n } \gamma_1 \sigma_1 \gamma_2 \otimes  
\gamma_1 \sigma_2 \gamma_2 \otimes  
 \gamma_1 \sigma_3 \gamma_2 , \; \s_1, \s_2,\s_3 \in S_n
\Big\}
\ee
The equivalence classes defining $ \cK_{ \ung} ( n)$ are the double cosets 
\bea 
\diag ( S_n ) \ses ( S_n \times S_n \times S_n ) / \diag ( S_n ) 
\eea 
It is simple to check that $\cK_{\ung}(n)$ is stable under multiplication.
The identity of $\cK_{\ung}(n)$ is $id \otimes id \otimes id$.
 The rest of required properties to make $\cK_{\ung}(n)$ a sub-algebra 
of $\mC(S_n)^{\otimes 3}$ can be easily verified. 

\begin{proposition}
$\cK_{\ung}(n)$  is an associative unital sub-algebra of $\mC(S_n)^{\otimes 3}$. 
\end{proposition}
In fact, one shows that the two algebras
$\cK(n)$ and $\cK_{\ung}(n)$ have the same dimension. 
The isomorphism between the basis elements stems
from a change or variable:
$\gamma_1 \to \gamma_2^{-1}\s_1^{-1}$, 
and then renaming $\s_1^{-1}\s_i$ as $\s_{i}$, $i=2,3,$
and $\g_2 \to \g$. Under
this change of variable we obtain 
\bea
\sum_{ \g_1, \g_2 \in S_n } \gamma_1 \sigma_1 \gamma_2 \otimes  
\gamma_1 \sigma_2 \gamma_2 \otimes  
 \gamma_1 \sigma_3 \gamma_2 
= \sum_{\g \in S_n} 
id  \otimes  
\gamma \sigma_2 \gamma^{-1} \otimes  
 \gamma \sigma_3 \gamma^{-1} 
\eea
and the r.h.s is clearly associated with the basis element 
$\sum_{\g \in S_n} \gamma \sigma_2 \gamma^{-1} \otimes  
 \gamma \sigma_3 \gamma^{-1}$ of $\cK(n)$. 
It is direct to get $\dim\cK_{\ung}(n) = \dim\cK(n)=Z_3(n)$. 
Finally, we will keep the name of ``graphs'' as elements of 
$\cK_{\ung}(n)$.

\

\noindent{\bf Fourier basis $Q_{\ung}$ -}
In this formulation in terms of triples of permutations, the
 basis of invariants of $\cK_{\ung}(n)$ is given by 
\be\label{qunbasis}
Q^{R,S,T}_{\ung;\tau_1,\tau_2}
= \kappa_{R,S,T} \sum_{\s_l \in S_n} \sum_{i_l,j_l} 
C^{R , S ; T , \tau_1  }_{ i_1 , i_2 ; i_3 } C^{R , S ; T , \tau_2  }_{ j_1 , j_2 ; j_3 } 
D^{R}_{ i_1 , j_1 } ( \sigma_1 )
 D^{S}_{ i_2, j_2 } ( \sigma_2 ) D^{T}_{ i_3 , j_3 } ( \sigma_3 )
\sigma_1 \otimes \sigma_2 \otimes \sigma_3 
\ee
with $\kappa_{R,S,T} = \frac{d(R)d(S)d(T)}{ (n!)^3}$.
The basis $\{Q^{R,S,T}_{\ung;\tau_1,\tau_2}\}$ 
is called $Q_{\ung}$-basis. 
Its elements are invariant under left
and right diagonal actions (see \eqref{inqUN2} 
for an intermediate step, appendix \ref{app:baskn}): 
\be\label{unginvariance}
(\gamma_1^{\otimes 3})\cdot Q^{R,S,T}_{\ung; \tau_1,\tau_2} \cdot
(\gamma_2^{\otimes 3})=
Q^{R,S,T}_{\ung; \tau_1,\tau_2}
\ee
and multiply like matrices (for a few details, see \eqref{qqUNmatrix} in appendix \ref{app:baskn}):
\bea
&&
Q^{R,S,T}_{\ung; \tau_1,\tau_2} Q^{R',S',T'}_{\ung; \tau_2,\tau_3}  = 
  \kappa_{R,S,T} \delta_{R,R'}\delta_{S,S'}\delta_{T,T'}  \cr\cr
&&\times 
\sum_{\s_l \in S_n} \sum_{i_l,j_l} 
\sum_{\s'_l \in S_n} \sum_{a_l} 
C^{R , S ; T , \tau_1  }_{ i_1 , i_2 ; i_3 } 
C^{R, S ; T , \tau_3  }_{ a_1 , a_2 ; a_3 } 
D^{R}_{ i_1 , a_1 } (\sigma_1) D^{S}_{ i_2, a_2 } ( \sigma_2)
 D^{T}_{ i_3 , a_3 } ( \sigma_3)\, 
\sigma_1\otimes \sigma_2\otimes \sigma_3
\cr\cr
&&
\times \sum_{j_l}
C^{R , S ; T, \tau_2  }_{ j_1 , j_2 ; j_3 } 
C^{R , S ; T , \tau_2  }_{ j_1 , j_2 ; j_3 } 
 = \delta_{R,R'}\delta_{S,S'}\delta_{T,T'}\,  Q^{R,S,T}_{\ung; \tau_1,\tau_3} 
\label{qungmatrix}
\eea
Computing the pairing of two $Q_{\ung}$'s yields
(for details see \eqref{qqUNin} in appendix \ref{app:baskn})  
\be
\bdel_3 (Q_{\ung;\tau_1,\tau_2}^{R,S,T}; Q_{\ung;\tau_1',\tau'_2}^{R',S',T'})
  =   \kappa_{R,S,T}\, d(T)^2\,
\delta_{RR'}\delta_{SS'}\delta_{TT'}
\delta_{\tau_1\tau'_1} \delta_{\tau_2\tau_2'}  
\label{qqUNin0} 
\ee
we infer that the basis $Q_{\ung}$ is orthogonal. Same comments
about making $Q_{\ung}$ orthonormal by appropriately
tuning the $\kappa_{R,S,T}$ normalization factor can be made
at this stage. 

\

\noindent{\bf The centre $\cZ(\cK_{\ung}(n))$ -}
We now investigate the centre $\cZ(\cK_{\ung}(n))$ of $\cK_{\ung} (n)$. 
Using the same strategy as in section \ref{sect:ZKn}, we construct now the basis of the centre by taking the trace of the matrices 
$Q_{\ung}$'s
\be
P^{R,S,T}_{\ung}  = \sum_{\tau}  Q^{R,S,T}_{\ung;\tau,\tau} 
\ee
We show that $P^{R,S,T}_{\ung}$ is commuting with any $Q^{R',S',T'}_{\ung; \tau_1,\tau_2}$: 
\bea
&&
P^{R,S,T}_{\ung}  \cdot Q^{R',S',T'}_{\ung;\tau_1,\tau_2} 
 = \sum_{\tau} Q^{R,S,T}_{\ung;\tau,\tau}  Q^{R',S',T'}_{\ung;\tau_1,\tau_2} \cr\cr
&&
 =\delta_{R,R'}\delta_{S,S'}\delta_{T,T'}  \sum_{\tau} \delta_{\tau,\tau_1} 
 Q^{R,S,T}_{\ung;\tau,\tau_2} 
 =  \delta_{R,R'}\delta_{S,S'}\delta_{T,T'} Q^{R,S,T}_{\ung;\tau_1,\tau_2} \cr\cr
&&
Q^{R',S',T'}_{\ung;\tau_1,\tau_2}  \cdot P^{R,S,T}_{\ung}
 = \delta_{R,R'}\delta_{S,S'}\delta_{T,T'}  \sum_{\tau} \delta_{\tau,\tau_2} 
 Q^{R,S,T}_{\ung;\tau_1,\tau}  = P^{R,S,T}_{\ung}  \cdot Q^{R',S',T'}_{\ung;\tau_1,\tau_2} 
\eea
Hence $P^{R,S,T}_{\ung} $ is in the centre of $\cK_{\ung}(n)$. 
Adapting the arguments of the proof of Proposition \ref{prop:PgenZ} in the present context, the next result can be deduced without
difficulties. 
\begin{proposition}
$\{P_{\ung}^{R,S,T}\}$ is a basis of $\cZ(\cK_{\ung}(n))$. 
\end{proposition}
The pairing of two $P_{\ung}$'s gives
\bea
&&
\bdel_{3}(P_{\ung} ^{R,S,T}; P_{\ung} ^{R',S',T'})
 = \sum_{\tau,\tau'} \bdel_{3}(Q^{R,S,T}_{\ung; \tau,\tau};Q^{R',S',T'}_{\ung;\tau',\tau'}) \cr\cr
&&
 =
 \kappa_{R,S,T} d(T)^2
\delta_{RR'}\delta_{SS'}\delta_{TT'}
 \sum_{\tau,\tau'} \delta_{\tau\tau'} 
 = \kappa_{R,S,T}\, d(T)^2 \sC(R,S,T)  \delta_{RR'} \delta_{SS'}
\delta_{TT'}  
\eea

\noindent{\bf Overcomplete bases of  $\cZ(\cK_{\ung}(n))$ -}
Given three Young diagrams $R_i$, $i=1,2,3$, 
and the central element $z_{R_i} = \sum_{\s}\chi^{R_i}(\s)\s$
of $\mC(S_n)$, we are interested by the element
\be
z_{\ung}^{R_1,R_2,R_3}= z_{R_1}\otimes z_{R_2}\otimes z_{R_2}= \sum_{\s_i \in S_n}  
\chi^{R_1}(\s_1)\chi^{R_2}(\s_2)\chi^{R_3}(\s_3)  \s_1 \otimes \s_2 \otimes \s_3 
\ee
that proves to belong to the centre of $\mC(S_n)^{\otimes 3}$.
It is sufficient to prove this claim for any basis element as  
\bea
&&
\g_1 \otimes \g_2 \otimes \g_3 \cdot z_{\ung}^{R_1,R_2,R_3}=  
 \sum_{\s_i \in S_n}
\chi^{R_1}(\s_1)\chi^{R_2}(\s_2)\chi^{R_3}(\s_3) 
 \g_1 \s_1 \otimes
 \g_2 \s_2  \otimes
 \g_3 \s_3  \cr\cr
&& = 
 \sum_{\s_i \in S_n}
\chi^{R_1}(\g_1^{-1}\s_1)\chi^{R_2}(\g_2^{-1}\s_2)\chi^{R_3}(\g_3^{-1}\s_3) 
  \s_1 \otimes \s_2  \otimes \s_3  \cr\cr
&& = 
 \sum_{\s_i \in S_n}
\chi^{R_1}(\s_1\g_1^{-1})\chi^{R_2}(\s_2\g_2^{-1})\chi^{R_3}(\s_3\g_3^{-1}) 
  \s_1 \otimes \s_2  \otimes \s_3   \cr\cr
&& = 
 \sum_{\s_i \in S_n}
\chi^{R_1}(\s_1)\chi^{R_2}(\s_2)\chi^{R_3}(\s_3) 
  \s_1\g_1 \otimes \s_2\g_2  \otimes \s_3 \g_3
\cr\cr
&&= z_{\ung}^{R_1,R_2,R_3} \cdot \g_1 \otimes \g_2 \otimes \g_3 
\eea
we used a change of variable $\s_i \to \g_i ^{-1}$. 

The following statement holds. 
\begin{proposition}
$\{z_{\ung}^{R_1,R_2,R_3} \}$ forms an overcomplete basis 
of the centre $\cZ(\cK_{\ung}(n))$. 
\end{proposition}
\proof 
We want to find an expansion 
\bea
P_{\ung}^{R_1,R_2,R_3} = \sum_{R_1',R_2',R_3'} \bdel_{3} (z_{\ung}^{R_1',R_2',R_3'};P_{\ung}^{R_1,R_2,R_3})  \,z_{\ung}^{R_1',R_2',R_3'} 
\eea
 with the coefficient $\bdel_{3} (z_{\ung}^{R'_1,R'_2,R'_3};P_{\ung}^{R_1,R_2,R_3})$.
That quantity  has been computed in \eqref{eq:PUNz}
of appendix \ref{app:Zbaskn} and one finds it as
\be
\bdel_{3} (z_{\ung}^{R'_1,R'_2,R'_3};P_{\ung}^{R_1,R_2,R_3})   =
d(R_3)\sC(R_1 , R_2 , R_3)\delta_{R_1R'_1}\delta_{R_2R'_2}\delta_{R_3R_3'}   
\ee
Now, for a given triple $(R_1 , R_2 , R_3)$  for which $P^{R_1 , R_2 , R_3}_{\ung}$ is not vanishing,
then the coefficient $\bdel_{3} (z_{\ung}^{R'_1,R'_2,R'_3};P_{\ung}^{R_1 , R_2 , R_3})$ is not vanishing. Therefore $P_{\ung}^{R_1 , R_2 , R_3}$ has an expansion in terms of the $z_{\ung}^{R_1,R_2,R_3}$'s. The cardinality of $\{z_{\ung}^{R_1,R_2,R_3} \}$ is $p(n)^3$, cube of the number of partitions of $n$, is larger  than the number of nonvanishing Kroneckers $\sC(R_1,R_2,R_3)$. 
The basis $\{z_{\ung}^{R_1,R_2,R_3} \}$ is therefore overcomplete. 

\qed 

\section{$\cK(n)$  as a  graph algebra  }\label{cKgraph} 

As already mentioned, 
to each element of $\cK_{\ung}(n)$ of the form 
$\sum_{\g_1,\g_2 \in S_n} \g_1 \sigma_1 \gamma_2  \otimes \gamma_1 \sigma_2 \gamma_2 \otimes \gamma_1  \sigma_3 \gamma_2  $, 
we associate a  tensor observable, determined by the triple
of permutations $(\s_1,\s_2,\s_3)$ subjected to the equivalence
$(\s_1,\s_2,\s_3) \sim   \gamma_1 \cdot (  \s_1,\s_2,\s_3)\cdot  \gamma_2$. 
We now investigate the algebra inherited on colored bipartite graphs induced
from the multiplication law of $\cK_{\ung}(n)$.  As observed in \cite{Sanjo} and
discussed earlier in this paper, the gauge-fixed formulation involves permutation pairs 
subject to simultaneous conjugation equivalence. These naturally correspond to ordinary bi-partite graphs (edges are not colored). A different algebra  structure on the space of bi-partite graphs has been considered in \cite{MMN12}. 

First, given the normalized graph elements of $\cK_{\ung}(n)$, labelled
by $\s_i\in S_n$, 
\be
A_{\s_1,\s_2,\s_3} = \frac{1}{(n!)^2}\sum_{\g_1,\g_2 \in S_n} \g_1 \sigma_1 \gamma_2  \otimes \gamma_1 \sigma_2 \gamma_2 \otimes \gamma_1  \sigma_3 \gamma_2 
\ee
 we write a product of two of these elements in $\cK_{\ung}(n)$ as
\be
A_{\s_1,\s_2,\s_3} A_{\s_4,\s_5,\s_6}  =  \frac{1}{(n!)^4}\sum_{\g_1,\g_2,\tau_1,\tau_2 \in S_n}  
 \g_1 \sigma_1 \gamma_2 \tau_1 \sigma_4 \tau_2 \otimes 
\gamma_1 \sigma_2 \gamma_2 \tau_1 \s_5 \tau_ 2 \otimes 
\gamma_1  \sigma_3 \gamma_2\tau_1 \s_6 \tau_2  
\ee
A change of variables $  \gamma_2\tau_1 \to \tau_1 $, and renaming of $\tau_2$
as $\g_2$ and $\tau_1$ as $\tau$, allow us to get
\bea\label{graphAlg}
A_{\s_1,\s_2,\s_3} A_{\s_4,\s_5,\s_6}  &=&
 \frac{1}{n!}
\sum_{\tau \in S_n}\Bigg[
 \frac{1}{(n!)^2}
 \sum_{\g_1,\g_2 \in S_n}  
 \g_1    (\sigma_1 \tau \sigma_4) \g_2 \otimes 
\g_1 (\sigma_2  \tau \s_5) \g_ 2 \otimes 
\g_1  (\sigma_3 \tau \s_6) \g_2 \Bigg] \crcr
&=&  \frac{1}{n!}
\sum_{\tau\in S_n}
A_{\sigma_1 \tau \sigma_4,\, \sigma_2 \tau \sigma_5,\,\sigma_3 \tau \sigma_6} \label{AB}
\eea
Thus, the product of two graphs can be written as a sum of graphs. 
There is a particular element such that
\bea
A_{\s_1,\s_2,\s_3} A_{id,id,id}  &=&
 \frac{1}{n!}
\sum_{\tau \in S_n}\Bigg[
 \frac{1}{(n!)^2}
 \sum_{\g_1,\g_2 \in S_n}  
 \g_1    (\sigma_1 \tau ) \g_2 \otimes 
\g_1 (\sigma_2  \tau ) \g_ 2 \otimes 
\g_1  (\sigma_3 \tau ) \g_2 \Bigg] \crcr
& = & \frac{1}{(n!)^2}
 \sum_{\g_1,\g_2 \in S_n}  
 \g_1    \sigma_1   \g_2 \otimes 
\g_1 \sigma_2  \g_ 2 \otimes 
\g_1   \sigma_3  \g_2
 =  A_{\s_1,\s_2,\s_3}
\eea
and 
similarly $A_{id,id,id}  A_{\s_1,\s_2,\s_3} = A_{\s_1,\s_2,\s_3}  $.
This shows that $A_{id,id,id}=E$ is 
a unit element of the graph algebra. 

In the gauge-fixed formulation, the graph multiplication takes the form: 
\bea
&&
B_{\s_1,\s_2} B_{\s_3,\s_4} 
 = \sum_{\g_1,\g_2} \g_1 \s_1 \g_1^{-1}   \g_2 \s_3  \g_2^{-1} \otimes \g_1 \s_2 \g_1^{-1}   \g_2 \s_4  \g_2^{-1} \cr\cr
 && 
 = \sum_{\tau,\g_2} \g_2\tau^{-1} \s_1  \tau   \s_3  \g_2^{-1} \otimes\g_2\tau^{-1} \s_2  \tau \s_4  \g_2^{-1} \cr\cr
 && = 
 \sum_{\tau} B_{\tau^{-1} \s_1  \tau   \s_3 ,\tau^{-1} \s_2  \tau \s_4} 
\eea
where we used a change of variables $\gamma_1 \to \g_2\tau^{-1}$ and omit normalization
factor, for simplicity. This relation can be also 
obtained from \eqref{AB} after some proper gauge fixing. 

Coming back to the formula \eqref{AB}, we can illustrate 
this in diagram 
\bea
\begin{tikzpicture}
\draw (-10,2.3) rectangle(-4,2) node {} ; 
\node (r1) at (-4.2,2.1)  {$\s_1$};  
\draw (-10,1.2) rectangle(-4,0.9) node {} ; 
\node (r2) at (-4.2,1)  {$\s_2$}; 
\draw (-10,0.1) rectangle(-4,-0.2) node {} ;
\node (r3) at (-4.2,-0.1)  {$\s_3$}; 
\draw[solid] (-9.5,2.8) circle (0.1);
\draw[solid] (-8.5,2.8) circle (0.1);
\draw[thick, dotted] (-8,2.8)  -- (-5.2,2.8) ;
\draw[solid] (-4.7,2.8) circle (0.1);
\draw[shorten <= 0.1cm, shorten >= 0.1cm] (-9.5,2.8) to[out=-180, in=90] (-9.8,2.2);
\draw[shorten <= 0.1cm, shorten >= 0.1cm,  dashed] (-9.5,2.8) to[out=0, in=90] (-9.2,2.2);
\draw[ dashed] (-9.2,2.2)-- (-9.2,0);
\draw[shorten <= 0.1cm, shorten >= 0.1cm] (-8.5,2.8) to[out=-180, in=90] (-8.8,2.2);
\draw[shorten <= 0.1cm, shorten >= 0.1cm,  dashed] (-8.5,2.8) to[out=0, in=90] (-8.2,2.2);
\draw[ dashed] (-8.2,2.2)-- (-8.2,0);
\draw[shorten <= 0.1cm, shorten >= 0.1cm] (-4.7,2.8) to[out=-180, in=90] (-5,2.2);
\draw[shorten <= 0.1cm, shorten >= 0.1cm,  dashed](-4.7,2.8) to[out=0, in=90] (-4.4,2.2);
\draw[ dashed] (-4.4,2.2)-- (-4.4,0);
\draw[ dash dot] (-9.5,2.7) -- (-9.5,1.2);
\draw[dash dot] (-8.5,2.7) -- (-8.5,1.2);
\draw[dash dot]  (-4.7,2.7) -- (-4.7,1.2);
\draw[shorten <= 0.1cm, shorten >= 0.1cm] (-9.5,-0.75) to[out=180, in=270] (-9.8, -0.1);
\draw[] (-9.8,-0.2)-- (-9.8,2);
\draw[shorten <= 0.1cm, shorten >= 0.1cm,  dashed] (-9.5,-0.75) to[out=0, in=270] (-9.2,-0.1);
\draw[shorten <= 0.1cm, shorten >= 0.1cm] (-8.5,-0.75) to[out=180, in=270] (-8.8,-0.1);
\draw[] (-8.8,-0.2)-- (-8.8,2);
\draw[shorten <= 0.1cm, shorten >= 0.1cm,  dashed] (-8.5,-0.75) to[out=0, in=270] (-8.2,-0.1);
\draw[shorten <= 0.1cm, shorten >= 0.1cm] (-4.7,-0.75) to[out=180, in=270] (-5,-0.1);
\draw[] (-5,-0.2)-- (-5,2);
\draw[shorten <= 0.1cm, shorten >= 0.1cm,  dashed] (-4.7,-0.75) to[out=0, in=270] (-4.4,-0.1);
\fill[color=black] (-9.5,-0.75) circle (0.1);
\fill[color=black] (-8.5,-0.75) circle (0.1);
\draw[thick,dotted] (-8,-0.75)  -- (-5.2,-0.75) ;
\fill[color=black] (-4.7,-0.75) circle (0.1);
%
%
%
\draw[dash dot] (-9.5,-0.75) -- (-9.5,0.9);
\draw[dash dot] (-8.5,-0.75) -- (-8.5,0.9);
\draw[dash dot]  (-4.7,-0.75) -- (-4.7,0.9);
\draw (-10,-2.3) rectangle(-4,-2) node {} ; 
\node (r4) at (-4.2,-2.2)  {$\s_4$};  
\draw (-10,-3.4) rectangle(-4,-3.1) node {} ; 
\node (r5) at (-4.2,-3.3)  {$\s_5$}; 
\draw (-10,-4.5) rectangle(-4,-4.2) node {} ;
\node (r6) at (-4.2,-4.4)  {$\s_6$}; 
\draw[solid] (-9.5,-1.5) circle (0.1);
\draw[solid] (-8.5,-1.5) circle (0.1);
\draw[thick,dotted] (-8,-1.5)  -- (-5.2,-1.5) ;
\draw[solid] (-4.7,-1.5) circle (0.1);
\fill[color=black] (-9.5,-5) circle (0.1);
\fill[color=black] (-8.5,-5) circle (0.1);
\draw[thick,dotted] (-8,-5)  -- (-5.2,-5) ;
\fill[color=black] (-4.7,-5) circle (0.1);

\draw[shorten <= 0.1cm, shorten >= 0.1cm] (-9.5,-1.5) to[out=-180, in=90] (-9.8,-2.1);
\draw[shorten <= 0.1cm, shorten >= 0.1cm,  dashed] (-9.5,-1.5) to[out=0, in=90] (-9.2,-2.1);
\draw[ dashed] (-9.2,-2.1)-- (-9.2,-4.3);
\draw[shorten <= 0.1cm, shorten >= 0.1cm] (-8.5,-1.5) to[out=-180, in=90] (-8.8,-2.1);
\draw[shorten <= 0.1cm, shorten >= 0.1cm,  dashed] (-8.5,-1.5) to[out=0, in=90] (-8.2,-2.1);
\draw[ dashed] (-8.2,-2.1)-- (-8.2,-4.3);
\draw[shorten <= 0.1cm, shorten >= 0.1cm] (-4.7,-1.5) to[out=-180, in=90] (-5,-2.1);
\draw[shorten <= 0.1cm, shorten >= 0.1cm,  dashed](-4.7,-1.5) to[out=0, in=90] (-4.4,-2.1);
\draw[ dashed] (-4.4,-2.1)-- (-4.4,-4.3);
\draw[dash dot] (-9.5,-1.6) -- (-9.5,-3.1);
\draw[dash dot] (-8.5,-1.6) -- (-8.5,-3.1);
\draw[dash dot]  (-4.7,-1.6) -- (-4.7,-3.1);
\draw[shorten <= 0.1cm, shorten >= 0.1cm] (-9.5,-5) to[out=180, in=270] (-9.8,-4.4);
\draw[] (-9.8,-4.5)-- (-9.8,-2.3);
\draw[shorten <= 0.1cm, shorten >= 0.1cm,  dashed] (-9.5,-5) to[out=0, in=270] (-9.2,-4.5);
\draw[shorten <= 0.1cm, shorten >= 0.1cm] (-8.5,-5) to[out=180, in=270] (-8.8,-4.4);
\draw[] (-8.8,-4.5)-- (-8.8,-2.3);
\draw[shorten <= 0.1cm, shorten >= 0.1cm,  dashed] (-8.5,-5) to[out=0, in=270] (-8.2,-4.5);
\draw[shorten <= 0.1cm, shorten >= 0.1cm] (-4.7,-5) to[out=180, in=270] (-5,-4.4);
\draw[] (-5,-4.5)-- (-5,-2.3);
\draw[shorten <= 0.1cm, shorten >= 0.1cm,  dashed] (-4.7,-5) to[out=0, in=270] (-4.4,-4.5);
\fill[color=black] (-9.5,-0.75) circle (0.1);
\fill[color=black] (-8.5,-0.75) circle (0.1);
\draw[thick,dotted] (-8,-0.75)  -- (-5.2,-0.75) ;
\fill[color=black] (-4.7,-0.75) circle (0.1);
%
%
%
\draw[ dash dot] (-9.5,-5) -- (-9.5,-3.4);
\draw[dash dot] (-8.5,-5) -- (-8.5,-3.4);
\draw[dash dot]  (-4.7,-5) -- (-4.7,-3.4);
\end{tikzpicture}
\qquad 
\begin{tikzpicture}
\draw (-10,2.3) rectangle(-4,2) node {} ; 
\node (r1) at (-4.2,2.1)  {$\s_1$};  
\draw (-10,1.2) rectangle(-4,0.9) node {} ; 
\node (r2) at (-4.2,1)  {$\s_2$}; 
\draw (-10,0.1) rectangle(-4,-0.2) node {} ;
\node (r3) at (-4.2,-0.1)  {$\s_3$}; 
\draw[solid] (-9.5,2.8) circle (0.1);
\draw[solid] (-8.5,2.8) circle (0.1);
\draw[thick, dotted] (-8,2.8)  -- (-5.2,2.8) ;
\draw[solid] (-4.7,2.8) circle (0.1);
\draw[shorten <= 0.1cm, shorten >= 0.1cm] (-9.5,2.8) to[out=-180, in=90] (-9.8,2.2);
\draw[shorten <= 0.1cm, shorten >= 0.1cm, thick,   dashed] (-9.5,2.8) to[out=0, in=90] (-9.2,2.2);
\draw[thick,   dashed] (-9.2,2.2)-- (-9.2,0);
\draw[thick,  dashed] (-9.2,-0.2)-- (-9.2,-1.3);
\draw[shorten <= 0.1cm, shorten >= 0.1cm] (-8.5,2.8) to[out=-180, in=90] (-8.8,2.2);
\draw[shorten <= 0.1cm, shorten >= 0.1cm, thick,  dashed] (-8.5,2.8) to[out=0, in=90] (-8.2,2.2);
\draw[thick,  dashed] (-8.2,2.2)-- (-8.2,0);
\draw[thick,  dashed] (-8.2,-0.2)-- (-8.2,-1.3);
\draw[shorten <= 0.1cm, shorten >= 0.1cm] (-4.7,2.8) to[out=-180, in=90] (-5,2.2);
\draw[shorten <= 0.1cm, shorten >= 0.1cm, thick,   dashed](-4.7,2.8) to[out=0, in=90] (-4.4,2.2);
\draw[thick,   dashed] (-4.4,2.2)-- (-4.4,0);
\draw[thick,  dashed] (-4.4,-0.2)-- (-4.4,-1.3);
\draw[thick,   dash dot] (-9.5,2.7) -- (-9.5,1.2);
\draw[thick,  dash dot] (-8.5,2.7) -- (-8.5,1.2);
\draw[thick,  dash dot]  (-4.7,2.7) -- (-4.7,1.2);
\draw[] (-9.8,-0.7)-- (-9.8,2);
\draw[] (-8.8,-0.7)-- (-8.8,2);
\draw[] (-5,-0.7)-- (-5,2);
%
%
%
\draw[thick,  dash dot] (-9.5,-1) -- (-9.5,0.9);
\draw[thick,  dash dot] (-8.5,-1) -- (-8.5,0.9);
\draw[thick,   dash dot]  (-4.7,-1) -- (-4.7,0.9);
\draw (-10,-2.3) rectangle(-4,-2) node {} ; 
\node (r4) at (-4.2,-2.2)  {$\s_4$};  
\draw (-10,-3.4) rectangle(-4,-3.1) node {} ; 
\node (r5) at (-4.2,-3.3)  {$\s_5$}; 
\draw (-10,-4.5) rectangle(-4,-4.2) node {} ;
\node (r6) at (-4.2,-4.4)  {$\s_6$}; 
\draw (-10,-0.9) rectangle(-4,-0.7) node {} ;
\node (r7) at (-4.2,-0.8)  {$\tau$};   
\draw (-10,-1) rectangle(-4,-1.2) node {} ;
\node (r7) at (-4.2,-1.1)  {$\tau$};    
\draw (-10,-1.3) rectangle(-4,-1.5) node {} ;  
\node (r7) at (-4.2,-1.4)  {$\tau$};   
\node(r10) at (-11.2,-1.1) {$\sum_{\tau}$};
\draw[thick,  dashed] (-9.2,-1.5)-- (-9.2,-4.3);
\draw[thick,  dashed] (-8.2,-1.5)-- (-8.2,-4.3);
\draw[thick,  dashed] (-4.4,-1.5)-- (-4.4,-4.3);
\draw[thick,  dash dot] (-9.5,-1.2) -- (-9.5,-3.1);
\draw[thick,  dash dot] (-8.5,-1.2) -- (-8.5,-3.1);
\draw[thick,  dash dot]  (-4.7,-1.2) -- (-4.7,-3.1);
\fill[color=black] (-9.5,-5) circle (0.1);
\fill[color=black] (-8.5,-5) circle (0.1);
\draw[thick,dotted] (-8,-5)  -- (-5.2,-5) ;
\fill[color=black] (-4.7,-5) circle (0.1);

\draw[shorten <= 0.1cm, shorten >= 0.1cm] (-9.5,-5) to[out=180, in=270] (-9.8,-4.4);
\draw[] (-9.8,-4.5)-- (-9.8,-2.3);
\draw[] (-9.8,-2)-- (-9.8,-0.9);
\draw[shorten <= 0.1cm, shorten >= 0.1cm, thick,  dashed] (-9.5,-5) to[out=0, in=270] (-9.2,-4.5);
\draw[shorten <= 0.1cm, shorten >= 0.1cm] (-8.5,-5) to[out=180, in=270] (-8.8,-4.4);
\draw[] (-8.8,-4.5)-- (-8.8,-2.3);
\draw[] (-8.8,-2)-- (-8.8,-0.9);
\draw[shorten <= 0.1cm, shorten >= 0.1cm, thick,  dashed] (-8.5,-5) to[out=0, in=270] (-8.2,-4.5);
\draw[shorten <= 0.1cm, shorten >= 0.1cm] (-4.7,-5) to[out=180, in=270] (-5,-4.4);
\draw[] (-5,-4.5)-- (-5,-2.3);
\draw[] (-5,-2)-- (-5,-0.9);
\draw[shorten <= 0.1cm, shorten >= 0.1cm, thick,  dashed] (-4.7,-5) to[out=0, in=270] (-4.4,-4.5);
%
%
%
%
\draw[thick,  dash dot] (-9.5,-5) -- (-9.5,-3.4);
\draw[thick,  dash dot] (-8.5,-5) -- (-8.5,-3.4);
\draw[thick,  dash dot]  (-4.7,-5) -- (-4.7,-3.4);
\end{tikzpicture}
\cr\cr
&&
\eea

\noindent{\bf Algebra $\cK_{\ung}(n=1)$ -}
Let us illustrate the formula \eqref{AB} at $n=1$. There is no 
choice here, we obtain $E^2 = E$. This is 
the unique invariant made by contraction of two tensors, that is $\sum_{n_1,n_2,n_3}T_{n_1,n_2,n_3} \bar T_{n_1,n_2,n_3} =E$. We consider 
$E $ as an idempotent or unit element of an 1 dimensional algebra $\{E\}$. 

\
 
\noindent{\bf Algebra $\cK_{\ung}(n=2)$ -}
We now examine $n=2$. There are 4 possible diagrams, see \eqref{graphs}. 
\bea\label{graphs}
&&
E \sim  
(\s_1 = id, \s_2 = id, \s_3 =id)  = 
\begin{tikzpicture}
\draw[solid] (1.5,0.5) circle (0.1);
\fill[fill=black] (2.25,0.5) circle (0.1);
\fill[fill=black] (1.5,1) circle (0.1);
\draw[solid] (2.25,1) circle (0.1);
\draw[shorten <= 0.1cm, shorten >= 0.1cm] (1.5,0.5) to[out=50, in=130] (2.25,0.5);
\draw[shorten <= 0.1cm, shorten >= 0.1cm] (1.5,0.5) to[out=-50, in=-130] (2.25,0.5);
\draw[shorten <= 0.1cm, shorten >= 0.1cm] (1.5,1) to[out=50, in=130] (2.25,1);
\draw[shorten <= 0.1cm, shorten >= 0.1cm] (1.5,1) to[out=-50, in=-130] (2.25,1);
\draw (1.6,0.5) -- (2.15,0.5) ;
\draw (1.6,1) -- (2.15,1) ;
\end{tikzpicture}
\crcr
&&
A_{id,id,(12)}=A \sim( \s_1 = id, \s_2 = id, \s_3 =(12)) 
 = \begin{tikzpicture}
\node (r1) at (0.3,0.75)  {3};  
\draw[solid] (0.5,0.5) circle (0.1);
\fill[fill=black] (1.25,0.5) circle (0.1);
\fill[fill=black] (0.5,1) circle (0.1);
\draw[solid] (1.25,1) circle (0.1);
\draw (0.5,0.6) -- (0.5,0.9); 
\draw (1.25,0.6) -- (1.25,0.9); 
 \draw[shorten <= 0.1cm, shorten >= 0.1cm] (0.5,0.5) to[out=50, in=130] (1.25,0.5);
\draw[shorten <= 0.1cm, shorten >= 0.1cm] (0.5,0.5) to[out=-50, in=-130] (1.25,0.5);
\draw[shorten <= 0.1cm, shorten >= 0.1cm] (0.5,1) to[out=50, in=130] (1.25,1);
\draw[shorten <= 0.1cm, shorten >= 0.1cm] (0.5,1) to[out=-50, in=-130] (1.25,1);
\end{tikzpicture}
\crcr
&&
A_{id,(12),id}=B \sim (\s_1 = id, \s_2 = (12), \s_3 =id) 
 = 
\begin{tikzpicture}
\node (r1) at (0.3,0.75)  {2};  
\draw[solid] (0.5,0.5) circle (0.1);
\fill[fill=black] (1.25,0.5) circle (0.1);
\fill[fill=black] (0.5,1) circle (0.1);
\draw[solid] (1.25,1) circle (0.1);
\draw (0.5,0.6) -- (0.5,0.9); 
\draw (1.25,0.6) -- (1.25,0.9); 
 \draw[shorten <= 0.1cm, shorten >= 0.1cm] (0.5,0.5) to[out=50, in=130] (1.25,0.5);
\draw[shorten <= 0.1cm, shorten >= 0.1cm] (0.5,0.5) to[out=-50, in=-130] (1.25,0.5);
\draw[shorten <= 0.1cm, shorten >= 0.1cm] (0.5,1) to[out=50, in=130] (1.25,1);
\draw[shorten <= 0.1cm, shorten >= 0.1cm] (0.5,1) to[out=-50, in=-130] (1.25,1);
\end{tikzpicture}
\crcr
&& 
A_{(12), id,id} =D \sim (\s_1 = (12), \s_2 = id, \s_3 =id)
= 
\begin{tikzpicture}
\node (r1) at (0.3,0.75)  {1};   
\draw[solid] (0.5,0.5) circle (0.1);
\fill[fill=black] (1.25,0.5) circle (0.1);
\fill[fill=black] (0.5,1) circle (0.1);
\draw[solid] (1.25,1) circle (0.1);
\draw (0.5,0.6) -- (0.5,0.9); 
\draw (1.25,0.6) -- (1.25,0.9); 
 \draw[shorten <= 0.1cm, shorten >= 0.1cm] (0.5,0.5) to[out=50, in=130] (1.25,0.5);
\draw[shorten <= 0.1cm, shorten >= 0.1cm] (0.5,0.5) to[out=-50, in=-130] (1.25,0.5);
\draw[shorten <= 0.1cm, shorten >= 0.1cm] (0.5,1) to[out=50, in=130] (1.25,1);
\draw[shorten <= 0.1cm, shorten >= 0.1cm] (0.5,1) to[out=-50, in=-130] (1.25,1);
\end{tikzpicture}
\eea 
where the labels $3,2,1$ denote a particular colored edge
which can be used as a label of the invariant. 
Note that due to the equivalence under left and right diagonal action, any other choice
reduces to one of the above. For example $\s_1 = id, s_2 = (12), s_3 =(12)
\sim (12) \cdot (\s_1 = id, s_2 = (12), s_3 =(12)) \cdot (12) 
= (\s_1 = (12), \s_2 =id, \s_3 =id) $. We then compute some products
(note that they are normalized by $1/(2!)^2$ and we use \eqref{graphAlg}): 

\bea
A \cdot  E  
 &=&  \frac{1}{2}(A_{id, id, (12)}  + A_{(12),(12),id}) 
 =  A_{id,id,(12)} = A 
 \crcr
\begin{tikzpicture}
\node (r1) at (0.1,0.75)  {3};  
\draw[solid] (0.3,0.5) circle (0.1);
\fill[fill=black] (1.05,0.5) circle (0.1);
\fill[fill=black] (0.3,1) circle (0.1);
\draw[solid] (1.05,1) circle (0.1);
\draw (0.3,0.6) -- (0.3,0.9); 
\draw (1.05,0.6) -- (1.05,0.9); 
 \draw[shorten <= 0.1cm, shorten >= 0.1cm] (0.3,0.5) to[out=50, in=130] (1.05,0.5);
\draw[shorten <= 0.1cm, shorten >= 0.1cm] (0.3,0.5) to[out=-50, in=-130] (1.05,0.5);
\draw[shorten <= 0.1cm, shorten >= 0.1cm] (0.3,1) to[out=50, in=130] (1.05,1);
\draw[shorten <= 0.1cm, shorten >= 0.1cm] (0.3,1) to[out=-50, in=-130] (1.05,1);
\fill[fill=black] (1.28,0.75) circle (0.05);
\draw[solid] (1.5,0.5) circle (0.1);
\fill[fill=black] (2.25,0.5) circle (0.1);
\fill[fill=black] (1.5,1) circle (0.1);
\draw[solid] (2.25,1) circle (0.1);
 \draw[shorten <= 0.1cm, shorten >= 0.1cm] (1.5,0.5) to[out=50, in=130] (2.25,0.5);
\draw[shorten <= 0.1cm, shorten >= 0.1cm] (1.5,0.5) to[out=-50, in=-130] (2.25,0.5);
\draw[shorten <= 0.1cm, shorten >= 0.1cm] (1.5,1) to[out=50, in=130] (2.25,1);
\draw[shorten <= 0.1cm, shorten >= 0.1cm] (1.5,1) to[out=-50, in=-130] (2.25,1);
\draw (1.6,0.5) -- (2.15,0.5) ;
\draw (1.6,1) -- (2.15,1) ;
\end{tikzpicture}
& =&    \begin{tikzpicture}
\node (r1) at (0.3,0.75)  {3};   
\draw[solid] (0.5,0.5) circle (0.1);
\fill[fill=black] (1.25,0.5) circle (0.1);
\fill[fill=black] (0.5,1) circle (0.1);
\draw[solid] (1.25,1) circle (0.1);
\draw (0.5,0.6) -- (0.5,0.9); 
\draw (1.25,0.6) -- (1.25,0.9); 
 \draw[shorten <= 0.1cm, shorten >= 0.1cm] (0.5,0.5) to[out=50, in=130] (1.25,0.5);
\draw[shorten <= 0.1cm, shorten >= 0.1cm] (0.5,0.5) to[out=-50, in=-130] (1.25,0.5);
\draw[shorten <= 0.1cm, shorten >= 0.1cm] (0.5,1) to[out=50, in=130] (1.25,1);
\draw[shorten <= 0.1cm, shorten >= 0.1cm] (0.5,1) to[out=-50, in=-130] (1.25,1);
\end{tikzpicture} 
\cr\cr
A \cdot A &=& 
 \frac{1}{2}(  E   + A_{(12),(12),(12)} )
= E  
\crcr
\begin{tikzpicture}
\node (r1) at (0.3,0.75)  {3};   
\draw[solid] (0.5,0.5) circle (0.1);
\fill[fill=black] (1.25,0.5) circle (0.1);
\fill[fill=black] (0.5,1) circle (0.1);
\draw[solid] (1.25,1) circle (0.1);
\draw (0.5,0.6) -- (0.5,0.9); 
\draw (1.25,0.6) -- (1.25,0.9); 
 \draw[shorten <= 0.1cm, shorten >= 0.1cm] (0.5,0.5) to[out=50, in=130] (1.25,0.5);
\draw[shorten <= 0.1cm, shorten >= 0.1cm] (0.5,0.5) to[out=-50, in=-130] (1.25,0.5);
\draw[shorten <= 0.1cm, shorten >= 0.1cm] (0.5,1) to[out=50, in=130] (1.25,1);
\draw[shorten <= 0.1cm, shorten >= 0.1cm] (0.5,1) to[out=-50, in=-130] (1.25,1);
\end{tikzpicture}\;
\begin{tikzpicture}
\fill[fill=black] (0,0.75) circle (0.05);
\node (r1) at (0.3,0.75)  {3};   
\draw[solid] (0.5,0.5) circle (0.1);
\fill[fill=black] (1.25,0.5) circle (0.1);
\fill[fill=black] (0.5,1) circle (0.1);
\draw[solid] (1.25,1) circle (0.1);
\draw (0.5,0.6) -- (0.5,0.9); 
\draw (1.25,0.6) -- (1.25,0.9); 
 \draw[shorten <= 0.1cm, shorten >= 0.1cm] (0.5,0.5) to[out=50, in=130] (1.25,0.5);
\draw[shorten <= 0.1cm, shorten >= 0.1cm] (0.5,0.5) to[out=-50, in=-130] (1.25,0.5);
\draw[shorten <= 0.1cm, shorten >= 0.1cm] (0.5,1) to[out=50, in=130] (1.25,1);
\draw[shorten <= 0.1cm, shorten >= 0.1cm] (0.5,1) to[out=-50, in=-130] (1.25,1);
\end{tikzpicture}
&=& 
 \begin{tikzpicture}
\draw[solid] (1.5,0.5) circle (0.1);
\fill[fill=black] (2.25,0.5) circle (0.1);
\fill[fill=black] (1.5,1) circle (0.1);
\draw[solid] (2.25,1) circle (0.1);
 \draw[shorten <= 0.1cm, shorten >= 0.1cm] (1.5,0.5) to[out=50, in=130] (2.25,0.5);
\draw[shorten <= 0.1cm, shorten >= 0.1cm] (1.5,0.5) to[out=-50, in=-130] (2.25,0.5);
\draw[shorten <= 0.1cm, shorten >= 0.1cm] (1.5,1) to[out=50, in=130] (2.25,1);
\draw[shorten <= 0.1cm, shorten >= 0.1cm] (1.5,1) to[out=-50, in=-130] (2.25,1);
\draw (1.6,0.5) -- (2.15,0.5) ;
\draw (1.6,1) -- (2.15,1) ;
\end{tikzpicture}
\cr\cr
A \cdot   B  &=& 
 \frac{1}{2}( A_{id,(12), (12)} + D  )
 = D 
\crcr
\begin{tikzpicture}
\node (r1) at (0.3,0.75)  {3};   
\draw[solid] (0.5,0.5) circle (0.1);
\fill[fill=black] (1.25,0.5) circle (0.1);
\fill[fill=black] (0.5,1) circle (0.1);
\draw[solid] (1.25,1) circle (0.1);
\draw (0.5,0.6) -- (0.5,0.9); 
\draw (1.25,0.6) -- (1.25,0.9); 
 \draw[shorten <= 0.1cm, shorten >= 0.1cm] (0.5,0.5) to[out=50, in=130] (1.25,0.5);
\draw[shorten <= 0.1cm, shorten >= 0.1cm] (0.5,0.5) to[out=-50, in=-130] (1.25,0.5);
\draw[shorten <= 0.1cm, shorten >= 0.1cm] (0.5,1) to[out=50, in=130] (1.25,1);
\draw[shorten <= 0.1cm, shorten >= 0.1cm] (0.5,1) to[out=-50, in=-130] (1.25,1);
\end{tikzpicture}\;
\begin{tikzpicture}
\fill[fill=black] (0,0.75) circle (0.05);
\node (r1) at (0.3,0.75)  {2};   
\draw[solid] (0.5,0.5) circle (0.1);
\fill[fill=black] (1.25,0.5) circle (0.1);
\fill[fill=black] (0.5,1) circle (0.1);
\draw[solid] (1.25,1) circle (0.1);
\draw (0.5,0.6) -- (0.5,0.9); 
\draw (1.25,0.6) -- (1.25,0.9); 
 \draw[shorten <= 0.1cm, shorten >= 0.1cm] (0.5,0.5) to[out=50, in=130] (1.25,0.5);
\draw[shorten <= 0.1cm, shorten >= 0.1cm] (0.5,0.5) to[out=-50, in=-130] (1.25,0.5);
\draw[shorten <= 0.1cm, shorten >= 0.1cm] (0.5,1) to[out=50, in=130] (1.25,1);
\draw[shorten <= 0.1cm, shorten >= 0.1cm] (0.5,1) to[out=-50, in=-130] (1.25,1);
\end{tikzpicture}
&= &
\begin{tikzpicture}
\node (r1) at (0.3,0.75)  {1};   
\draw[solid] (0.5,0.5) circle (0.1);
\fill[fill=black] (1.25,0.5) circle (0.1);
\fill[fill=black] (0.5,1) circle (0.1);
\draw[solid] (1.25,1) circle (0.1);
\draw (0.5,0.6) -- (0.5,0.9); 
\draw (1.25,0.6) -- (1.25,0.9); 
 \draw[shorten <= 0.1cm, shorten >= 0.1cm] (0.5,0.5) to[out=50, in=130] (1.25,0.5);
\draw[shorten <= 0.1cm, shorten >= 0.1cm] (0.5,0.5) to[out=-50, in=-130] (1.25,0.5);
\draw[shorten <= 0.1cm, shorten >= 0.1cm] (0.5,1) to[out=50, in=130] (1.25,1);
\draw[shorten <= 0.1cm, shorten >= 0.1cm] (0.5,1) to[out=-50, in=-130] (1.25,1);
\end{tikzpicture}
\cr\cr
A \cdot D &= &
\frac{1}{2}( A_{(12),id, (12)} + B  )
 =B 
 \crcr
\begin{tikzpicture}
\node (r1) at (0.3,0.75)  {3};   
\draw[solid] (0.5,0.5) circle (0.1);
\fill[fill=black] (1.25,0.5) circle (0.1);
\fill[fill=black] (0.5,1) circle (0.1);
\draw[solid] (1.25,1) circle (0.1);
\draw (0.5,0.6) -- (0.5,0.9); 
\draw (1.25,0.6) -- (1.25,0.9); 
 \draw[shorten <= 0.1cm, shorten >= 0.1cm] (0.5,0.5) to[out=50, in=130] (1.25,0.5);
\draw[shorten <= 0.1cm, shorten >= 0.1cm] (0.5,0.5) to[out=-50, in=-130] (1.25,0.5);
\draw[shorten <= 0.1cm, shorten >= 0.1cm] (0.5,1) to[out=50, in=130] (1.25,1);
\draw[shorten <= 0.1cm, shorten >= 0.1cm] (0.5,1) to[out=-50, in=-130] (1.25,1);
\end{tikzpicture}\;
\begin{tikzpicture}
\fill[fill=black] (0,0.75) circle (0.05);
\node (r1) at (0.3,0.75)  {1};   
\draw[solid] (0.5,0.5) circle (0.1);
\fill[fill=black] (1.25,0.5) circle (0.1);
\fill[fill=black] (0.5,1) circle (0.1);
\draw[solid] (1.25,1) circle (0.1);
\draw (0.5,0.6) -- (0.5,0.9); 
\draw (1.25,0.6) -- (1.25,0.9); 
 \draw[shorten <= 0.1cm, shorten >= 0.1cm] (0.5,0.5) to[out=50, in=130] (1.25,0.5);
\draw[shorten <= 0.1cm, shorten >= 0.1cm] (0.5,0.5) to[out=-50, in=-130] (1.25,0.5);
\draw[shorten <= 0.1cm, shorten >= 0.1cm] (0.5,1) to[out=50, in=130] (1.25,1);
\draw[shorten <= 0.1cm, shorten >= 0.1cm] (0.5,1) to[out=-50, in=-130] (1.25,1);
\end{tikzpicture}
&= & 
\begin{tikzpicture}
\node (r1) at (0.3,0.75)  {2};   
\draw[solid] (0.5,0.5) circle (0.1);
\fill[fill=black] (1.25,0.5) circle (0.1);
\fill[fill=black] (0.5,1) circle (0.1);
\draw[solid] (1.25,1) circle (0.1);
\draw (0.5,0.6) -- (0.5,0.9); 
\draw (1.25,0.6) -- (1.25,0.9); 
 \draw[shorten <= 0.1cm, shorten >= 0.1cm] (0.5,0.5) to[out=50, in=130] (1.25,0.5);
\draw[shorten <= 0.1cm, shorten >= 0.1cm] (0.5,0.5) to[out=-50, in=-130] (1.25,0.5);
\draw[shorten <= 0.1cm, shorten >= 0.1cm] (0.5,1) to[out=50, in=130] (1.25,1);
\draw[shorten <= 0.1cm, shorten >= 0.1cm] (0.5,1) to[out=-50, in=-130] (1.25,1);
\end{tikzpicture}
\cr\cr
B \cdot D  &=&
\frac{1}{2}( A_{(12),(12),id} + A    )
 =A   
\crcr
\begin{tikzpicture}
\node (r1) at (0.3,0.75)  {2};   
\draw[solid] (0.5,0.5) circle (0.1);
\fill[fill=black] (1.25,0.5) circle (0.1);
\fill[fill=black] (0.5,1) circle (0.1);
\draw[solid] (1.25,1) circle (0.1);
\draw (0.5,0.6) -- (0.5,0.9); 
\draw (1.25,0.6) -- (1.25,0.9); 
 \draw[shorten <= 0.1cm, shorten >= 0.1cm] (0.5,0.5) to[out=50, in=130] (1.25,0.5);
\draw[shorten <= 0.1cm, shorten >= 0.1cm] (0.5,0.5) to[out=-50, in=-130] (1.25,0.5);
\draw[shorten <= 0.1cm, shorten >= 0.1cm] (0.5,1) to[out=50, in=130] (1.25,1);
\draw[shorten <= 0.1cm, shorten >= 0.1cm] (0.5,1) to[out=-50, in=-130] (1.25,1);
\end{tikzpicture}\;
\begin{tikzpicture}
\fill[fill=black] (0,0.75) circle (0.05);
\node (r1) at (0.3,0.75)  {1};   
\draw[solid] (0.5,0.5) circle (0.1);
\fill[fill=black] (1.25,0.5) circle (0.1);
\fill[fill=black] (0.5,1) circle (0.1);
\draw[solid] (1.25,1) circle (0.1);
\draw (0.5,0.6) -- (0.5,0.9); 
\draw (1.25,0.6) -- (1.25,0.9); 
 \draw[shorten <= 0.1cm, shorten >= 0.1cm] (0.5,0.5) to[out=50, in=130] (1.25,0.5);
\draw[shorten <= 0.1cm, shorten >= 0.1cm] (0.5,0.5) to[out=-50, in=-130] (1.25,0.5);
\draw[shorten <= 0.1cm, shorten >= 0.1cm] (0.5,1) to[out=50, in=130] (1.25,1);
\draw[shorten <= 0.1cm, shorten >= 0.1cm] (0.5,1) to[out=-50, in=-130] (1.25,1);
\end{tikzpicture}
&= & 
\begin{tikzpicture}
\node (r1) at (0.3,0.75)  {3};   
\draw[solid] (0.5,0.5) circle (0.1);
\fill[fill=black] (1.25,0.5) circle (0.1);
\fill[fill=black] (0.5,1) circle (0.1);
\draw[solid] (1.25,1) circle (0.1);
\draw (0.5,0.6) -- (0.5,0.9); 
\draw (1.25,0.6) -- (1.25,0.9); 
 \draw[shorten <= 0.1cm, shorten >= 0.1cm] (0.5,0.5) to[out=50, in=130] (1.25,0.5);
\draw[shorten <= 0.1cm, shorten >= 0.1cm] (0.5,0.5) to[out=-50, in=-130] (1.25,0.5);
\draw[shorten <= 0.1cm, shorten >= 0.1cm] (0.5,1) to[out=50, in=130] (1.25,1);
\draw[shorten <= 0.1cm, shorten >= 0.1cm] (0.5,1) to[out=-50, in=-130] (1.25,1);
\end{tikzpicture}
\eea
Other products behave like, in loose notations, $BE=B$, $DE=D$, $B^2 = E, D^2 =E,AB=BA=D, AD=DA=B, BD=DB=A.$ Thus $E$ is the unit
element of the multiplication law. Furthermore, it is simple to check
that the law its associative $(AB)D= D^2 = E = A^2=A(BD),$
$(AB)B=  DB = A =(AE) = A(B^2)$, commutative and any element of the
graph basis is its own inverse.  As expected $ \cK ( 2) = \mC ( S_2 ) \otimes \mC ( S_2) $:  
 the diagonal conjugation action which defines $\cK (2)$ leaves the permutation pairs invariant.

\

\noindent{\bf Algebra $\cK_{\ung}(n=3)$ -}
The number of invariants is $Z_3(3)=11$
and this makes the multiplication table more complicated.  
We have listed the products in  appendix \ref{app:11m}.
In fact, 21 products involving the  unit $E  = \idtr$ are known.    

From the multiplication table, we see that $\cK_{\ung}(3)$ is commutative,
that some basis elements can be factorized. 
We illustrate the product for a non trivial situation obtained 
by taking the product of the following elements
$A_{(12),(123),id}$ and $A_{(123),(123),(12)}$ depicted as: 
\bea
A_{(12),(123),id}  = \scalebox{2}{{\ac{$1$}}} \qquad \qquad
A_{(123),(123),(12)} = \scalebox{2}{\dtwo{$3$}}
\eea
Then we get: 
\bea
 \scalebox{2}{{\ac{$1$}}}  \; \;\cdot\scalebox{2}{\dtwo{$3$}}
=\frac{2}{3}
\scalebox{2}{{\ac{$2$}}} + \frac13 \scalebox{2}{\dtwo{$2$}}
 = \scalebox{2}{\dtwo{$3$}}\; \;\cdot\scalebox{2}{{\ac{$1$}}} 
\eea

\section{PCAs and correlators }
\label{sect:PCAandCorr}

In this section, we undertake the analysis of correlators of Gaussian tensor models, building on the permutation description for tensor model observables introduced in \cite{Sanjo} and reviewed earlier. 
We start with one-point functions of tensor model 
observables corresponding to central elements in $ \cK ( n)$. These observables are labelled 
by triples of Young diagrams and are sums of permutation basis operators weighted by characters.  As explained in section \ref{sec:PCA} such sums of permutations weighted by characters lead to an overcomplete basis for the centre in $ \cK ( n )$. 
Correlators parametrized by Young diagrams using characters  have also been highlighted in \cite{DR1706,MM1706}.  
Analogous correlators at higher $d$ are expressed in terms of 
sums of products of Kronecker coefficients. In section \ref{sect:corrYD},
we use known results on Kronecker coefficients to give  explicit formulae  for several families of correlators.  In section  \ref{sect:CCorr} we consider normal ordered 2-point correlators, which we have briefly discussed in \cite{Sanjo}.  We show that the 
tensor model observables corresponding to the WA basis for $ \cK ( n )$ 
discussed in section \ref{sec:PCA} provide an orthogonal basis for these 2-point functions. 
This orthogonality property has also been considered in \cite{DR1706,DGT1707}.

\subsection{Correlators for central observables }
\label{sect:corrFT}

We start our analysis with  correlators  of general observables at $d=3$, parametrised by 
permutations, corresponding to general elements of $ \cK (n )$. We then specialize to central observables labelled by triples of Young diagrams: as we saw in section \ref{sec:PCA} triples of projectors labelled by Young diagrams lead to an overcomplete basis for the centre $ \cZ ( \cK (n) ) $.  We extend the discussion to any $d$. 

\

\noindent{\bf Rank $d=3$ correlator -}
In rank $d=3$  tensor models, 
consider a general  observable $\cO_{ \sigma_1 , \sigma_2 , \sigma_3 }$ defined
by three permutations $\s_i$, $i=1,2,3$. 
The expectation  value $ \langle  \cO_{ \sigma_1 , \sigma_2 , \sigma_3 } \rangle $
evaluates in the Gaussian measure, using appendix \ref{app:correlators}.   
We  write: 
\bea \label{six9}
&& \langle  \cO_{ \sigma_1 , \sigma_2 , \sigma_3 } \rangle = 
\sum_{ \gamma } N^{ \cy( \gamma \sigma_1 ) + \cy( \gamma \sigma_2 ) + \cy( \gamma \sigma_3)  } \\  
&& = \sum_{ \gamma } \sum_{  \alpha_1 , \alpha_2 , \alpha_3 } 
N^{ \cy( \alpha_1) + \cy( \alpha_2 ) + \cy( \alpha_3)  } 
\delta ( \gamma \sigma_1  \alpha_1 ) \delta ( \gamma \sigma_2  \alpha_2 ) \delta ( \gamma \sigma_3  \alpha_3 )   \cr 
&& = \sum_{ \gamma } \sum_{    \alpha_l } 
\sum_{    R_l } 
\frac{d(R_1) d(R_2) d(R_3)}{(n!)^3 } 
N^{ \cy(\alpha_1) + \cy( \alpha_2 ) + \cy(\alpha_3)  } 
 \chi^{ R_1} ( \gamma \sigma_1 \alpha_1 ) \chi^{ R_2} ( \gamma \sigma_2 \alpha_2 ) 
  \chi^{ R_3} ( \gamma \sigma_3 \alpha_3 ) \nonumber
\eea
where we expand the $\delta$'s over $S_n$ using characters as in \eqref{deltasym} of 
appendix \ref{sapp:syrep}. 
Now we use three facts:  (1) $ \sum_{ \alpha } N^{ \cy_\alpha } \alpha $ is a central element 
 in $\mC(S_n)$, since    $\cy(g\alpha\g^{-1})
= \cy(\alpha)$,  (2) if $B$ is a central element,  characters factorize as
$\chi(AB) = \frac{1}{d(R)} \chi^R(A)\chi^R(B)$,  see \eqref{chiAB}, appendix \ref{sapp:syrep}, 
and  (3) that characters extend by linearity over $\mC(S_n)$,  
$\chi^R(\sum_{\g} c_{\g} \g) = \sum_{\g} c_{\g} \chi^R(\g)$,  to write \eqref{six9} as
\bea
&& \langle  \cO_{ \sigma_1 , \sigma_2 , \sigma_3 } \rangle = 
\sum_{ \gamma } 
\sum_{    R_l } 
{  1 \over (n!)^3 } 
d(R_1)d(R_2)d(R_3)\cr\cr
&& \times 
 \chi^{ R_1} \Big( \gamma \sigma_1 (\sum_{\alpha_1} N^{ \cy( \alpha_1) }\alpha_1) \Big) 
\chi^{ R_2} \Big( \gamma \sigma_2  (\sum_{\alpha_2} N^{ \cy( \alpha_2) }\alpha_2) \Big) 
  \chi^{ R_3} \Big( \gamma \sigma_3  (\sum_{\alpha_3} N^{ \cy( \alpha_3) }\alpha_3) \Big) 
\cr\cr
&& 
  = 
 \sum_{ \gamma } 
\sum_{    R_l } 
{  1 \over (n!)^3 } \chi^{ R_1} ( \gamma \sigma_1 )
\chi^{ R_2} ( \gamma \sigma_2 )
\chi^{ R_3} ( \gamma \sigma_3 )
 \sum_{\alpha_l} N^{ \cy( \alpha_1) +\cy( \alpha_2)+ \cy( \alpha_3)}
  \chi^{ R_1}(\alpha_1) \chi^{ R_2}(\alpha_2)\chi^{ R_3}(\alpha_3) 
\cr\cr
&&
= \sum_{ \gamma } 
\sum_{    R_l } 
\Big[\prod_{l=1}^3 \Dim_N(R_l)\Big] 
\chi^{ R_1} ( \gamma \sigma_1)
\chi^{ R_2} ( \gamma \sigma_2 )
\chi^{ R_3} ( \gamma \sigma_3  ) 
\label{Osss}
\eea
where,  in the last stage, we use \eqref{chiN} of appendix \ref{sapp:syrep}. 
$\Dim_N(R)$ is the dimension of the representation of the unitary group $U(N)$
determined by the Young tableau $R$.  
 Consider sums of $ \cO_{ \sigma_1 , \sigma_2 , \sigma_3}$ weighted by 
characters with Young diagrams $S_l \vdash n$, $l=1,2,3$,  and define the function
\bea 
 \langle \cO_{ S_1 , S_2 , S_3  }  \rangle  
= { 1 \over (  n!)^3 }  \sum_{ \sigma_l \in S_n }
\chi^{ S_1} ( \sigma_1 ) \chi^{S_2} ( \sigma_2 ) \chi^{S_3}  ( \sigma_3 )
\langle \cO_{\sigma_1,\s_2,\s_3 }  \rangle  
\eea
These observables correspond to central elements in $ \cK ( n )$ by the map 
(\ref{mapPermsObs}). 
We can use character orthogonality (see \eqref{cscs}
and \eqref{dims}) 
\bea 
\sum_{ \sigma  } \chi^S ( \sigma ) \chi^{R} ( \sigma \gamma ) = { n! \over d (R)} \delta^{ R,S} \chi^R ( \gamma )  
\eea
to write 
\bea 
&&  \langle \cO_{ S_1 , S_2 , S_3  }  \rangle 
=   \Big[\prod_{l=1}^3 \frac{\Dim_N(S_l)}{ d(S_l) } \Big]
\sum_{ \gamma } \chi^{ S_1} ( \gamma) \chi^{ S_2} ( \gamma ) \chi^{ S_3} ( \gamma ) 
=n!
\Big[\prod_{l=1}^3 \frac{\Dim_N (S_l)}{ d(S_l) } \Big]~ \sC(S_1,S_2,S_3)
\cr\cr 
&& = \frac{1}{(n!)^2}
\Big[\prod_{l=1}^3 f_N(S_l)\Big]~ \sC(S_1,S_2,S_3) 
\label{O1sss}
\eea  
Thus the correlators $\langle \cO_{ S_1 , S_2 , S_3}  \rangle $   are proportional
to the Kronecker coefficients. The factors $ f_N(S_i) $ are products of box weights of the Young diagrams \eqref{fN}. In a large $N$ limit where we are considering tensor invariants of degree $n$, hence Young diagrams with $n$ boxes, where $n$ is kept fixed and $N$ is taken to 
infinity, the $f$-factors behave like $ N^n $ at leading order. The relative magnitudes of the correlators in this limit is determined purely by the Kronecker coefficients. 
At finite $N$, since we are dealing with a theory where the tensor indices are taking $N$ possible values, the Young diagrams are cut-off to have no more than $N$ rows.

\

\noindent{\bf Rank $d$ correlator -}
The above formula \eqref{Osss} can be generalized at any rank $d$. 
Rank $d$ Gaussian correlators of 
a generic observables $ \cO_{ \sigma_1 , \sigma_2 , \dots , \sigma_d } $ labelled by $d$ permutations $\sigma_l$, $l=1,2,\dots,d$. 
We will sketch the above analysis for  $ \cO_{ \sigma_1 , \sigma_2 ,  \dots , \sigma_d } $: 
\be \label{Os1s}
 \langle  \cO_{ \sigma_1 , \sigma_2 ,  \dots , \sigma_d} \rangle = 
\sum_{ \gamma } N^{ \cy( \gamma \sigma_1 ) + \cy( \gamma \sigma_2 ) +\dots + \cy( \gamma \sigma_d ) }  
 = \sum_{ \gamma }  \sum_{R_l}
{  1 \over (n!)^d } \Big[\prod_{l=1}^d d( R_l) \chi^{ R_l}\Big( \gamma \sigma_l \sum_{ \alpha_l}N^{\cy(\alpha_l)} \alpha_l \Big) \Big]
\ee
Then, using the same technique, we arrive at
 \be
 \langle \cO_{ \sigma_1 , \sigma_2 ,  \dots , \sigma_d } \rangle
=\sum_{ \gamma }  \sum_{ R_l } \Big[ \prod_{l=1}^d  \Dim_N (R_l) \chi^{ R_l} ( \gamma \sigma_l )\Big]  
\ee

We calculate the Fourier transform of $ \cO_{ \sigma_1 , \sigma_2 , \dots , \sigma_d}$ weighted by 
characters. Let $S_l$, $l=1,\dots,d$, partitions of $n$, 
\bea 
 \langle  \cO_{ S_1 , S_2 , \dots , S_d  }   \rangle  
& =& \sum_{ \sigma_1 , \sigma_2 , \dots , \sigma_d } { 1 \over (  n!)^d } 
\chi^{ S_1} ( \sigma_1 ) \chi^{S_2} ( \sigma_2 ) \dots \chi^{S_d}  ( \sigma_d ) 
\langle\cO_{ \sigma_1 , \sigma_2 , \dots , \sigma_d }  \rangle  \cr 
&=&  
\Big[\prod_{l=1}^d \frac{\Dim_N(S_l)}{ d(S_l) } \Big] \sum_{ \gamma } 
\Big[\prod_{l=1}^d \chi^{ S_l} ( \gamma) \Big]  
\eea 
Introducing $\sC_d ( S_1 , S_2 , \dots , S_d )={ 1 \over n! }  \sum_{ \gamma } 
\Big[\prod_{l=1}^d \chi^{ S_l} ( \gamma) \Big]$,  the number of invariants 
in $ S_1 \otimes S_2 \otimes \dots \otimes S_d $,  we can write 
\bea \label{corrFourier}
  \langle \cO_{ S_1, S_2, \dots , S_d  } \rangle
= n!  \Big[\prod_{l=1}^d \frac{\Dim_N(S_l)}{ d(S_l) } \Big]  \sC_d ( S_1 , S_2 , \dots , S_d ) 
\eea
  As an illustration, 
restricting to rank $d=4$, and using the relation \eqref{c4=c32}, that is $\sC_4 ( S_1 , S_2 , S_3 , S_4 ) =  \sum_{ S } \sC ( S_1 , S_2 , S ) \sC ( S , S_3 , S_4 ) $ counting the number of invariants in $ S_1 \otimes S_2 \otimes S_3 \otimes S_4 $, we have  
\be \label{corrFourier4}
  \langle \cO_{ S_1, S_2, S_3 , S_4  } \rangle
= n!  \Big[\prod_{l=1}^4 \frac{\Dim_N(S_l)}{ d(S_l) } \Big]  
 \sum_{ S } \sC ( S_1 , S_2 , S ) \sC ( S , S_3 , S_4 )  
\ee
Note that 
$\sC_d(S_1,\dots, S_d)$ can be decomposed 
as a sum of Kronecker coefficients convoluted 
in one of their indices. A possible 
sequence of such a convolution could be $\sum_{\bar S_l}
\sC(S_1,S_2,\bar S_1)$ $\sC(\bar S_1,S_3, \bar S_2)\sC(\bar S_2,S_4,\bar S_3)\sC(\bar S_3,S_5,\bar S_4) \dots$. 
Any permutation over $S_i$'s giving a different sequence should give the same answer $\sC_{d}(S_1,\dots,S_d)$. 
Then, we observe that there is graphical way to encode the expansion 
of $ \langle \cO_{ S_1, S_2, \dots , S_d  } \rangle$ as
 a convoluted sum 
of Kronecker coefficient $\sC(S_a,S_b,S_c)$. 
Reminiscent of Feynman 
rules, we associate $\sC(S_a,S_b,S_c)$ with a trivalent graph vertices
and half edges labelled by $S_a$, $S_b$ and $S_c$, each symbol $S$ summed over between two 
Kroneckers $\sC(S_a,S_b,S)$ and $\sC(S,S_{b'},S_{c'})$ 
is associated with an edge between the vertices $\sC(S_a,S_b,S)$ and $\sC(S,S_{b'},S_{c'})$. It is not hard 
to realize that the corresponding graph is always a  
tree graph with vertex set with vertices of degree 3 and $d$ 
half-edges.
Therefore, each correlator $\langle \cO_{ S_1 , S_2 , \dots , S_d  }\rangle$ 
is associated with a decomposition in several tree graphs
the half edges of which are labelled by 
 $S_1,\dots, S_d$.  Any of these tree graphs 
to which we finally give a weight $n!  \Big[\prod_{l=1}^d \frac{\Dim_N(S_l)}{ d(S_l) } \Big]$ is a valid representative
of $ \langle \cO_{ S_1, S_2, \dots , S_d  } \rangle$. 
For example,  
the correlator  \eqref{corrFourier4} is associated with any of
the following trees: 
 \bea 
\begin{tikzpicture}
\draw (0,0)  -- (1,0) -- (1.5,0.5) ; 
\draw(-0.5,-0.5) -- (0,0) -- (-0.5,0.5) ; 
\draw (1.5,-0.5) -- (1,0) ;
\draw[color=black] (0.5,0.2) node {$S$};
\draw[color=black] (-0.6,-0.6) node {$S_1$};
\draw[color=black] (-0.6,0.7) node {$S_2$};
\draw[color=black] (1.7,0.7) node {$S_3$};
\draw[color=black] (1.7,-0.6) node {$S_4$};
 \fill[fill=black] (0,0) circle (0.1);
\fill[fill=black] (1,0) circle (0.1);
\end{tikzpicture}\qquad 
\begin{tikzpicture}
\draw (0,0)  -- (1,0) -- (1.5,0.5) ; 
\draw(-0.5,-0.5) -- (0,0) -- (-0.5,0.5) ; 
\draw (1.5,-0.5) -- (1,0) ;
\draw[color=black] (0.5,0.2) node {$S$};
\draw[color=black] (-0.6,-0.6) node {$S_1$};
\draw[color=black] (-0.6,0.7) node {$S_3$};
\draw[color=black] (1.7,0.7) node {$S_2$};
\draw[color=black] (1.7,-0.6) node {$S_4$};
 \fill[fill=black] (0,0) circle (0.1);
\fill[fill=black] (1,0) circle (0.1);
\end{tikzpicture}\qquad
\begin{tikzpicture}
\draw (0,0)  -- (1,0) -- (1.5,0.5) ; 
\draw(-0.5,-0.5) -- (0,0) -- (-0.5,0.5) ; 
\draw (1.5,-0.5) -- (1,0) ;
\draw[color=black] (0.5,0.2) node {$S$};
\draw[color=black] (-0.6,-0.6) node {$S_1$};
\draw[color=black] (-0.6,0.7) node {$S_4$};
\draw[color=black] (1.7,0.7) node {$S_2$};
\draw[color=black] (1.7,-0.6) node {$S_3$};
 \fill[fill=black] (0,0) circle (0.1);
\fill[fill=black] (1,0) circle (0.1);
\end{tikzpicture}
\eea
At order $5$, there is a unique unlabelled tree  configuration 
(see Figure \ref{fig:tree} A) 
yielding 15 different tree labellings of half edges (or leaves) and, at order 6, 
there are 2 unlabelled tree configurations (see, Figure \ref{fig:tree} B1 and B2) giving 120 different tree labellings of half edges (B1 yields 90, and B2,  30). 
\begin{figure}[h]
\centering
\begin{tikzpicture}
\draw (0,0)  -- (1,0) -- (1.5,0.5) ;
\draw (1,0)  -- (2,0) ; 
\draw(-0.5,-0.5) -- (0,0) -- (-0.5,0.5) ; 
\draw (2.5,-0.5) -- (2,0) ;
\draw (2.5,0.5) -- (2,0) ;
 \fill[fill=black] (0,0) circle (0.1);
\fill[fill=black] (1,0) circle (0.1);
\fill[fill=black] (2,0) circle (0.1);
\end{tikzpicture}
A
\qquad 
\begin{tikzpicture}
\draw (0,0)  -- (1,0) -- (1.5,0.5) ;
\draw (1,0)  -- (2,0) -- (3,0); 
\draw(-0.5,-0.5) -- (0,0) -- (-0.5,0.5) ; 
\draw (2.5,-0.5) -- (2,0) ;
\draw (3.5,-0.5) -- (3,0) ;
\draw (3.5,0.5) -- (3,0) ;
 \fill[fill=black] (0,0) circle (0.1);
\fill[fill=black] (1,0) circle (0.1);
\fill[fill=black] (2,0) circle (0.1);
\fill[fill=black] (3,0) circle (0.1);
\end{tikzpicture}
B1\qquad  
\begin{tikzpicture}
\draw (0,0)  -- (1,0) -- (2,0) -- (2.5,0.5) ; 
\draw(-0.5,-0.5) -- (0,0) -- (-0.5,0.5) ; 
\draw (2.5,-0.5) -- (2,0) ;
\draw (1,0) -- (1,0.5) ;
\draw (0.5,1) -- (1,0.5) ;
\draw (1.5,1) -- (1,0.5) ;
  \fill[fill=black] (0,0) circle (0.1);
\fill[fill=black] (1,0) circle (0.1);
\fill[fill=black] (2,0) circle (0.1);
\fill[fill=black] (1,0.5) circle (0.1);
\end{tikzpicture}
B2
\label{fig:tree}
\end{figure}
The counting of that type of trees is 
{\it the counting of 3-regular (or binary) trees 
with $d$ leaves and $d-2$ vertices (and so $2d-3$ edges)}. 
This will involve a mixture of a counting of the so-called
binary beanstalk (A and B1)  but also more general terms. 
For $d=3,4,5,6$, we have the sequence
\bea
1,3,15,120, 
\eea
respectively, which should be completed at any $d$. 

\subsection{Correlators and Kronecker coefficients: Explicit examples}
\label{sect:corrYD}

To illustrate the above formula \eqref{O1sss}, we evaluate correlators of rank 3 
tensor models as a function of $N>1$, and $n\ge 0$, for some particular Young diagrams. 

For any $S_2$ and $S_3$, and for $S_1=[n]=\underbrace{\scriptsize \begin{ytableau}
 \empty &\empty  &
\end{ytableau} \dots
\begin{ytableau}
\empty &
\end{ytableau}}_{n-{\text{boxes}}}\,$, then $\sC(S_1,S_2,S_3)=1$ and 
so $\langle \cO_{ S_1 , S_2 , S_3  }  \rangle  
= [\prod_{l=1}^3 f_N(S_l)]/(n!)^2$, from  \eqref{O1sss}. That 
computes to 
\be\label{cofNfN}
 \langle \cO_{ S_1 , S_2 , S_3  }  \rangle  
 =  (N-2)(N-3)\dots (N-1-n) \frac{f_N(S_2)f_N(S_3)}{(n!)^2}
\ee
Note that, in the following, we consider that $N$ is large enough compared to $n$. 
Specifying $S_2$ and $S_3$ to give a more precise formula
for the correlator. For all $n$, consider the Young diagrams defined by 
\bea
S_l = [n]=\underbrace{\scriptsize \begin{ytableau}
 \empty &\empty  &
\end{ytableau} \dots
\begin{ytableau}
\empty &
\end{ytableau}}_{n-{\text{boxes}}}\; , \;\, l=1,2,3,
\eea
such that from \eqref{cofNfN}, one gets
\be
  \langle \cO_{ S_1 , S_2 , S_3  }  \rangle  
=   \frac{[(N-2)(N-3)\dots (N-1-n)]^3}{(n!)^2} 
=  \frac{[(N-2)!]^3}{(n!)^2[(N-2-n)!]^3} 
\ee
Varying the order of the symmetric group, that is varying $n=1,2,3,\dots,10$ gives Table \ref{taboOsym}. 

\begin{table}
\begin{center}
\begin{tabular}{|c||c|c|c|c|c|}
\hline 
             & n=1  & n=2   & n=3           & n=4        & n= 5   \\ 
$N=2 $  & 0      &  0       & -              & -             & -       \\ 
$N=3 $  & 1      &  0       &  0            &  -             & -        \\ 
$N=4 $  & 8      &  2       &  0            &  0             &  -       \\ 
$N=5 $  &  27   &  54      &  6            &  0            &  0       \\ 
$N=6 $  & 64    &  432    &  384        &  24           &  0       \\ 
$N=7 $  & 125  &  2000   & 6000       &  3000       &  120       \\ 
$N=8 $  &  216  & 6750   & 48000     &  81000     &   25920      \\ 
$N=9 $  &  343  & 18522 & 257250    &  1029000 &    1111320     \\ 
$N=10 $ & 512  & 43904  & 1053696  &  8232000 &     21073920    \\ 
\hline
\end{tabular}
\caption{Evaluation of $\langle \cO_{ [n] , [n], [n]  }  \rangle$.}
\label{taboOsym}
\end{center}
\end{table}

\

Let us introduce the notation $[n-k,1,\dots,1] = [n-k, 1^k]$, 
where $k$ is the number of $1$ appearing in the dots. 
For all $n$, $\sC([n],[n-k,1^k],[n-k,1^k])= 1$. 
Hence, for $k\in \{ 0, 1, \cdots   , n-1 \} $, 
\be
S_1= [n] 
 = \underbrace{\scriptsize \begin{ytableau}
 \empty &\empty  &
\end{ytableau} \dots
\begin{ytableau}
\empty &
\end{ytableau}}_{n-{\text{boxes}}}\,, 
\quad \; 
S_2=S_3=[n-k,1^k]=
\overbrace{
\scriptsize  \begin{ytableau}
 \empty &\empty  & \\
\vdots \\
\vdots \\
\empty \\
\end{ytableau} \dots
\begin{ytableau}
\empty &
\end{ytableau}
}^{n-k-{\text{boxes}}} 
\ee
where the dots in the 1st vertical column refers to $k$-times a block of size 1. 
Hence for this class of Young diagrams, the correlator calculation can be easily made. 
\begin{align}\label{Os1sksk}
&  \langle \cO_{  [n]  , [ n-k,1^k] ,  [ n-k ,1^k]  }  \rangle  
= (N-2)(N-3)\dots (N-1-n) \crcr
& \times \frac{[(N-2)(N-3)^2
  \dots (N-1- k)^2 (N-k-2)^2  (N-k-3) \dots (N-1-(n-k))]^2}{(n!)^2} \crcr
&  = 
\frac{[(N-2)(N-3)\dots (N-1-(n-k))]^3 }{(n!)^2}\crcr
&\times   (N-3)^2(N-4)^2 \dots (N-2-k)^2 (N-n+k-2) (N-n+k-3)\dots (N-1-n) \crcr
&
 = \frac{[(N-2)!]^3\Big[ \prod_{l=1}^{k} (N-2-l)^2\Big]\Big[ \prod_{l=1}^{k} (N-n+k-1-l)\Big] }{(n!)^2[(N-n+k-2)!]^3} 
\end{align}
It turns out that $\langle \cO_{ [n]  , [ n-k, 1^k ]  , [ n - k , 1^k ]  }  \rangle $ is not
necessarily an integer for any values of $k$. One can check this by direct evaluation 
for instance using $N=7, n=3, \langle \cO_{ [3] , [2,1], [2,1] }\rangle= \frac{32000}{3}$. 
In any case, we further restrict to the case $k=1$ and give $\langle \cO_{ [n] , [n-1,1], [n-1,1]  }  \rangle$
for different values of $N$ and $n\geq 2$, in Table \ref{tabOsym1}. 

\begin{table}
\begin{center}
\begin{tabular}{|c||c|c|c|c|c|}
\hline 
             & n=2  & n=3                       & n=4           & n=5        & n= 6     \\
$N=2 $  & 0      &  -                          & -              & -             & -          \\
$N=3 $  & 0      &  0                          &  -             &  -             & -           \\
$N=4 $  & 2      &  0                          &  0              &  -             &  -        \\
$N=5 $  & 54    &  24                        &  0               &  0            &  -       \\
$N=6 $  & 432    &  864                      & 216             &  0          &  0        \\
$N=7 $  & 2000  &  $\frac{32000}{3}$   & 12000       &  1920       &  0  \\
$N=8 $  & 6750  &  75000                  & 225000       &  162000       &  18000\\
\hline
\end{tabular}
\caption{Evaluation of $\langle \cO_{ [n] , [n-1,1], [n-1,1]  }  \rangle$.}
\label{tabOsym1}
\end{center}
\end{table}

Next, we relax the assumption that $S_1$ is 
the symmetric representation. 
Avoiding  trivial cases, 
consider $n \geq 2$ and $k\leq { n \over 2 } $, then we consider  the following 
\bea
S_1=S_2=S_3= [n-k,k]
 = \overbrace{\scriptsize \begin{ytableau}
 \empty &\empty  & \\
\empty & \empty & \dots & \\
\end{ytableau} \dots
\begin{ytableau}
\empty &
\end{ytableau}}^{n-k-{\text{boxes}}}
\eea
where of course the second row has $k$ boxes. 

According to a computation using SAGE mathematical 
software up to order $n=25$,  for all $n \ge 3k $, the Kronecker coefficient of the two-row Young diagrams 
is given by 
\bea  
\qquad \sC([n-k,k],[n-k,k],[n-k,k]) = \Big\lfloor \frac{k}{2} \Big\rfloor +1 
\eea
where $\lfloor \cdot  \rfloor$ denotes the floor function. 
Therefore, from \eqref{O1sss}, we have   
\be\label{Osksksk}
\langle \cO_{ [n-k,k] , [n-k,k] , [n-k,k]  }  \rangle  
= \frac{\Big[ \Big(\prod_{l=1}^{n-k}(N-1-l)\Big)
\Big(\prod_{l=1}^{k}( N-2-l)\Big) \Big]^3}{(n!)^2} \left (  \Big\lfloor \frac{k}{2} \Big\rfloor +1    \right ) 
\ee
for $ n \le 25 $ and $ k \le { n \over 3 } $: while we have checked for $n \le 25$, we expect this will hold for higher $n$ as well. 
General stability properties of Kronecker coefficients are described in \cite{vallejo}. 
This correlator \eqref{Osksksk} is not integral in general. 
We obtain  Table \ref{tabOsym2}  listing some values of $\langle \cO_{ [n-1,1]  , [n-1,1] , [n-1,1] }  \rangle  $
(at $k=1$). Note that for this table, the column $n=2$ coincides with the column $n=2$ of Table \ref{tabOsym1} as expected from the 
correlator formulas \eqref{Osksksk} and \eqref{Os1sksk}
at $n=2$ and $k=1$. 
\begin{table}
\begin{center}
\begin{tabular}{|c||c|c|c|c|c|}
\hline 
             & n=2  & n=3                    & n=4                       & n=5                 & n= 6     \\
$N=2 $  & 0      &  -                       & -                            &  -                    & -          \\
$N=3 $  & 0      &  0                       &  -                           &  -                     & -           \\
$N=4 $  & 2      &  $\frac{2}{9}$      & 0                            &  -                     &  -        \\
$N=5 $  & 54      &  48                       &  3                       &  0           &  -       \\
$N=6 $  & 432    &  1296                      & 648                  &  $\frac{648}{25}$   & 0     \\
$N=7 $  & 2000  &  $\frac{128000}{9}$   & 24000              &  7680                     &  $\frac{640}{3}$  \\
$N=8 $  & 6750    &  93750                   & 375000             & 405000                   &  90000  \\
$N=9$  & 18522   &  444528                   & 3472875          &8890560                 &  6667920 \\
\hline
\end{tabular}
\caption{Evaluation of $\langle \cO_{ [n-1,1] , [n-1,1], [n-1,1]  }  \rangle$.}
\label{tabOsym2}
\end{center}
\end{table}

Finally, let us consider three Young diagrams with rectangular shape. Consider $n$ to be divisible such that $n=jk$, with $j$ and $k$ integers: 
\bea
S_1 = S_2 = S_3 = [ k^j ] = 
\begin{array}{c}
\overbrace{\scriptsize  
\begin{ytableau}
 \empty &\empty  & \dots & \empty &\empty \\
\vdots  &\empty  & \dots & \empty &\empty \\
\vdots  &\empty  & \dots & \empty &\empty  \\
\empty &\empty  & \dots & \empty &\empty \\ 
\end{ytableau} 
}^{k-{\text{boxes}}}
\end{array} 
\eea 
For simplicity, let us assume that $j\leq k < N-1$,  
\bea
\langle \cO_{ [k^j ]  , [ k^j ] ,  [ k^j ]   }  \rangle  
=  \frac{1}{(n!)^2} \Big[ \prod_{l=1}^{k}  \frac{ (N-1-l)!} {(N-j-1-l)!} \Big]^3 \sC([k^j],[k^j],[k^j])
\eea
which can be again computed. We will restrict to the lowest
orders in $k$: 

- If the number of rows is $j=2$, such that $S_l = [k^2]$, for
$k$ even, we have $\sC([k^2],[k^2],[k^2])=1$ and for $k$ odd,  $\sC([k^2],[k^2],[k^2])=0$. 
In this case, the correlator values coincide with columns $n=2$ and  $4$
of Table \ref{tabOsym2} but, at order $n=8$, they start to differ. 
In this case the correlator is also not an integer. 
We obtain the l.h.s of Table \ref{tabOrect}. 

\begin{table}
\begin{center}
\begin{tabular}{|c||c|c|c|c|}
\hline 
             & n=2       & n=4                      & n=6     & n =8    \\
$N=2 $  & -         & -                            &  0         & -   \\
$N=3 $  & 0         & -                            &  0         & -    \\
$N=4 $  & 2          & 0                          &  0         &  -       \\
$N=5 $  & 54          & 3                         &  0         &  -       \\
$N=6 $  & 432      & 648                    & 0          &   -    \\
$N=7 $  & 2000   & 24000                  &  0        &  -  \\
$N=8 $  & 6750       & 375000             & 0       & $\frac{2430000}{49}$ \\
$N=9$  & 18522               & 3472875      &0   &   17010000 \\
\hline
\end{tabular}
\qquad \qquad \quad
\begin{tabular}{|c||c|c|}
\hline 
             & n=3               & n=6                   \\
$N=2 $  & -                    & -                             \\
$N=3 $  & -                     & -                                 \\
$N=4 $  & $\frac29$         &-                                        \\  
$N= 5 $  & 6                    & -                                  \\
$N=6 $  & 48                   & -                                   \\
$N=7 $  & $\frac{2000}{9}$   & $\frac{80}{3}$                \\
$N=8 $  & 750               & $\frac{1250}{3}$            \\
$N=9$  & 2058               & $\frac{15435}{4}$       \\
\hline
\end{tabular}
\caption{Evaluation of $\langle \cO_{ [k^2] , [k^2], [k^2]  }  \rangle$, for $n=2k$ (left) and of $\langle \cO_{ [k^3] , [k^3], [k^3]  }  \rangle$, with $n=3k$ (right).}
\label{tabOrect}
\end{center}
\end{table}

- If the number of rows is $j=3$, such that $S_l=[k^3]$, 
we have, for $k\leq 30$ and  for $k$ even, $\sC([k^3],[k^3],[k^3])=k/2$ and for $k$ odd, $\sC([k^3],[k^3],[k^3])=\lfloor k/2 \rfloor +1$. 
We obtain the r.h.s of Table \ref{tabOrect}.  

\subsection{Orthogonality of two-point functions and WA basis for  $ \cK (n) $  } 
\label{sect:CCorr}

If we consider correlators of normal-ordered operators (see appendix \ref{app:correlators}), then we can write them in terms of the product in $ \cK ( n ) $ and the delta function 
\be \label{ooc}
\langle \cO_{ \sigma_1 , \sigma_2 , \sigma_3 } \cO_{ \tau_1 , \tau_2 , \tau_3 } \rangle 
= \sum_{ \gamma_1 , \gamma_2 } N^{ 3 n } \bdel _3[(  \sigma_1 \otimes \sigma_2 \otimes \sigma_3 ) \gamma_1^{ \otimes 3 }   (  \tau_1 \otimes \tau_2 \otimes \tau_3 ) \gamma_2^{ \otimes 3 }
( \Omega_1 \otimes \Omega_2 \otimes \Omega_3 ) ]
\ee
with $\Omega_i =\sum_{\alpha_i\in S_n}N^{\cy(\alpha_i)-n}\alpha_i$. An important property of the $\Omega_i$'s  is that they are central elements of $ \mC ( S_n)$. Indeed,
using $\cy(\g \alpha \g^{-1})=\cy(\alpha)$, 
one finds that $\g \Omega_i \g^{-1}= \Omega_i$. 
The tensor product 
of $ \Omega $ factors defines a central element in $ \cK ( n )$, hence the product in \eqref{ooc} involves the triples of permutations defining the two observables and the central elements  defined by the $\Omega$-factors. The above  equation \eqref{ooc} was derived as eq. (42) in \cite{Sanjo} (or (102) of the arXiv version)). There is a mistake in a following equation  ((105) of the arXiv)
which should be
\bea 
\langle \cO_{ \sigma_1 , \sigma_2 , \sigma_3 } \cO_{ \tau_1 , \tau_2 , \tau_3 } \rangle  =
n! \sum_{ \mu \in S_n } \bdel_{ 2 } [ (  \beta_2^{-1} \mu^{-1} \alpha_2 \mu \Omega_2) \otimes ( \beta_3^{-1} \mu^{-1} \alpha_3 \mu \Omega_3 ) \Delta ( \Omega_1 ) ]    
\eea 
where $\alpha_2 =\s_1^{-1}\s_2$, $\alpha_3 =\s_1^{-1}\s_3$, 
$\beta_2 =\tau_1^{-1}\tau_2$, $\beta_3 =\tau_1^{-1}\tau_3$,
and 
\be
\Delta ( \Omega ) = \sum_{ \alpha } N^{ \cy( \alpha )} 
\alpha \otimes \alpha  
\ee

The Fourier basis (or representation basis ) of operators is defined by 
\bea 
\cO^{ R , S , T }_{ \tau_1 , \tau_2} && = \sum_{ \sigma_1 , \sigma_2 , \sigma_3 } 
\bdel_3  ( Q^{ R , S , T}_{\ung; \tau_1 , \tau_2 } \sigma_1^{-1} \otimes \sigma_2^{-1} \otimes \sigma_3^{-1} ) 
\cO_{ \sigma_1 , \sigma_2 , \sigma_3 }  \\
&&  =\kappa_{R,S,T} \sum_{ \sigma_l } \sum_{  i_l ,  j_l } 
C^{ R , S ;T , \tau_1 }_{ i_1 , i_2 ; i_3 } C^{ R , S ; T , \tau_2}_{ j_1 , j_2 ; j_3  } 
D^{ R}_{ i_1 j_1 } ( \sigma_1 ) D^{ S}_{ i_2 j_2 } ( \sigma_2 ) D^{ S}_{ i_3  j_3 } ( \sigma_3 ) \cO_{ \sigma_1 , \sigma_2 , \sigma_3 } \nonumber
\eea
The conjugate operator is 
\bea 
&&
\overline{ \cO^{ R , S , T }_{ \tau_1 , \tau_2 } }
= 
\kappa_{R,S,T} \sum_{ \sigma_l } \sum_{  i_l ,  j_l } 
C^{ R , S ;T , \tau_1 }_{ i_1 , i_2 ; i_3 } C^{ R , S ; T , \tau_2}_{ j_1 , j_2 ; j_3  } 
D^{ R}_{ i_1 j_1 } ( \sigma_1 ) D^{ S}_{ i_2 j_2 } ( \sigma_2 ) D^{ S}_{ i_3  j_3 } ( \sigma_3 ) \cO_{ \sigma_1 , \sigma_2 , \sigma_3 } 
\cr\cr
&&
 = 
\kappa_{R,S,T} \sum_{ \sigma_l } \sum_{  i_l ,  j_l } 
C^{ R , S ;T , \tau_1 }_{ i_1 , i_2 ; i_3 } C^{ R , S ; T , \tau_2}_{ j_1 , j_2 ; j_3  } 
D^{ R}_{ i_1 j_1 } ( \sigma_1 ) D^{ S}_{ i_2 j_2 } ( \sigma_2 ) D^{ S}_{ i_3  j_3 } ( \sigma_3 ) \cO_{ \sigma_1^{-1} , \sigma_2 ^{-1} , \sigma_3^{-1} }   \cr\cr
&&
 = 
\kappa_{R,S,T} \sum_{ \sigma_l } \sum_{  i_l ,  j_l } 
C^{ R , S ;T , \tau_1 }_{ i_1 , i_2 ; i_3 } C^{ R , S ; T , \tau_2}_{ j_1 , j_2 ; j_3  } 
D^{ R}_{ j_1  i_1} ( \sigma_1 ) D^{ S}_{ j_2 i_2 } ( \sigma_2 ) D^{ S}_{ j_3  i_3 } ( \sigma_3 ) \cO_{ \sigma_1 , \sigma_2 , \sigma_3 } 
\cr\cr
&& = 
 \cO^{ R , S , T }_{ \tau_2 , \tau_1 } 
\eea 
where in the last stage of the equality, we simply rename $i_l \to j_l$
and vice-versa. 

The two-point correlator evaluates as: 
\be\label{2ptmeas}
\langle \cO^{ R_1 , S_1 , T_1 }_{ \tau_1 , \tau_1' } \overline { \cO^{ R_2 , S_2 , T_2 }_{ \tau_2 , \tau_2' } }  \rangle 
= \sum_{ \gamma_1 , \gamma_2 } N^{ 3n } \bdel_3[   Q^{ R_1 , S_1 , T_1 }_{ \ung;\tau_1 , \tau_1'} \gamma_1^{ \otimes 3 }
 Q^{ R_2 , S_2 , T_2 }_{\ung; \tau_2' , \tau_2} \gamma_2^{ \otimes 3} (\Omega_1 \otimes \Omega_2 \otimes \Omega_3 )]
\ee
This shows that the  inner product on   tensor model observables, given by the 
Gaussian integral (with a normal ordering prescription which excludes Wick contractions within the observable), is proportional to the group theoretic  inner product  on $ \cK ( n) $ 
with the insertion of the $ \Omega_1 \otimes \Omega_2 \otimes \Omega_3 $ factor. 
The invariance property of the $Q_{\ung}$-operators \eqref{unginvariance} gives 
\bea 
\langle \cO^{ R_1 , S_1 , T_1 }_{ \tau_1 , \tau_1' } \overline { \cO^{ R_2 , S_2 , T_2 }_{ \tau_2 , \tau_2' } }  \rangle  
= (n!)^2 N^{ 3n } \bdel_3 (   Q^{ R_1 , S_1 , T_1 }_{ \ung ;\tau_1 , \tau_1'}
 Q^{ R_2 , S_2 , T_2 }_{\ung ; \tau_2' , \tau_2}  \Omega_1 \otimes \Omega_2 \otimes \Omega_3 ) 
\eea
Use the matrix-multiplication property of the $Q_{\ung}$-operators 
from \eqref{qungmatrix} to write: 
\bea \label{interm}
\langle \cO^{ R_1 , S_1 , T_1 }_{ \tau_1 , \tau_1' } \overline { \cO^{ R_2 , S_2 , T_2 }_{ \tau_2 , \tau_2' } }  \rangle   
= (n!)^2 N^{ 3n } \delta_{ R_1  , R_2} \delta_{ S_1 , S_2 } \delta_{ T_1 , T_2} \delta_{ \tau_1' , \tau_2'} 
 \bdel_3 ( Q^{ R_1 , S_1 , T_1 }_{ \ung; \tau_1 , \tau_2 } \Omega_1 \otimes \Omega_2 \otimes \Omega_3 ) 
\eea

Since $ \Omega_i$ are central elements in $ \mC ( S_n)$, we have 
 \bea 
Q^{ R_1, R_2, R_3 }_{\ung; \tau_1 , \tau_2 } \Omega_1 \otimes \Omega_2 \otimes \Omega_3  
& =& { \chi^{ R_1 } ( \Omega_1 ) \over{ d(R_1 )} } { \chi^{ R_2} ( \Omega_2 ) \over {d(R_2 )} } { \chi^{ R_3 } ( \Omega_3 )  \over {d (R_3)} }  
Q^{ R_1 , R_2 , R_3 }_{\ung;  \tau_1 , \tau_2 }\cr 
& = & N^{ - 3n }  \Dim_N (R_1)  \Dim_N (R_2)  \Dim_N(R_3) Q^{ R_1 , R_2 , R_3 }_{ \tau_1 , \tau_2 }
\eea
where,  recalling the form of $\chi^{R}(\Omega) = 
\sum_\alpha N^{\cy(\alpha)-n}\chi^{R}(\alpha)$, 
we then use \eqref{chiN2} in appendix \ref{sapp:syrep}. 

We also have (as we prove shortly) 
\bea 
&& 
 \bdel_3  ( Q_{ \ung; \tau_1 , \tau_2}^{ R , S , T } ) 
= \kappa_{R,S,T} \sum_{i_l,j_l} C^{R,S;T,\tau_1}_{i_1,i_2;i_3}
C^{R,S;T,\tau_2}_{j_1,j_2;j_3} \delta_{i_1,j_1} \delta_{i_2,j_2} \delta_{i_3,j_3} = \kappa_{R,S,T} d(T) \delta_{ \tau_1 , \tau_2 } 
\eea 
The $Q$'s behave like elementary matrices of a matrix algebra. The $\bdel_3 $ behaves like the trace. 

Using \eqref{interm} and the above, the correlator becomes 
\bea
&&
\langle \cO^{ R_1 , S_1 , T_1 }_{ \tau_1 , \tau_1' } \overline { \cO^{ R_2 , S_2 , T_2 }_{ \tau_2 , \tau_2' } }  \rangle 
= \crcr
&& \delta_{ R_1 , R_2 } \delta_{ S_1 , S_2 } \delta_{ T_1 , T_2 } 
\delta_{ \tau_1 ,\tau_2 }\delta_{ \tau_1'  ,\tau_2' } 
 \kappa_{R_1,S_1,T_1} d(T_1) 
 \Dim_N (R_1) \Dim_N (S_1)  \Dim_N (T_1) 
\eea
and this shows that  $\{\cO^{R,S,T}_{\tau,\tau'}\}$ forms
an orthogonal  basis for Gaussian correlators (with normal ordered prescription) 
arising directly from the $Q^{ R , S , T }_{\ung;  \tau , \tau'} $, which are the 
representation theoretic  WA 
basis of $\cK  ( n)$,  via the map \eqref{mapPermsObs}. 

\subsection{$ \cK^{ \infty} $  and correlators }\label{KinfCorr}  

We have emphasized, in the bulk of the paper, the role of the 
algebra $ \cK ( n)$ in organizing several aspects of correlators. 
Similar algebras arise in the context of matrix theory problems. 
For the half-BPS sector, we have the centre of  $\mC( S_n )$, denoted by 
$ \cZ ( \mC ( S_n )  )$.   A basis at fixed $n$ is labelled by 
partitions, which in turn can be used to organize multi-trace holomorphic functions of 
 a single matrix of degree $n$ (i.e. a multi-trace containing $n$ copies of the matrix $Z$).
 The multiplication of two multi-traces of degrees $n_1$ and $ n_2$ produces a 
 multi-trace of degree $ n_1 + n_2$. Using the map to central elements of 
 $ \cZ ( \mC ( S_n ) )  $, this corresponds to an outer product which takes central 
 elements $ T_1 \in  \cZ ( \mC ( S_{n_1}  ) )$ and $ T_2 \in  \cZ ( \mC ( S_{n_1}  ) )$ 
to get something in $ \cZ ( \mC ( S_{ n_1 + n_2  } ) )$ followed by 
a projection to the centre. The direct sum 
\bea 
\bigoplus_{ n =0}^{ \infty } \cZ ( \mC (  S_{ n } )  ) 
\eea
provides a natural setting for correlators. 

 For the 2-matrix problem, relevant to the quarter BPS sector of
$ N=4$ SYM, there is a PCA $ \cA ( m , n ) $ corresponding to multi-traces 
with $m$ copies of $X $ and $n$ copies of $Y$. In this case too, it is natural to consider a direct sum over all $ m , n \ge 0$.

For the problem which is our present main focus, rank-$3$ complex  tensors, 
the analogous infinite dimensional  associative algebra is 
\bea 
\cK^{ \infty} = \bigoplus_{ n=0}^{ \infty } \cK ( n ) 
\eea
There are in fact two associative products on this vector space. 
The product at fixed $n$ has been the main subject of this paper. The WA decomposition at fixed $n$ is related to the formula for $ \cK ( n )$ as a sum 
of squares of $ \sC ( R , S , T )$. There is also an outer product on $ \cK^{ \infty} $ 
which we now decribe.  
Given two permutation pairs in $ \cK ( n_1 ) $ and $ \cK ( n_2)$ respectively,  
\bea 
&& ( \overline \sigma_1  , \overline \sigma_2 ) =  \sum_{ \gamma_1 \in S_{ n_1} } ( \gamma_1 \sigma_1 \gamma_1^{-1} , \gamma_1 \sigma_2 \gamma_1^{-1} ) \cr 
&& ( \overline \tau_1 , \overline \tau_2 )  
  = \sum_{ \gamma_2 \in S_{ n_2} } 
   ( \gamma_2 \tau_1 \gamma_2^{-1} , \gamma_2 \tau_2 \gamma_2^{-1} ) 
\eea
we have an outer product 
\bea 
&& \circ : \cK ( n_1 ) \otimes \cK ( n_2 ) \rightarrow \cK ( n_1 + n_2 ) \cr 
&& ( \overline \sigma_1  , \overline \sigma_2 ) \circ 
 ( \overline \tau_1 , \overline \tau_2 ) = \sum_{ \gamma \in S_{ n_1 + n_2 } } 
 ( \gamma (\overline \sigma_1 \circ \overline \tau_1) \gamma^{-1} , \gamma (\overline \sigma_2 \circ \overline \tau_2) \gamma^{-1} ) 
\eea
This outer product to the multiplication of the gauge-invariant 
functions of $ \Phi , \bar \Phi $ of degrees $n_1$ and $ n_2 $ to 
give a function of degree $ n_1 + n_2 $. 
This  outer product is related to the ring structure 
which has been described in detail, using the representation basis in \cite{DGT1707}. 

As a generalization of normal ordered two-point functions,
\bea 
\langle { \mathcal { N}}  ( \cO_{ \sigma_1 , \sigma_2 } )   \cN ( \cO_{ \tau_1 , \tau_2 } ) \rangle 
\eea
which is expressed using the product in $ \cK ( n)$, we may consider 
\bea 
\langle { \mathcal { N}}  ( \cO_{ \sigma_1 , \sigma_2 } \cO_{ \tau_1 , \tau_2   }  )    \cN ( \cO_{ \sigma_3 , \sigma_4 } ) \rangle 
\eea
Here $ \sigma_3 , \sigma_4 $ are in $ S_{ n_1 + n_2 } $ and the correlator 
 can be expressed in terms of 
\bea 
( ( \overline \sigma_1 , \overline \sigma_2 ) \circ  ( \overline \tau_1 , \overline \tau_2 ) ) \cdot  ( \overline \sigma_3 , \overline  \sigma_4 ) 
\eea
which involves the commutative outer product followed by the non-commutative product 
within $ \cK ( n_1 + n_2)$. This interplay between the two products in the context of correlators of one and two-matrix models has been described in \cite{PCA1601}.  

Infinite dimensional algebras constructed as direct sums of the above kind,
with more than one associative product, in some cases with a co-product and Hopf algebra structure, have been studied in the subject of combinatorial Hopf algebras, with applications in diverse areas of combinatorics (see e.g. \cite{AgSot05}). It would be interesting to explore the application of such combinatorial Hopf algebras in providing  a  mathematical framework for the computation of   correlators in matrix/tensor models.

\section{Correlators, permutation-TFT2 and covers of 2-complexes}
\label{sec:tft2}

In  this section, we will show how correlators in tensor models can be expressed as 
observables in topological field theory of $S_n$ flat connections on appropriate 2-complexes. This $S_n$ description is closely related to covering spaces of the 2-complexes. 
The powers of $N$ are shown to be related to the Euler character of the cover. This generalizes to tensor models analogous results obtained for single-matrix models and  quiver matrix models \cite{Gal,quivcalc}. Unlike the case of matrix models, where the 2-complexes are cell-decompositions of smooth 2-manifolds (possibly with boundaries and possibly equipped with line defects), here the 2-complexes do not discretize smooth 2-dimensional spaces.

Consider \eqref{Os1s} that we re-express:  
\bea 
&& \langle \cO_{ \sigma_1 , \sigma_2 , \cdots , \sigma_d} \rangle 
= \sum_{\gamma ,  \alpha_1 , \alpha_2 , \cdots, \alpha_d \in S_n } 
 \delta ( \alpha_1^{-1} \gamma \sigma_1 ) \delta ( \alpha_2^{-1}  \gamma \sigma_2 ) 
 \cdots \delta ( \alpha_d^{ -1} \gamma \sigma_d ) N^{ \cy( \alpha_1) + \cy( \alpha_2 ) 
  + \cdots + \cy( \alpha_d)  }   \cr 
 && = \sum_{ \alpha_1 , \alpha_2, \cdots, \alpha_d } 
 \delta ( \alpha_2 \alpha_1 \sigma_1^{-1} \sigma_2 ) \delta ( \alpha_3 \alpha_1 \sigma_1^{-1} \sigma_3) \cdots \delta ( \alpha_d \alpha_1 \sigma_1^{-1} \sigma_d ) 
\eea
In the last line,  we did the sum over $ \gamma $ by using the first delta function. 
Defining $ \tau_1 = \sigma_1^{-1} \sigma_2, \tau_2 = \sigma_1^{-1} \sigma_3, \cdots 
\tau_{ d-1} = \sigma_1^{-1} \sigma_2 $, we have 
\be
\langle \cO_{ \sigma_1 , \sigma_2 , \cdots , \sigma_d )} = 
\langle \cO_{ 1, \tau_1 , \tau_2 , \cdots , \tau_{ d-1} } \rangle 
= \sum_{ \alpha_l } 
 \delta ( \alpha_2 \alpha_1 \tau_1  ) \delta ( \alpha_3 \alpha_1 \tau_2 ) \cdots \delta ( \alpha_d \alpha_1 \tau_{ d-1} ) N^{ \cy( \alpha_1) + \cy( \alpha_2) + \cdots + \cy( \alpha_{ d} ) } 
\ee
It is instructive to normalize the observables by including factors 
\bea 
\prod_{ i=1}^{ d-1} N^{ ( \cy( \tau_i) -3n  )   }
 = N^{ -2n ( d-1) } \prod_{ i=1}^{  d-1 } N^{ ( \cy( \tau_i) - n  )   } 
 \eea 
 so that the normalized observable takes the form
 \be 
 \widetilde \cO_{ 1 , \tau_1 , \tau_2 , \cdots , \tau_{d-1} } 
 = \left ( N^{ -2 n ( d-1) } \prod_{ i=1}^{  d-1 } N^{ ( \cy( \tau_i) - n  )   } \right ) 
 \cO_{ 1 , \tau_1 , \tau_2 , \cdots , \tau_{d-1} }  
 \ee
 We then write the above normalized correlator as
\bea 
&&\langle \widetilde \cO_{ 1, \tau_1 , \tau_2 , \cdots , \tau_{ d-1} } \rangle =  \cr 
&& \sum_{ \alpha_l } 
 N^{ -n(d-1) + \sum_{ i =1}^{ d -1} ( \cy(\tau_i) - n )  + \sum_{ i=1}^d ( \cy( \alpha_{ i } ) - n )  } \delta ( \alpha_2 \alpha_1 \tau_1  ) \delta ( \alpha_3 \alpha_1 \tau_2 ) \cdots \delta ( \alpha_d \alpha_1 \tau_{ d-1} )
\eea

\begin{figure}
  \centering
 \includegraphics[width=11cm,height=3cm]{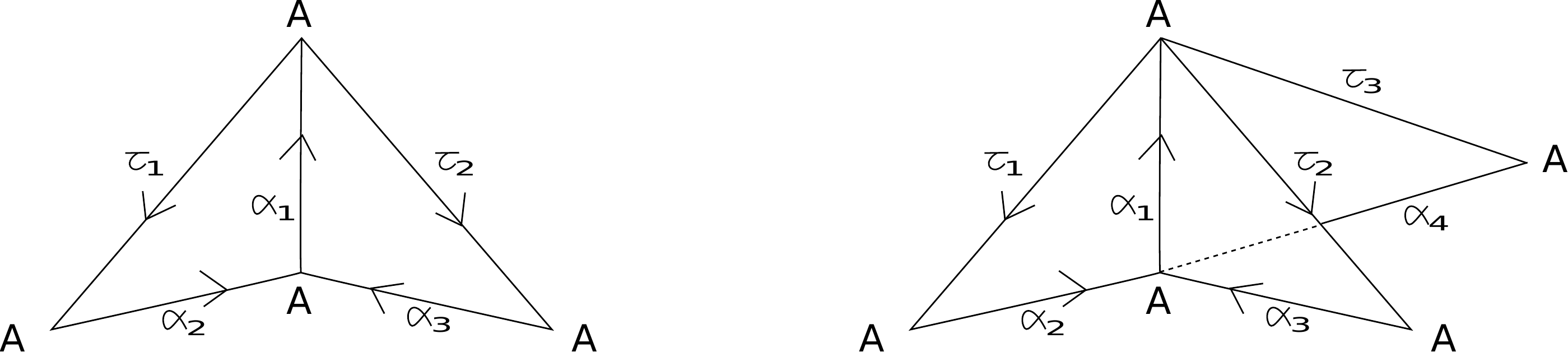}
  \caption{2-cell for $S_n$ TFT2 interpretation of correlator at $d=3$
(left) and $d=4$ (right). 
The 2-cells in the 2-complexes are $(\alpha_1\tau_l\alpha_l)$, $l=2,3,4$.}
  \label{fig:2celld=3}
\end{figure}

These sums over delta functions can be  interpreted as partition functions of two-dimensional $S_n$ 2D-topological field theory (TFT2) on 2-complexes. These $S_n$ TFT2 are 
lattice gauge theories with a simple topological plaquette action consisting of 
delta functions for  the product of permutations along the edges of the 2-cell.  
For a review of these TFT2 and their applications to correlators in quiver gauge theories 
and Feynman graph combinatorics see \cite{quivcalc,FeynCount}. 

$S_n$ TFT2 are also closely related to branched covers, a fact which has applications 
in the string theory of 2dYM \cite{grosstaylor,CMR}. 
It is interesting that it is possible to choose the normalizations of the operators 
and an overall normalization in terms of powers of $N$, such that the amplitude 
has a power of $N$ which is exactly equal to the Euler character of the covering 2-complex. 
The appropriate 2-complexes have a single vertex and loops corresponding to the 
permutations, and 2-cells (plaquettes) corresponding to the delta functions.  
Figure \ref{fig:2celld=3}
shows the appropriate 2-cells for the cases 
$d=3,4$. The vertices, labelled $A$ are all identified to a single point. The $ \alpha_i$ 
as well as the $ \tau_i$ therefore correspond to closed loops. The triangles are the 2-cells. In the covering space interpretation, there is an $n$-fold cover of the 
2-complex. The permutations $ \alpha_i$ describe the monodromy of the sheets of the covering as the different cycles are traverses from $A$ back to $A$. The terms $ ( \cy( \sigma ) - n)  $  is the contribution to the Euler character of a 2-surface whose boundary covers a 
circle on the target space with monodromy $ \sigma $ (see for example the review on 
branched covers in \cite{CMR,CMR2}). Hence we have these factors for each of the $ \alpha$'s and $ \tau$'s. The 2-complex has one vertex, $2(d-1) +1 = 2d-1 $ edges, and $ (d-1)$ faces. 
The Euler characteristic  $ V -E+F  = -d +1$. According to the  correlator formula above, when the permutations are all equal to the identity, the power is $-n( d-1)$, which is the correct Euler characteristic of the  trivial $n$-fold cover. 

The weight $ N^{ \chi }$ is expected from a string theory where $g_{st} = N^{-1}$. 
The above formula for the correlators, as a sum over coverings of a 2-complex by a 2-complex, suggests an interpretation where the covering 2-complex is viewed as a ``string worldsheet'' and the target 2-complex as a target space for the string. Interestingly the 
2-complexes for general $d$ have $d-1$ 2-cells joined at the vertical edge. So these 2-complexes are not cell decompositions of smooth two-dimensional surfaces. They can be 2-skeletons for cell decompositions of $3$-manifolds. 

\begin{figure}
  \centering
\includegraphics[width=10cm,height=3.5cm]{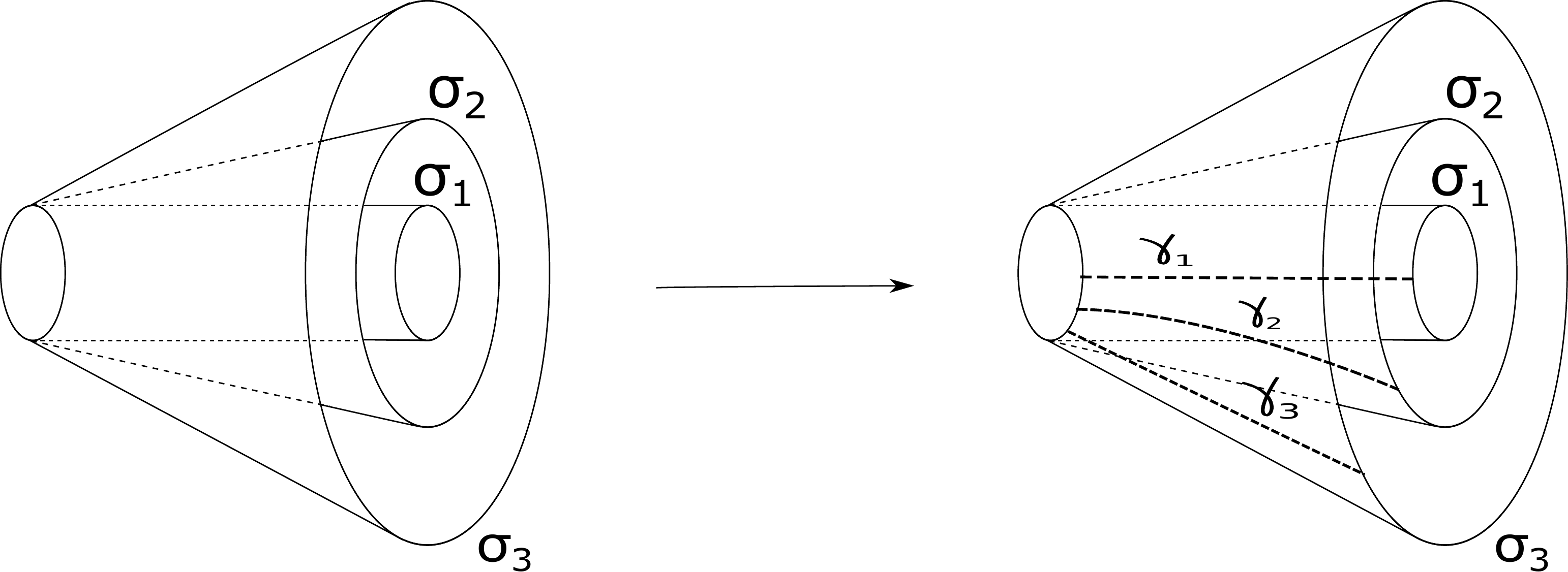}
  \caption{ $S_n$ TFT2 for Kronecker coefficients: permutation basis }
  \label{fig:PFK1}
\end{figure}

The $S_n$-TFT2 picture also gives a geometrical picture for correlators in the Fourier basis. Take for example the correlator \eqref{O1sss} that is: 
\be
\langle \cO_{ R_1 , R_2 , R_3 } \rangle = 
n!\, \sC ( R_1 , R_2 , R_3 ) \prod_{ i=1}^3 { \Dim_N (R_i) \over d(R_i) }  
\ee 
Consider the partition function of $S_n$ TFT2 on the 2-complex shown in Figure \ref{fig:PFK1}
\bea 
&& Z (\sigma_1 , \sigma_2 , \sigma_3 ) 
= \sum_{ \sigma_0 , \gamma_1 , \gamma_2 , \gamma_3 \in S_n } 
\delta ( \sigma_0 \gamma_1 \sigma_1 \gamma_1^{-1} ) 
\delta ( \sigma_0 \gamma_2 \sigma_2 \gamma_2^{-1} ) 
\delta ( \sigma_0 \gamma_3 \sigma_3 \gamma_3^{-1} ) \cr 
&& = n!  \sum_{ \sigma_1 , \sigma_2 , \sigma_3 , \gamma_1 , \gamma_2 } 
\delta ( \sigma_1 \gamma_1 \sigma_2 \gamma_1^{-1} ) 
\delta ( \sigma_1 \gamma_2 \sigma_3 \gamma_2^{-1} )  
\eea
Considering the central correlators in the Fourier basis 
\be
 Z ( P_{ R_1} , P_{ R_2} , P_{ R_3} ) 
= n!\, \sC ( R_1  , R_2 , R_3) 
\ee
Inserting the central elements $ \Omega = \sum_{ \sigma \in S_n } N^{ \cy( \sigma )  } \sigma $ we have 
\be
Z ( { N^n \over n! } \Omega P_{ R_1} , { N^n\over n! }  \Omega P_{ R_2} , { N^n\over n! }  \Omega P_{ R_3} ) 
= n! { \Dim_N(R_1) \over d(R_1) }  { \Dim_N( R_2) \over d(R_2) }{ 
\Dim_N( R_3) \over d(R_3) }
 \sC ( R_1  , R_2 , R_3) 
\ee
Similarly, for $ d=4$, 
\be
Z ( P_{ R_1} , P_{ R_2} , P_{ R_3} , P_{ R_4}  ) 
= n! \sC ( R_1  , R_2 , R_3 , R_4 ) 
\ee
and 
\bea
&& Z ( { N^n \over n! } \Omega P_{ R_1} , { N^n \over n! } \Omega P_{ R_2} , { N^n \over n! } \Omega P_{ R_3} , { N^n \over n! }  \Omega P_{ R_4}  ) \cr \cr
&& = n! { \Dim_N(R_1) \over d(R_1) }  { \Dim_N( R_2) \over d(R_2) }{ 
\Dim_N( R_3) \over d(R_3) }{ \Dim_N(R_4) \over d(R_4) }
 \sC ( R_1  , R_2 , R_3 , R_4  )  
\eea

\begin{figure}
  \centering
\includegraphics[width=15cm,height=3.5cm]{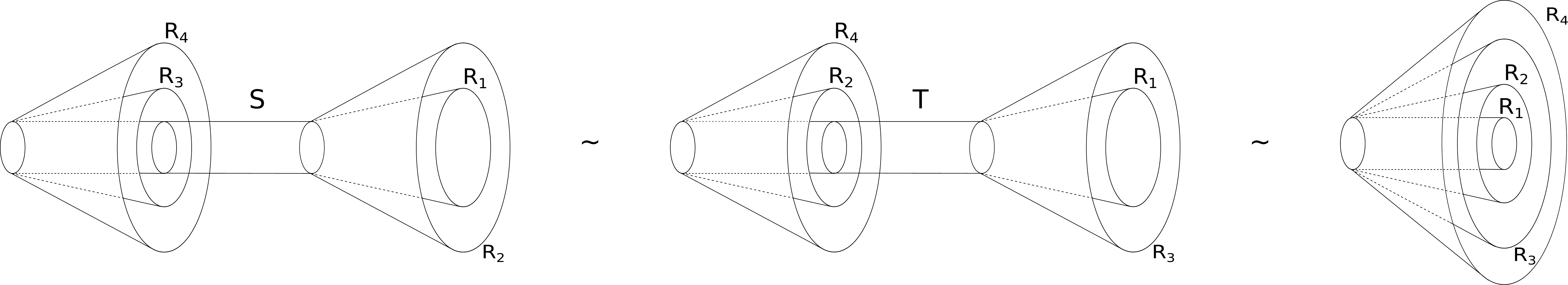}
  \caption{ $S_n$ TFT2 for Kronecker coefficients: Factorization equation 
\eqref{c4=c32} or \eqref{corrFourier4}.}
  \label{fig:PFK4}
\end{figure}

The different ways of factorizing this 4-point function corresponds to 
the different topologically equivalent ways of resolving the 
4 copies of incident circles into successive 3-fold incidences. This is illustrated in 
Figure \ref{fig:PFK4}.

\begin{figure}
  \centering
\includegraphics[width=12cm,height=3.5cm]{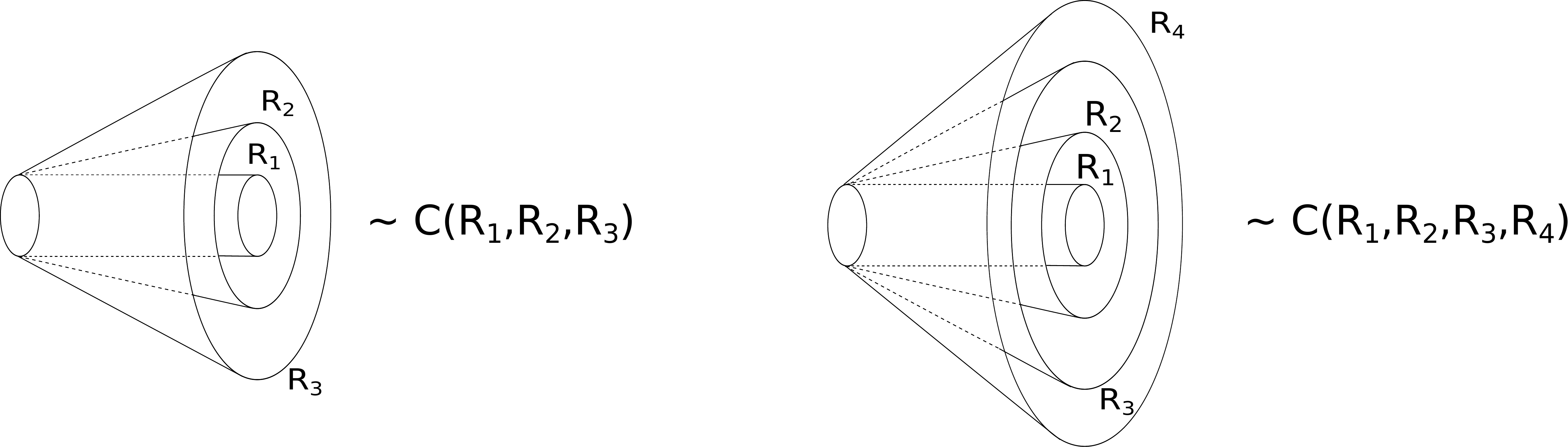}
  \caption{ $S_n$ TFT2 for Kronecker coefficients: central in rep basis 
in $d=3$ (left) and $d=4$ (right). }
  \label{fig:PFK2}
\end{figure}

\section{$S_3$-color exchange symmetry }
\label{sect:colsym}

 The  standard Gaussian 
integral  over  tensor fields is symmetric under exchange of 
the $d $ colors, and as such is expected to provide selection rules for correlators. 
This is also expected in any interacting model, which is obtained by adding to a Gaussian  term an interaction which is invariant under color-exchange. 
 It is therefore natural to consider the implications of 
the $S_d$ permutation group symmetry for the correlators of the tensor model 
and for  the algebra $ \cK(n)$, which has been shown to be intimately related to 
correlators in previous sections.  Colored
symmetric tensor model observables have been enumerated
in \cite{Sanjo} using group algebra techniques. Color-symmetric interactions have also played a 
distinguished role as interactions  in  renormalizable tensor field theories \cite{BGR,TFT,tensortrack}.

The counting of color-symmetric observables in \cite{Sanjo} was done in terms of 
sums over the conjugacy classes of $S_d$. For each conjugacy class $ p $ in $S_d$, we had expressions $ S^{(d)}_{ p } $ 
which were  sums over partitions of $S_n$. These were themselves observed to have integrality properties. Explicit formulae 
for $ S^{(d)}_{ p } $ were given for $ d $ up to $4$, and they have been computed to high orders in OEIS \cite{OEIS-color-symm}. 

In this section, we will give formulae in terms of character sums for $ S^{(d)}_{ p  } (n) $ for 
general $d$ and $ p  \vdash d $. We will consider the decomposition of $ \cK ( n , d  )$ in terms of symmetry types of $ S_d$, labelled by Young diagrams $Y$ of $S_d$. The integrality of 
these will be used to prove that the integrality of $ S^{(d)}_{ p } (n)  $. The completely symmetric  Young diagram $ Y = [d] \equiv Y_0 $ corresponds to the counting of color-symmetrized graphs. 

Focusing on the case $ d=3$, where we have 3 colors, we describe selection rules for the multiplication in $ \cK ( n)$ which follow  from color-exchange symmetry.
We  observe that the subspace $ \cK_{Y_0} (n) $ forms a sub-algebra of $ \cK ( n)$. 
Like $ \cK ( n)$ it inherits a non-degenerate pairing from $ \mC ( S_n ) \otimes \mC ( S_n )$. The WA theorem implies therefore that it has a decomposition 
 as a direct sum of matrix algebras. This in turn implies that we should be able to 
 write a formula for the dimension of $ \cK_{ Y_0} ( n )$ as a sum of squares. We find such a formula  and explain how to construct a basis for $ \cK_{ Y_0} ( n)$ which matches 
 the counting. We give group theoretic formulae for the dimensions of the other $S_3$ invariant subspaces of $ \cK ( n)$, namely $ \cK_{ Y_1 } ( n), \cK_{ Y_2 } ( n)$ 
 where $ Y_1 = [2,1] , Y_2 = [3]$ are the Young diagrams for mixed symmetry and for  the antisymmetric representation of $S_3$.

\subsection{Counting observables in the rank $d=3$ case} 

Color-symmetrised graphs are defined \cite{Sanjo} 
by imposing an equivalence under $S_3$ permutations 
of the permutation triples describing the graph
\be
 ( \s_1  , \s_2 , \s_3 ) \sim ( \s_2  , \s_1 , \s_3 ) \sim ( \s_1 , \s_3 , \s_2 ) \sim \dots 
\label{sss123}
\ee 
These $S_3$ permutations commute with the diagonal left $S_n$ action and the diagonal right multiplication, which are used to get colored graphs from the permutation triples.  
Elements of  $( \mC(S_n))^{ \otimes 3 } $ invariant under this $S_3$ action are  
\be
[\s_{1}  \s_{2}  \s_{3} ]:=\sum_{\alpha \in S_3} \s_{\alpha(1)} \otimes \s_{\alpha(2)} \otimes
\s_{\alpha(3)} \; \in\;  \IC(S_n)^{\otimes 3}  
\label{symel}
\ee
The left and right $S_n$ equivalences are imposed on these $S_3$ symmetric triples 
\be\label{quoc3}
[\s_{1}  \s_{2}  \s_{3} ] \sim 
[\gamma_1^{\otimes 3} ][\s_{1}  \s_{2}  \s_{3} ]
[\gamma_2^{\otimes 3} ]
=\sum_{\alpha \in S_3} 
\gamma_1\s_{\alpha(1)}\gamma_2 \otimes \gamma_1\s_{\alpha(2)}\gamma_2 \otimes
\gamma_1\s_{\alpha(3)}\gamma_2 
\ee
The color-symmetric subspace of $ \cK (n)$, denoted $ \cK_{ Y_0} ( n) $  has a dimension given by 
\bea
&& 
\dim ( \cK_{ Y_0 } ( n ) ) =  { 1 \over 6 n ! } 
\sum_{ \gamma \in S_n} \sum_{\s_2, \s_3 \in S_n}\delta(\gamma^{-1} \,\s_{2}\gamma\,\s_{2}^{-1})
\delta(\gamma^{-1} \,\s_{3}\gamma\,\s_{3}^{-1})  \crcr 
&&+   { 1 \over 2 n! }  \sum_{ \gamma \in S_n } \sum_{ \sigma \in S_n } \delta ( \gamma^2 \sigma \gamma^{-2}\sigma^{-1} ) 
+   { 1 \over 3 n! } \sum_{ \gamma  , \sigma \in S_n  } \delta ( \gamma^3 \sigma^3 ) \cr\cr
&&=  { 1 \over 6 n! } \sum_{ p  \vdash n }   \Sym ( p  ) 
   +  { 1 \over 2 n! }  \sum_{ \gamma \in S_n } \sum_{ \sigma \in S_n } 
\delta ( \gamma^2 \sigma \gamma^{-2}\sigma^{-1} ) + { 1 \over 3 n! } \sum_{ \gamma  , \sigma \in S_n  } \delta ( \gamma^3 \sigma^3 )  \cr 
&& = { 1 \over 6 } S^{(3)}_{ [1^3] } (n) + { 1 \over 2 } S^{(3)}_{ [2,1] } (n) + 
{ 1 \over 3 } S^{(3)}_{ [3]} (n) 
\label{s1s2s3}
\eea
This was denoted $  Z_{3;\, \col}  ( n )$  in \cite{Sanjo}. 
Since the actions of $ S_3$ commutes with $S_n \times S_n$, we in fact have 
an action of $ S_3 \times S_n \times S_n$ and we can equally first apply the 
$ S_n \times S_n $ equivalence in $ \mC ( S_n ) \otimes \mC ( S_n ) \otimes  \mC (  S_n ) $ to get 
$ \cK ( n ) $ and subsequently project to the invariants of the $S_3$ action.  
Following the steps in \cite{Sanjo} (see equation (59) and following there) we see that the three terms 
can be written as 
\be
S^{(3)}_{[1^3]  }  =   \tr_{ \cK ( n ) } ( ( 1 )(2) ( 3) ) 
 \,, 
\qquad
S^{(3)}_{[2,1]  }  =   \tr_{ \cK ( n ) } ( (12) )\,, 
\qquad 
S^{(3)}_{ [3 ] }  =   \tr_{ \cK ( n ) } (  ( 123) )
\ee
where $\tr_{\cK(n)}(\cdot)$ is a trace over  the vector space $ \cK ( n)$. 
We thus have 
\bea 
 \dim ( \cK_{ Y_0 } ( n ) ) = { 1 \over 6 } \tr_{ \cK ( n ) } ( ( 1 )(2) ( 3) ) 
 + { 1\over 3}  \tr_{ \cK ( n ) } ( (12) ) 
  +  { 1 \over 2 } \tr_{ \cK ( n ) } (  ( 123) )
\eea 
The Burnside lemma calculation in \cite{Sanjo} can be regarded
as the application of the normalized projector for $ S_n \times S_n \times S_3$ 
acting on  $ \mC ( S_n ) \otimes \mC ( S_n ) \otimes \mC ( S_n ) $ followed by taking the 
trace. This is because under   inner product on  $ \mC ( S_n)$ given by the delta function, 
the permutations form an orthonormal basis. 

\subsection{The algebra $\cK_{Y_0}(n)$ }
\label{ksc}

We define the algebra $\cK_{Y_0 }(n)$ the left and right invariant 
and colored symmetric sub-algebra of $\mC(S_n)^{\otimes 3}$ 
as
\be\label{Kcolsym}
\cK_{\col}(n) = {\rm Span}_{\mC}\Big\{
\sum_{\g_1,\g_2,\g_3 \in S_n,\tau \in S_3} 
\gamma_1\s_{\tau(1)}\gamma_2 \otimes \gamma_1\s_{\tau(2)}\gamma_2 \otimes
\gamma_1\s_{\tau(3)}\gamma_2 \,, \; \s_1,\s_2, \s_3 \in S_n \Big\} 
\ee
It can be checked that the product of two basis elements of $\cK_{\col}(n)$, $A =
\sum_{\tau\in S_3, \gamma_i\in S_n } ( 
\gamma_1\s_{\tau(1)}\gamma_2 \otimes 
\gamma_1\s_{\tau(2)}\gamma_2 \otimes
\gamma_1\s_{\tau(3)}\gamma_2) $ and $B
=\sum_{\tau' \in S_3,\gamma_i' \in S_n }
  \gamma_1'\s_{\tau'(1)}\gamma'_2 \otimes
 \gamma_1'\s_{\tau'(2)}\gamma'_2 \otimes
\gamma_1'\s_{\tau'(3)}\gamma'_2 $ belongs to $\cK_{\col}(n)$. We
have
\bea
&&
AB  = \\
&&
\sum_{\tau,\tau' \in S_3}\sum_{ \gamma_i, \gamma_i' \in S_n } ( 
\gamma_1\s_{\tau(1)}\gamma_2 \otimes 
\gamma_1\s_{\tau(2)}\gamma_2 \otimes
\gamma_1\s_{\tau(3)}\gamma_2) 
(  \gamma_1'\s_{\tau'(1)}\gamma'_2 \otimes
 \gamma_1'\s_{\tau'(2)}\gamma'_2 \otimes
\gamma_1'\s_{\tau'(3)}\gamma'_2 ) \crcr
&& = n! 
\sum_{\tau',\gamma_2\in S_3} \Big\{
\sum_{\tau\in S_3}\sum_{ \gamma_1, \gamma_2' \in S_n } 
\gamma_1 (\s_{\tau(1)}\gamma_2 \s_{\tau'(1)})\gamma'_2\otimes 
\gamma_1(\s_{\tau(2)}\gamma_2\s_{\tau'(2)})\gamma'_2 \otimes
\gamma_1(\s_{\tau(3)}\gamma_2 \s_{\tau'(3)})\gamma'_2 \Big\} \nonumber
\eea
where we successively used $\gamma_2 \gamma_1' \to \gamma_2$, 
$\tau' \to \tau'\tau^{-1}$. Then, the product $AB$ is a sum 
of basis elements of $\cK_{\col}(n)$ hence belongs to $\cK_{\col}(n)$. 

\

\subsection{Decomposition of $\cK(n)$ into representations of $ S_3$}
\label{sect:grading}

The group algebra $ \mC ( S_n ) \otimes \mC ( S_n ) \otimes \mC ( S_n ) $
is a representation of $ S_n \times S_n \times S_3$.  
One $S_n $ acts on the diagonally on the left, the other acts diagonally on the right, 
 and the $S_3$ permutes the three factors. These three actions commute. 
 Once we have projected to the invariant subspace of $S_n \times S_n$ to get $ \cK ( n )$ 
 we still have an $S_3$ action. We can decompose $\cK ( n ) $ into subspaces which
 transform as irreducible representations of $S_3$.  
We have a decomposition of $ \cK ( n ) $ into 
\bea 
\cK ( n ) = \cK_{ Y_0}  ( n ) \oplus \cK_{ Y_1}  ( n ) \oplus \cK_{ Y_2}  ( n ) 
\eea
$ Y_0=  [3] $ is the one-dimensional trivial rep of $S_3$, corresponding to a
 Young diagram with a single row of length $3$. 
$ Y_1 = [2,1]$ is the two-dimensional irrep of $S_3$. 
 $Y_2 = [1^3] $ is the anti-symmetric irrep of $S_3$. 
In other words the vector space of 3-colored graphs is a representation of $S_3$. 
This space can be decomposed into a direct sum of the trivial along with the $[2,1]$ and $[1^3]$
representations, which will all appear, generically with multiplicities.

Using the standard  formula for the projector $P_Y$ in the group algebra of $ S_d$
\bea 
P_{ Y } = { d(Y)  \over d! } \sum_{ \alpha \in S_d } \chi^{ Y } ( \alpha  )~  \alpha 
\eea
and specializing to $ d=3$, we can write 
\bea\label{dimKY} 
\dim ( ( \cK_{Y_i} ( n  )  )  ) = { 1 \over 6 }  \sum_{ \alpha \in S_3 }    d(Y_i)  \chi^{Y_i} ( \alpha  )\, \tr_{ \cK( n )  } ( \alpha )  \,, \quad  i=0,1,2  
\eea
We have 
\bea\label{DimKsym} 
\dim ( \cK_{ Y_0}  ( n  ) ) = {1 \over 6 } S^{(3)}_{ [1^3] } (n ) + { 1 \over 2 }  S^{(3)}_{ [2,1] } (n  ) + { 1 \over 3 }  S^{(3)}_{ [3] } ( n  ) 
\eea
and 
\bea\label{DimKforms}  
&& \dim ( \cK_{ Y_1}  ( n  ) ) = {2 \over 3 } S^{(3)}_{ [1^3] } ( n  ) -  { 2 \over 3 }  S^{(3)}_{ [3] } ( n  ) \cr 
&& \dim  ( \cK_{Y_2}   ( n ) ) =  {1 \over 6 } S^{(3)}_{ [1^3] } ( n  ) -  { 1 \over 2 }  S^{(3)}_{ [2,1] } ( n  ) +{ 1 \over 3 }  S^{(3)}_{ [3] } ( n  ) 
\eea
Taking $ a^{(0)}, a^{(1)} , a^{(2)}  $ to belong  to subspaces of $ \cK ( n ) $
labelled by the three Young diagrams $ Y_0 = \{ 3 \}  , Y_1 = \{ 2,1 \}  , Y_2 = \{ 1,1,1 \} $ of $ S_3$, we will have the following selection rules 
\bea\label{multselrules}  
a^{(0) } . a^{(i) } \in \cK_{Y_i}  ( n ) \nnm \\
a^{(1) } . a^{(1) } \in \cK_{ Y_0} ( n )  \oplus \cK_{ Y_1} ( n )\oplus \cK_{ Y_2}  ( n )\nnm \\
a^{(1) } . a^{(2) } \in  \cK_{Y_1} ( n ) \nnm \\
a^{(2) } . a^{(2) } \in  \cK_{ Y_0} ( n )
\eea
These follow from the corresponding tensor product decompositions of $ S_3 $ representations. From these equations we see that $ \cK_{Y_0} ( n )$ is  a sub-algebra of $\cK(n)$. $ \cK_{ Y_1}   ( n )$ and $ \cK_{Y_2}   ( n )$ are modules for the algebra $\cK_{ Y_0} (n)$. $\cK_{ Y_0} \oplus \cK_{ Y_2}$ is also a closed sub-algebra of $ \cK ( n )$. 

An interesting consequence of the above decomposition of $ \cK (n) $ is that we can use it 
to prove  the integrality of the separate terms $ S^{(3)}_{ [1^3] } ,  S^{(3)}_{[2,1]}  , S^{(3)}_{ [3]} $. A similar argument can be made when we have $d$ colors instead of $3$. We will in fact 
present the argument in this generality.  Take a vector space $V$, which is a representation of $S_d$. The multiplicity of an irrep labelled by   Young diagram $Y$  (partition of $d$) is given by 
\be 
m_{V}^{Y}  =  { 1 \over d!  } \sum_{ p \vdash d  }  |T_p| ~  \chi^Y ( \sigma_p ) \, \tr_{ V } ( \sigma_p)  = \sum_{ p \vdash d } { 1 \over | \Sym  ( p )|  }  \chi^Y ( \sigma_p )  \, \tr_{ V } ( \sigma_p)   = \tr_{ V } \left ( { P_{ Y } \over d(Y)  }  \right)
\ee
$ \sigma_p $ is a permutation with cycle decomposition  given by the partition $p$, i.e. some 
fixed permutation in the conjugacy class $T_p$.  We will use 
$ \wTp$  to denote the sum of all the permutations in the conjugacy class $T_p$.   
The subspace of $V$ transforming in the irrep $Y$, denoted $V_Y$, has dimension 
\be 
\dim  V_Y = \tr_V(P_Y)= m_{V}^{Y}\, d(Y) 
\ee 
These $m_{V}^{Y} $ are natural numbers (zero or positive  integers). 
The quantities we called 
$ S^{ (d) }_{ p } $ in \cite{Sanjo}  are 
\be
S^{ (d)}_{ p} = \tr_V ( \sigma_p )  = { 1 \over |T_p| } \tr_V ( \wTp )  
\ee 
for the case where $ V = \cK ( n )$. 

$\wTp$ is a central element in the group algebra of $S_n$ and has an expansion in projectors
\bea
\wTp  = \sum_{ Y } { \chi^Y  ( \wTp  ) \over d(Y) } P_Y
\eea
This can be seen by writing 
\bea 
\wTp  = \sum_{ Y' } a_{Y'} P_{Y'} 
\eea
and then taking the trace in irrep $Y$ to find 
\be
 \chi^Y ( \wTp  ) = a_Y d(Y)\,,  \qquad 
a_Y = { \chi^Y ( \wTp ) \over d(Y) } 
\ee
Now write 
\be
\tr_V ( \wTp  )   =   \sum_{ R }  {  \chi^Y ( \wTp  ) \over d(Y) }  \tr_V (P_Y)  
  =  \sum_R {  \chi^Y  ( \wTp  ) \over d(Y) }  d(Y) m_V^Y 
  =  \sum_R  \chi^Y  ( \wTp  )  m_V^Y  
\ee
and observe 
\bea 
{ 1 \over |  T_p | } \tr_V ( \wTp  ) =  \sum_Y  {  \chi^Y  ( \wTp  ) \over |T_p| } m_V^Y 
= \sum_R {  \chi^Y  ( \wTp  ) \over |T_p| }  \tr_{ V } \left ( { P_{ Y } \over d(Y ) }  \right)
\eea
It turns out that, for symmetric groups, we know that the characters  $ {  \chi^Y ( \wTp  ) \over |T_p|  } =  \chi^Y ( \sigma_p) $ are integers (this follows for example from the Murnaghan-Nakayama lemma for computing the characters).
This proves that the $ S^{ (d) }_{ p } $ - for all $d$ and $p$ -   in our previous paper are integers. 
From the above equation it is not clear they are positive, but we also know from before that they are sums of 
delta functions. Combining these two facts, we conclude that they are indeed positive integers - as we found to high orders in \cite{Sanjo} and in \cite{OEIS-color-symm}.

As a consistency check of the above (\ref{DimKforms}) 
  we can construct the sequence for the $Y_1$ sector  $  { 1 \over d(Y_1) } \dim ( \cK_{Y_1} ( n  ) ) = 
  ( S^{(3)}_{ [1^3]} - S^{(3)}_{ [3] } ) /3 $ - using  the $S^{(3)}_p $ 
are found to be 
\bea 
0,1,3,13, 52, 296, 1850, 14386, 126082, 1247479
\eea
so indeed integral.
Also for $  \dim ( \cK_{ Y_2}  ( n  ) )$, we find 
\bea 
0,0,0, 2,13, 110 , 810 , 6796 , 61693 , 618880 
\eea
Again these are indeed integral as expected. 

\subsection{Color symmetrisation using Fourier basis}

 We can better understand the connection between the $Q^{ R_1 , R_2 , R_3 }_{ \tau_1 , \tau_2 } $ and the $S_3$  invariant  subspaces,   
$ \cK_{Y} ( n )$ by doing the color  symmetrisation directly in the  Fourier basis for $ \mC ( S_n ) \otimes \mC ( S_n) \otimes \mC ( S_n )$ 
\bea\label{tensprod3}  
Q^{ R_1}_{ i_1 , j_1 } \otimes Q^{ R_2 }_{ i_2  , j_2 } \otimes Q^{ R_3}_{ i_3 , j_3 } 
\eea
Important properties (see appendix \ref{app:baskn}) are 
\be 
Q^R_{ ij}   =    \sqrt{ d ( R ) \over n! } \sum_{ \sigma \in S_n } D^R_{ ij} ( \sigma ) \sigma \,,
 \quad  
\tau ~  Q^R_{ ij} = \sum_{ p }  D^R_{ pi} ( \tau ) ~   Q^R_{ p j} \,,\quad
Q^R_{ ij} ~  \tau   =  \sum_{ q }   Q^R_{ i  q } ~ D^R_{ j  q } ( \tau ) 
\ee
and the orthonormality property 
\bea 
  \cr 
  \delta ( Q^{R}_{ ij } , Q^{ S}_{ kl } )  = \delta_{ R , S } \delta_{ ik} \delta_{ jl} 
\eea
In order to project  \eqref{tensprod3}  to $ \cK_Y (n)$ we  apply the normalized projectors 
\bea 
P^{ (S_3) } P^{ (S_n)  }_{ L } P_{R}^{ (S_n) }   = { 1 \over 6 (n!)^2 } \sum_{ \sigma_1 \in S_n }\sum_{ \sigma_2  \in S_n }  \sum_{ \alpha \in S_3 }  \alpha   \rho_L ( \sigma_1 ) \rho_{ R } ( \sigma_2 )  
\eea  
where $  \rho_L ( \sigma_1 ) $ indicates the left diagonal action, $ \rho_R ( \sigma_2 )$ is the right diagonal action 
and $ \alpha $ acts by swapping the tensor slots. We will calculate 
\bea 
\dim ( \cK_{ Y_0 } ( n ) ) = \tr_{ \mC ( S_n ) ^{\otimes 3} }  ( P^{ (S_3) }  P^{ (S_n)  }_{ L } P_{R}^{ (S_n) } ) 
 = \tr_{ \cK ( n ) }  ( P^{ (S_3) } ) 
\eea
We have 
\bea 
&& \sum_{ \sigma_1 \in S_n }\sum_{ \sigma_2  \in S_n }  \sum_{ \alpha \in S_3 }  \alpha   \rho_L ( \sigma_1 ) \rho_{ R } ( \sigma_2 )   ~~
Q^{ R_1}_{ i_1 , j_1 } \otimes Q^{ R_2 }_{ i_2  , j_2 } \otimes Q^{ R_3}_{ i_3 , j_3 } \cr 
&& =  \sum_{ \sigma_1 \in S_n }\sum_{ \sigma_2  \in S_n }  \sum_{ \alpha \in S_3 } \sum_{  p_l ,  q_l }  D^{ R_{ 1} }_{ p_1 i_{ 1 } } ( \sigma_1 )  D^{ R_{ 2} }_{ p_2 i_{ 2 } } ( \sigma_1 ) 
  D^{ R_{ 3 } }_{ p_3 i_{ 3 } } ( \sigma_1 ) \cr 
 && ~~~  D^{ R_{ 1 } }_{ j_{ 1 } q_1  } ( \sigma_2 )   D^{ R_{ 2 } }_{j_{ 2 } q_2  } ( \sigma_2 )  D^{ R_{ 3} }_{ j_{ 3 } q_3 }( \sigma_2 ) \cr 
 && ~~ Q^{ R_{ \alpha(1)}}_{ p_{ \alpha(1) } , q_{ \alpha ( 1) }  }  \otimes Q^{ R_{ \alpha(2)}}_{ p_{ \alpha ( 2 )}  , q_{ \alpha ( 2 ) }  } \otimes Q^{ R_{ \alpha(3)} }_{ p_{ \alpha ( 3 ) }  , q_{ \alpha ( 3 ) }  }
\eea
To compute the trace, pair this with $  Q^{ R_1}_{ i_1 , j_1 } \otimes Q^{ R_2}_{ i_2 , j_2 } \otimes Q^{ R_3}_{ i_3 , j_3 } $ and sum over $ R_l ,   i_l ,  j_l $ giving  
\bea 
&& \sum_{ R_l } \sum_{ \sigma_l  \in S_n }  \sum_{ \alpha \in S_3 } \sum_{  i_l , j_l , p_l ,  q_l } 
D^{ R_{ 1} }_{ p_1 i_{ 1 } } ( \sigma_1 )  D^{ R_{ 2} }_{ p_2 i_{ 2 } } ( \sigma_1 ) 
  D^{ R_{ 3} }_{ p_3 i_{ 3  } } ( \sigma_1 ) \cr 
 && ~~~  D^{ R_{  1} }_{ j_{ 1  } q_{ 1   }   } ( \sigma_2 )   D^{ R_{ 2 } }_{j_{ 2 } q_{  2  }   } ( \sigma_2 )  D^{ R_{ 3 } }_{ j_{ 3 } q_{  3 }  }( \sigma_2 ) \cr 
&& ~~~  \delta^{ R_1 , R_{ \alpha(1)} } \delta_{ i_1 , p_{ \alpha ( 1) }  } \delta_{  j_1  , q_{ \alpha(1) } } \delta^{ R_2 , R_{ \alpha(2)} } \delta_{ i_2 , p_{ \alpha(2)}  } \delta_{  j_2 , q_{ \alpha(2)} } 
 \delta^{ R_3 , R_{ \alpha(3) } } \delta_{ i_3 , p_{ \alpha(3) } } \delta_{ j_3 , q_{ \alpha(3) }  } \cr \cr
 && = \sum_{ R_1 , R_2 , R_3 }  
  \sum_{ \sigma_1 \in S_n }\sum_{ \sigma_2  \in S_n }  \sum_{ \alpha \in S_3 } \sum_{  i_l ,  j_l } D^{ R_{ 1} }_{  i_{ \alpha(1) i_1 } } ( \sigma_1 )  D^{ R_{ 2 } }_{ i_{ \alpha(2) i_2  } } ( \sigma_1 ) 
  D^{ R_{ 3} }_{ i_{ \alpha(3)  i_3 } } ( \sigma_1 ) \cr 
 && ~~~  D^{ R_{ 1} }_{ j_1 j_{ \alpha(1) }  } ( \sigma_2 )   D^{ R_{ 2} }_{j_2 j_{ \alpha(2) }   } ( \sigma_2 )  D^{ R_{ 3} }_{ j_3 j_{ \alpha(3) }  }( \sigma_2 )  \delta^{ R_1 , R_{ \alpha(1)} }\delta^{ R_2 , R_{ \alpha(2)} }   \delta^{ R_3 , R_{ \alpha(3)} }
 \label{3terms}
\eea

There are three types of terms. If $ \alpha $ is the identity then we have 
\bea 
&& { 1 \over 6 (n!)^2  } \sum_{ \sigma_1 , \sigma_2 } \chi^{ R_1} ( \sigma_1)  \chi^{ R_2} ( \sigma_1)  \chi^{ R_3} ( \sigma_1) 
 \chi^{ R_1} ( \sigma_2)  \chi^{ R_2} ( \sigma_2)  \chi^{ R_3} ( \sigma_2)   \delta^{ R_1 , R_{ \alpha(1)} }\delta^{ R_2 , R_{ \alpha(2)} }   \delta^{ R_3 , R_{ \alpha(3)} }  \cr 
 && = { 1 \over 6 } \sum_{ R_1 , R_2 , R_3 }  ( \sC ( R_1 , R_2 , R_3) )^2
\eea
When $ \alpha = (1,2)$ we get 
\bea\label{21Fourier}  
&& { 1 \over 6 (n!)^2 } \sum_{ R_1 , R_3 }  \sum_{ \sigma_1 , \sigma_2 }
  \chi^{ R_1 } ( \sigma_1^2 )   \chi^{ R_3 } ( \sigma_1 )  \chi^{ R_1 } ( \sigma_2^2 )   \chi^{ R_3 } ( \sigma_2 )  \cr 
&& = { 1 \over 6 (n!)^2 }  \sum_{ R_1 } \sum_{ \sigma_1 , \sigma_2 }  \chi^{ R_1 } ( \sigma_1^2 )   \chi^{ R_1 } ( \sigma_2^2 )  
\sum_{ \gamma } \delta ( \sigma_1 \gamma \sigma_2^{-1}  \gamma^{-1} )  \cr 
&& = { 1 \over 6 (n!) } \sum_{ R_1 }   \sum_{ \sigma_1}  \chi^{ R_1 } ( \sigma_1^2 )  
\chi^{ R_1 } ( \sigma_1^2 )   
\eea
The other permutations in the same conjugacy class give the same factor. So we get 
\bea 
{ 1 \over 2 (n!) } \sum_{ R } \sum_{ \sigma }   \chi^{ R } ( \sigma^2  )  \chi^{ R } ( \sigma^2 )   
\eea
If $ \alpha  $ has cycle structure $[3]$, then we have 
\bea 
{ 1 \over 3 ( n! )^2  } \sum_{ R } \sum_{ \sigma_1 , \sigma_2}  \chi^R ( \sigma_1^3 )  \chi^R ( \sigma_2^3 )  
\eea
 So we recover the counting we previously had, from working with the delta functions over the group algebra. 
 
 These three expressions above
 \bea 
&&  S^{(3)}_{ p = [1^3] } ( n )  =   \sum_{ R_1 , R_2 , R_3 } (\sC ( R_1 , R_2 , R_3 ))^2 \cr 
&&   S^{ (3)}_{ p=[2,1]} ( n )  = { 1 \over n! } \sum_{ R }  \sum_{ \sigma } \chi^R ( \sigma^2 ) \chi^R ( \sigma^2 )  \cr 
&&  S^{ (3)}_{ p=[3]}  ( n )  = { 1 \over (n!)^2 }  \sum_{ R }\sum_{ \sigma_1 , \sigma_2} 
  \chi^{ R } ( \sigma_1^3 ) \chi^R ( \sigma_2^3 ) 
 \eea
  are directly related to the delta functions 
  derived in  \cite{Sanjo}  by starting from Burnside lemma.  The advantage of the present method is that the generalization to general $ S^{ (d)}_{p} (n)  $ is easily done. We will describe this generalization in section \ref{sec:gensolsymd} of these formulae in terms of character sums. 

\subsubsection{Subspace of $ \cK ( n ) $ with  $ R_{1,2,3}$ all different} 

Fixing three different representations of $ S_n$, let us call them 
$ \{ R , S , T \}$, there are subspaces of $ \cK_{Y_0} ( n ) , \cK_{Y_1} ( n )  ,\cK_{Y_2} ( n )$
which come from the six different assignments of the list
 $ [ R_1 , R_2 , R_3 ]  $ from the set  $ \{ R , S , T \} $. From the equations above we see that in this case, we only get contributions from $ \alpha = (1) ( 2) ( 3) $.  Using  \eqref{3terms} we find 
\bea\label{WARST} 
\dim ( \cK_{Y_0}^{ ( R  , S , T ) }  (n) ) = ( \sC ( R , S , T ))^2  \cr 
\eea 
for the color-symmetrized subspace and for the subspaces transforming according to 
the other irreps of $S_3$
\bea 
\dim ( \cK_{Y_1}^{ ( R  , S , T ) }  (n) ) = 2 ( \sC ( R , S , T ) )^2  \cr 
\dim ( \cK_{Y_2}^{ ( R  , S , T ) }  (n) ) = ( \sC ( R , S , T ))^2  
\eea 

Focusing on $ \cK_{Y_0}^{ ( R , S , T ) }  ( n )$, which forms an algebra of $ \cK ( n)$, we can write a basis for this algebra as
\bea 
\sum_{i_1 , i_2 , i_3 , j_1 , j_2 , j_3} \sum_{ \alpha \in S_3 } 
C_{ i_1 , i_2 ; i_3 }^{ R , S ; T , \tau_1 } C_{ i_1 , i_2 ; i_3 }^{ R , S ; T , \tau_2 } 
Q^{ \alpha (  R )  }_{ i_{ \alpha (1 )}  , j_{ \alpha ( 1 ) }  } 
\otimes Q^{ \alpha ( S )   }_{ i_{ \alpha ( 2) }, j_{ \alpha ( 2) }  }
\otimes Q^{ \alpha ( T )  }_{ i_{ \alpha ( 3) } , j_{ \alpha ( 3)  }}  
\eea
Since $ \tau_1 , \tau_2 $ run over a range of $ \sC ( R , S , T )$ 
the counting of these basis elements agrees precisely with the dimension of 
this subspace.

\subsubsection{Subspace of $ \cK ( n )$ where two $S_n$ irreps are equal and different from third  } 

Consider the subspace of $ \cK ( n )$ where two of the $ R_1 , R_2 , R_3 $ 
are equal to $R$ and a third is equal to $S$. The $ \alpha = ()$ term will now 
contribute $ 3 /6 (  \sC ( R , R , S ) )^2 = { 1 /2} ( \sC ( R , R , S ) )^2 $. The factor of $3$ comes from the choice of which of the $R_i$ is equal to $S$.

Consider for concreteness the term (\ref{21Fourier}), for case where $ R_1 = R_2 = R , R_3 = S$.  We will re-write this in a number of ways.

Let us return to  $ S_{ [ 2,1] }$
\bea 
S_{ [ 2,1] } (n)  = { 1 \over (n!)^2 } \sum_{ R , S }  \sum_{ \sigma_1 , \sigma_2 } \chi^R ( \sigma_1^2 ) \chi^S ( \sigma_1 )   \chi^R ( \sigma_2^2 ) \chi^S ( \sigma_2 )  
\eea
This sum contains cases where $ R=S $ as well as $ R \ne S$. Focusing on the $ R \ne S$ case, define 
\bea 
S_{ [2,1] }^{ ( R , R , S ) } (n)  =  { 1 \over (n!)^2 } \sum_{ \sigma_1 , \sigma_2 } \chi^R ( \sigma_1^2 ) \chi^S ( \sigma_1 )   \chi^R ( \sigma_2^2 ) \chi^S ( \sigma_2 )
\eea
The multiplicity of  $ S$  in the symmetric part of $ V_R \otimes V_R$ is  
\bea 
&& \Mult ( \Sym^2  (V_R) , V_S ) = 
 { 1 \over (n!) }  \sum_{ \sigma } \chi^S ( \sigma  )\, \tr_{ R \otimes R } \left (  ( \sigma \otimes \sigma )  {  1 + ( 12)  \over 2 } \right ) \cr 
&& = { 1 \over 2 n! } \sum_{ \sigma } \chi^S ( \sigma ) \left (  \chi^R ( \sigma^2 ) +  \chi^R ( \sigma ) \chi^R ( \sigma ) \right ) \cr 
&& = { 1 \over 2 } \sC (  R, R , S ) + { 1 \over 2 n! } \sum_{ \sigma }  \chi^S ( \sigma ) \chi^R ( \sigma^2 )
\eea
The multiplicity of $S $ in the anti-symmetric part of $ V_R \otimes V_R$ is  
\bea
 \Mult ( \Lambda^2 (V_R) , V_S ) =  { 1 \over 2 } \sC (  R, R , S )  -  { 1 \over 2 n! } \sum_{ \sigma }   \chi^S ( \sigma ) \chi^R ( \sigma^2 )
\eea
So we learn that 
\bea 
{ 1 \over   n! } \sum_{ \sigma }  \chi^S ( \sigma ) \chi^R ( \sigma^2 ) = \Mult ( \Sym^2  (V_R) , V_S ) -  \Mult ( \Lambda^2 (V_R) , V_S ) 
\eea
So we can re-write 
\bea 
S_{ [ 2,1] }^{ ( R , R , S ) }   =   \left  ( \Mult ( \Sym^2  (V_R) , V_S ) -  \Mult ( \Lambda^2 (V_R) , V_S )    \right )^2 
\eea
From the above of course 
\bea 
\sC ( R , R , S ) = \Mult ( \Sym^2  (V_R) , V_S ) + \Mult ( \Lambda^2 (V_R) , V_S )   
\eea
The two terms which contribute to the dimension of $ \cK ( n ) $ when we restrict the list $ [ R_1 , R_2 , R_3 ] $ to take values in $ \{  R , R , S \} $ are 
\bea 
{ 1 \over 2 } \sC ( R , R , S )^2 + { 1 \over 2 } \sC ( R , R , S ) = { 1 \over 2 } \sC ( R , R , S ) ( \sC ( R , R , S ) +1 ) 
\eea
This shows that the restriction  of $ \cK_{Y_0}  ( n ) $ to the sector where two of the irreps are equal is integer. 

Another  very instructive way to write the dimension of this subspace of $ \cK_{Y_0} ( n ) $ is 
\bea\label{WARRS} 
&& { 1 \over 2 } \sC ( R , R , S )^2  + { 1 \over 2 } \left  ( \Mult ( \Sym^2  (V_R) , V_S ) -  \Mult ( \Lambda^2 (V_R) , V_S )    \right )^2  \cr 
&& = { 1 \over 2 }  \left ( \Mult ( \Sym^2  (V_R) , V_S ) + \Mult ( \Lambda^2 (V_R) , V_S )  \right )^2 \crcr
&& 
   +  { 1 \over 2 } \left  ( \Mult ( \Sym^2  (V_R) , V_S ) -  \Mult ( \Lambda^2 (V_R) , V_S )    \right )^2  \cr 
   && =(  \Mult ( \Sym^2  (V_R) , V_S )  )^2  + ( \Mult ( \Lambda^2  (V_R) , V_S ) )^2  
\eea 
This formula gives an expression for the dimension of the $ ( R , R , S ) $ subspace of $ \cK_{Y_0} ( n ) $
as a sum of squares. That is, we have identified the WA blocks of the decomposition. 

Based on this counting, we can write down the basis elements for $ \cK_{Y_0} ( n ) $ 
in the subspace where two of the $ R_i $ are equal to $R$ and the third is $S$. 
We define Clebsch-Gordan coefficients
\bea 
&& C^{ R , R , S ; [2] , \tau_1 }_{ i_1 , i_2 , i_3 } \cr 
&& C^{ R , R , S ; [1^2] , \tau_2 }_{ i_1 , i_2 , i_3 } 
\eea
The first are the Clebsch-Gordan coefficients coupling $ \Sym^2 ( R ) \otimes S $ 
to the trivial representation. The second are the Clebsch-Gordan coefficients coupling $ \Lambda^2 ( R ) \otimes S $ 
to the trivial representation. The   basis  elements in $ \cK_{Y_0} ( n )$ are of two types 
\bea
&&
C^{ R , R , S ; [2] , \tau_1 }_{ i_1 , i_2 , i_3 }
  C^{ R , R , S ; [2] , \tau_2 }_{ j_1 , j_2 , j_3 } \times  \crcr
&&
\left ( 
Q^{ R }_{ i_1 , j_1 } \otimes Q^{ R }_{ i_2 , j_2 } \otimes Q^{ S}_{ i_3 , j_3 } 
+  Q^{ R }_{ i_1 , j_1 } \otimes Q^{ S}_{ i_3 , j_3 } \otimes Q^{ R }_{ i_2 , j_2 } 
+ Q^{ S}_{ i_3 , j_3 } \otimes Q^{ R }_{ i_1 , j_1 } \otimes Q^{ R }_{ i_2 , j_2 } 
\right )
\eea
and 
\bea
&&
 C^{ R , R , S ; [1^2] , \tau_1 }_{ i_1 , i_2 , i_3 }
 C^{ R , R , S ; [1^2] , \tau_2 }_{ j_1 , j_2 , j_3 } \times  \crcr
&&
  \left ( 
Q^{ R }_{ i_1 , j_1 } \otimes Q^{ R }_{ i_2 , j_2 } \otimes Q^{ S}_{ i_3 , j_3 } 
+  Q^{ R }_{ i_1 , j_1 } \otimes Q^{ S}_{ i_3 , j_3 } \otimes Q^{ R }_{ i_2 , j_2 } 
+ Q^{ S}_{ i_3 , j_3 } \otimes Q^{ R }_{ i_1 , j_1 } \otimes Q^{ R }_{ i_2 , j_2 } 
\right )
\eea
These agree with the correct counting in \eqref{WARRS}.
Note that this is different from the naive guess of $ \sC ( R , R , S )^2 $ 
for this sector.

\subsubsection{Subspace of  $\cK(n)$ with all equal $ R_1 = R_2 = R_3 = R$           }

The above allows us to guess what will happen when we consider the subspace of $ \cK_{Y_0} ( n )$ corresponding to $ \{  R , R , R \} $. 
We should be able to write the dimension of that as 
\bea 
( \Mult ( P_{ [3] } V_R^{ \otimes 3 } , V_0   ) )^2   + (  \Mult ( P_{ [2,1] } V_R^{ \otimes 3 } , V_0   ) )^2  + ( \Mult ( P_{ [1^3 ] } V_R^{ \otimes 3 } , V_0   ) )^2  
\eea
The subscripts on the $P$'s are Young diagrams with three boxes. Maybe we need a factor of $2$ or $4$ in front of the second term. 
This would be the WA decomposition of the $\cK_{Y_0} ( n )$ projected to the the $(R,R,R)$ sector.
 
We now prove this.  So we apply the projector 
 \bea 
&& { 1 \over 6 (d!)^2 }  \sum_{ \alpha \in S_{ 3 } } \sum_{ \sigma_1, \sigma_2  \in S_n } 
\alpha \rho_L ( \sigma_1 ) \rho_R ( \sigma_2 )  Q^{R}_{ i_1 , j_1 } \otimes Q^{R}_{i_2 j_2 } \otimes Q^{R}_{i_3 j_3  }  \cr 
&& = { 1 \over 6 (d!)^2 }  \sum_{ \alpha \in S_{ 3 } } \sum_{ \sigma_1, \sigma_2  \in S_n }  D^{R}_{  p_1 i_1 } ( \sigma_1 )  D^{R}_{  p_2 i_2 } ( \sigma_1)  D^{R}_{  p_3 i_3 } ( \sigma_1 ) 
        D^{R}_{  j_1 q_1 } ( \sigma_2 )  D^{R}_{  j_2 q_2  } ( \sigma_2 )  D^{R}_{ j_3 q_3 } ( \sigma_2 ) \cr\cr
        &&\times 
        Q^{R}_{ p_{ \alpha(1) } , q_{\alpha(1)}  } \otimes Q^{R}_{ p_{\alpha(2)}   q_{\alpha(2)}  } \otimes Q^{R}_{p_{ \alpha(3)}  q_{ \alpha(3)}    } 
\cr 
        && 
 \eea
Pair this with $ Q^{R}_{ i_1 , j_1 } \otimes Q^{R}_{i_2 j_2 } \otimes Q^{R}_{i_3 j_3  }$ and sum    over $ R , p_l$, and $q_l $ to find 
\begin{align}
&  { 1 \over 6 (d!)^2 }  \sum_{ \alpha \in S_{ 3 } , \sigma_l  \in S_n } \sum_{  p_l ,  q_l}  D^{R}_{  p_1  p_{ \alpha(1) } } ( \sigma_1 )  D^{R}_{  p_2 p_{\alpha(2)}   } ( \sigma_1)  D^{R}_{  p_3 p_{ \alpha(3)}   } ( \sigma_1 ) 
        D^{R}_{ q_{\alpha(1)}    q_1 } ( \sigma_2 )  D^{R}_{  q_{\alpha(2)}  q_2  } ( \sigma_2 )  D^{R}_{  q_{\alpha(3)}  q_3 } ( \sigma_2 ) \crcr
& = \sum_{ \alpha }  \sum_{ \sigma_1 \in S_n } \sum_{ \sigma_2 \in S_n }  \tr_{ R^{ \otimes 3 } } ( \sigma_1 \alpha )  \tr_{ R^{ \otimes 3 } } ( \sigma_2 \alpha )  = \sum_{ \alpha } \sum_{ \sigma_1 \in S_n } \sum_{ \sigma_2 \in S_n } \tr_{ R^{ \otimes 3 } \otimes R^{ \otimes 3 } }  ( \sigma_1 \otimes \sigma_2 ) ( \alpha \otimes \alpha ) 
\end{align}

This can be understood in terms of representation theory of $ S_n \times S_3 $
acting on $ V_R^{ \otimes 3 } $. 
\bea\label{tensRcubed} 
V_R^{ \otimes 3 } = \bigoplus_{ \Lambda_1 \vdash n }  \bigoplus_{ \Lambda_2 \vdash 3 } V_{ \Lambda_1 }^{ (S_n)}  \otimes V_{ \Lambda_2}^{(S_3)} 
 \otimes V_{R :  \Lambda_1 , \Lambda_2 }  
\eea
We take two copies of these tensor products 
\bea\label{LRVR3}  
V_R^{ \otimes 3 } \otimes V_{R}^{ \otimes 3 } = 
 \bigoplus_{ \Lambda_1 , \Lambda_1'  \vdash n }  \bigoplus_{ \Lambda_2 , \Lambda_2' \vdash 3 } V_{ \Lambda_1 }^{ (S_n)}  \otimes V_{ \Lambda_2}^{(S_3)} 
 \otimes V_{R :  \Lambda_1 , \Lambda_2 } \otimes V_{ \Lambda_1'}^{ (S_n)}  \otimes V_{ \Lambda_2'}^{ (S_3)}  \otimes V_{ R :  \Lambda_1' , \Lambda_2' }  
\eea
The sums over $ \sigma_1 , \sigma_2 \in S_n $  project to  $ \Lambda_1 = \Lambda_1' = [ n ] $. We further need to restrict to the trivial of the diagonal $ S_3$. 
This trivial rep occurs once inside $ V_{ \Lambda_2 } \otimes V_{ \Lambda'_2 } $ whenever $ \Lambda_2 = \Lambda'_2 $.
The multiplicity is 
\bea\label{WARRR}  
\sum_{ \Lambda \vdash 3 } ( \dim  V_{ R :  [ n ] , \Lambda  } )^2  
\eea
This gives the WA decomposition of 
$ \cK_{Y_0} ( n ) $ projected by overlapping the permutation triples with $ P_{ R } \otimes P_{ R } \otimes P_R $.

Corresponding to the decomposition (\ref{tensRcubed}) there are Clebsch-Gordan coefficients 
for the change of basis from  an orthogonal basis 
of states $ | R, R , R ; i_1 , i_2 , i_3 \rangle $ to another orthogonal basis of 
 states  $| \Lambda_1  , \Lambda_2 , \tau_{ \Lambda_1 , \Lambda_2 } ; m_{\Lambda_1} , m_{\Lambda_2} \rangle  $ as 
\bea 
C^{ R,R, R ; \Lambda_1 , \Lambda_2 , \tau_{ \Lambda_1 , \Lambda_2 }}_{ i_1 , i_2 , i_3 ; m_{ \Lambda_1}  , m_{ \Lambda_2}  }  ={  \langle \Lambda_1 , \Lambda_2 , \tau_{ \Lambda_1 , \Lambda_2 } ; m_{ \Lambda_1 }  , m_{ \Lambda_2}  | } R, R , R ; i_1 , i_2 , i_3 \rangle
\eea
For the case $ \Lambda_1 = [n] , \Lambda_2 = \Lambda $, we have 
\bea\label{ClbschRRRn} 
C^{ R,R, R ; [n] , \Lambda , \tau_{ [n] ,\Lambda }}_{ i_1 , i_2 , i_3 ;  m_{ \Lambda }  }  = { \langle [n] , \Lambda, \tau_{ [n] , \Lambda} ;  m_{ \Lambda} }| R, R , R ; i_1 , i_2 , i_3 \rangle
\eea
The irrep $[n]$ is one-dimensional so has no corresponding state label.

The elements of the WA basis for $ \cK_{ Y_0} ( n)$ will be labelled by $ (R ,\Lambda ,
 \tau_{ [n] ,  \Lambda}^{(l)}  , \tau_{[n],  \Lambda}^{(r)} )  $ where  $ \tau_{ [n] ,  \Lambda}^{(l)} , \tau_{[n],  \Lambda}^{(r)} $ label the multiplicity of 
$ V_{ [ n ] } \otimes V_{ \Lambda }  $ in
the left and right $V_{R}^{ \otimes 3 } $  of (\ref{LRVR3}). 
Using the Clebsch-Gordan coefficients (\ref{ClbschRRRn}) we can therefore  write the WA basis 
\bea 
Q^{R ,  \Lambda , \tau_{[n] ,  \Lambda}^{(l)}   , \tau_{ [n] , \Lambda }^{(r)}  } = \sum_{  i_l ,  j_l , m_{ \Lambda }  }
C_{i_1 , i_2 , i_3 }^{ R , R , R ; \Lambda  , m_{ \Lambda } , \tau_{ [n],  \Lambda }^{(l)}
 } 
C_{j_1 , j_2 , j_3 }^{ R , R , R ;  \Lambda  , m_{ \Lambda } , \tau_{ [n],  \Lambda}^{(r)}   }
    Q^{ R}_{ i_1 j_1} \otimes Q^{ R}_{ i_2 j_2} \otimes Q^{ R}_{ i_3 j_3} 
\eea
which matches the counting of states in this sector. Following steps 
similar to the ones for $ Q^{ R  , S , T ; \tau_1 , \tau_2 } $ in the case of $ \cK ( n )$, 
the orthogonality properties of these Clebsch-Gordan coefficients can be used to show that 
these form a basis of matrix units for $ \cK_{Y_0} ( n )$ in the subspace corresponding to 
$ R = S = T$.

 A consequence of the above discussion (in particular equations (\ref{WARST}), (\ref{WARRS}) and (\ref{WARRR})) 
  is that we can write the dimension of the algebra $ \cK_{Y_0} (n)$ as a sum of squares, corresponding to the WA decomposition 
\bea\label{WAK0}  
&& \dim ( \cK_{Y_0} ( n ) ) = \sum_{ R \ne S \ne T } ( \sC ( R , S , T ) )^2  + \sum_{ R \ne S } ( \Mult ( \Sym^2 ( R ) , S ) )^2 + ( \Mult ( \Lambda^2 ( R ) , S ) )^2 \cr 
&& ~~~ + ~~~~ \sum_R \sum_{ \Lambda } ( \Mult ( R^{ \otimes 3 } ,  [ n ] \otimes \Lambda  )  )^2 
\eea
This color-symmetrized analog of 
(\ref{dimKKron}) is a key result of the present paper.

\subsubsection{More on $ S^{(3)}_{ [2,1]} (n)  $ and the character table of $S_n$.   }

A useful fact is that for any irrep $R$ of $S_n$, the trivial appears in the symmetric part of $ V_R \otimes V_R $ with multiplicity one.  This is related to the reality of these representations (eq. 5.82 of \cite{Hammermesh}).
It leads to the identify 
\bea 
{ 1 \over n! } \sum_{ \sigma} \tr_{ R \otimes R } ( \sigma { ( 1 + (12) ) \over 2 } ) ={ 1 \over n! }  \sum_{ \sigma} \tr_{ R \otimes R } ( \sigma  )
\eea  
The r.h.s side  counts the number of times the trivial representation appears in $ R \otimes R $. 
 The l.h.s counts the number of times it appears in the symmetric part. 
 This leads to 
 \bea 
 { 1 \over 2 n! } \sum_{ \sigma \in S_n } \left ( ( \chi^R ( \sigma) )^2 + \chi^R ( \sigma^2 ) \right ) 
 =  { 1 \over n! } \sum_{ \sigma \in S_n } ( \chi^R ( \sigma ) )^2 
 \eea
 This in turn implies that 
 \bea 
\sum_{ \sigma \in S_n } ( \chi^R ( \sigma ) )^2  =  \sum_{ \sigma \in S_n } ( \chi^R ( \sigma^2 ) )
 \eea
 It also follows that  
\bea 
\sum_{ \sigma \in S_n  } \delta ( \tau^{-1}  \sigma^{2 }  )  = \sum_{  S} \chi^S ( \tau  )  
\eea
The number of permutations which squares to $ \tau $ can be written as a sum of 
characters in all irreps. To see this, use 
\bea 
&& \sum_{ \sigma \in S_n  } \delta ( \tau^{-1}  \sigma^{2 }  )  
 = {1 \over n! } \sum_{ \sigma \in S_n } \sum_{ S \vdash n } 
  \chi^S( \tau ) \chi^S  ( \sigma^2 ) \cr 
  && = {1 \over n! } \sum_{ \sigma \in S_n } \sum_{ S \vdash n } 
  \chi^S ( \tau ) \chi^S  ( \sigma ) \chi^S ( \sigma ) \cr 
  && = \sum_{ S } \chi^S ( \tau ) 
\eea

We can also write 
\bea 
&& { 1 \over n! } \sum_{ R , \sigma } \chi^R ( \sigma^2 )  \chi^R ( \sigma^2 )  = { 1 \over n! } 
 \sum_{ R , \sigma , \tau } \chi^R ( \tau )  \chi^R ( \tau  ) \delta  ( \tau^{-1}  \sigma^2 ) \cr 
&& ={ 1 \over n! }  \sum_{ R , S , \tau } \chi^R ( \tau )  \chi^R ( \tau  ) \chi^S ( \tau  )   \cr 
&& =\sum_{ R ,  S }  \sC( R , R , S ) 
\eea

So the second contribution 
\bea 
S_{ [ 2,1 ] } = \sum_{ R  , S } \sC ( R  , R , S ) 
\eea

Also if we do the sum over $R$, we  get 
\bea 
S_{ [ 2,1 ] } = { 1 \over n! } \sum_{ S }\sum_{ \tau }  \chi^S ( \tau ) \Sym ( \tau )  = \sum_{ p }\sum_{ S } \chi^{ S }  ( \tau_p ) 
\eea
We are summing over irreps $S$ and conjugacy classes. The weight is the character of a permutation  in the specified conjugacy class, here denoted $ \tau_p $ 
 for conjugacy class specified by $p$. Indeed  OEIS recognizes $S_{[2,1]} $ as the sum of entries of the character table of $S_n$.  The refinement of $S_{[2,1]} $
 parametrized by $S$ (where we drop the sum over $S$)  
 \bea 
  \sum_{ p } \chi^S ( \tau_p ) 
 \eea
is the subject of an open question posed by Stanley (problem 12 in \cite{Stanleyproblems}): to find a combinatoric construction which makes the positive integrality manifest. The positive integrality is manifest 
because 
 \bea 
 \sum_{ R } \sC ( R , R , S ) = \sum_{  p } \chi^{ S }  ( \tau_p ) 
 \eea
 (an identity that has been used above) but this is a representation theoretic argument, not a purely combinatoric one. 
 Tensor invariants at large $N$ (or equivalently colored graphs) provide a combinatoric interpretation of the sum of 
squares of the Kronecker coefficients. It would be interesting to investigate whether refined consideration of colored graphs can provide an approach to this question of Stanley. 

\subsection{$S_3$-refinement for  $ \cK ( n )$ } 

We have given above the dimension of the color-symmetrized subspace $ \cK_{ Y_0} ( n ) $  
as a sum over representation theoretic data. The expression is  a sum of squares as expected 
from the WA decomposition. This shows that the representation theoretic construction perspective based on permutations and Fourier transforms naturally leads to 
the explicit form of the Matrix blocks of the WA decomposition. The expression is a sum over three types of terms, which we may describe as $ ( R , S, T ) $ types involving three distinct Young diagrams, the $ ( R , R , S )$ which involves two distinct types 
and the $ ( R ,R , R )$ which involves one type. Here we give the dimensions of 
\bea 
\cK^{ ( R , S , T ) }_{ Y } , \,\cK^{ ( R , R, S  ) }_{ Y } , \,\cK^{ ( R , R , R )  }_{ Y } 
\eea
general Young diagram  $Y$ of $S_3$. 

 For the $ ( R , S , T )$ sector, the answer is easy when we consider the restriction of 
  the trace  
 \bea\label{tracePY} 
 \tr_{ \mC ( S_n ) ^{ \otimes 3 }  } ( P_{Y} P_{ L}^{ S_n } P_{R}^{ S_n} ) 
  = \tr_{ \cK ( n) } ( P_{ Y} ) 
 \eea 
 to the subspace of $ \mC ( S_n ) \otimes \mC ( S_n ) \otimes \mC ( S_n )$
  to the Fourier basis states $ Q^{ R_1}_{ i_1 , j_1} \otimes Q^{ R_2}_{ i_2 , j_2} 
 \otimes Q^{ R_3}_{ i_3 , j_3} $ where $ R_1 , R_2 , R_3 $ are all different 
 and take values in the set $ ( R , S , T ) $. There are $6$ choices, which add up 
 to $ 6 \sC ( R , S , T )^2 $. In the expansion of $P_Y$ in terms of permutations only
 the identity permutation contributes and we have 
 \bea 
 \cK^{ ( R , S , T ) }_{ Y } = d(Y)^2 \sC ( R , S , T )^2 
 \eea
Hence we have 
\bea 
&& \dim \cK^{ (R,S,T)}_{ [3]} = \sC ( R , S , T )^2 \cr 
 && \dim \cK^{ (R,S,T)}_{ [2,1]} = 4 \sC ( R , S , T )^2 \cr 
 && \dim \cK^{ (R,S,T)}_{ [1^3]} =  \sC ( R , S , T )^2  
\eea
 
 For the  $ ( R , R , S ) $ sector, we find 
  \bea\label{RRSY}  
 &&  \dim ( \cK^{ ( R , R , S ) }_{ Y=[3] } ) 
   = ( \Mult ( \Sym^2 ( V_R ) , V_S  ) )^2  + ( \Mult ( \Lambda^2 ( V_R ) , V_S ) )^2 \cr 
&& \dim ( \cK^{ ( R , R , S ) }_{ Y={[2,1]}  }  ) = 2 \sC ( R , R , S )^2  \cr 
&&  \dim ( \cK^{ ( R , R , S ) }_{ Y={[1^3 ]}  } )  = 2  \Mult ( \Sym^2 ( V_R ) , V_S  )\Mult ( \Lambda^2 ( V_R ) , V_S )
  \eea
 
For the $ ( R , R , R )$ sector, we find
\bea
&&  
\dim ( \cK^{ ( R , R , R ) }_{  Y } ( n ) ) 
=\crcr
&& \sum_{ Y \vdash 3 }~ d(Y) ~ \sC ( Y_1 , Y_2 , Y ) ~  
 \Mult ( V_R^{ \otimes 3 } , V^{ (S_3)}_{ \Lambda_1} \otimes V^{ (S_n)}_{ [n]} )
  ~ \Mult ( V_{ R}^{ \otimes 3 } , V_{ \Lambda_2 }^{ (S_3)} \otimes V_{ [n]}^{ (S_n)} )
  \eea  
  $ \sC( Y_1 , Y_2 , Y )$ is the Kronecker coefficient which counts the number of 
  invariants of $S_3$ in the tensor product $ Y_1 \otimes Y_2 \otimes Y$. 

The derivation of these formulae proceeds by unravelling the 
the equation (\ref{tracePY})in each of these case. 
Some interesting consistency checks of these formulae can be easily given. 
We have the identity
\bea 
d(Y) \sC ( Y_1 , Y_2 , Y ) = \sum_{ Y  } \sum_{ \sigma }  { 1 \over d! } \chi^{ Y_1  }  ( \sigma )  \chi^{ Y_2} ( \sigma ) \chi^{ Y  } ( \sigma ) = d( Y_1) d(Y_2) 
\eea  
  Doing the sum over irreps $ Y$ gives a delta function. 
For the $ (R, R , R) $ case, therefore  we have 
\bea 
&& \sum_{ Y } \dim ( \cK^{ ( R , R , R ) }_{  Y } ( n ) ) \crcr
&& 
= \sum_{ Y_1 , Y_2 } d( Y_1) d( Y_2) 
\Mult ( V_{R}^{ \otimes  3} , V_{ Y_1}^{ (S_3)} \otimes V_{ [n]}^{S_n }  ) \Mult ( V_R^{ \otimes 3} , V_{ Y_2}^{(S_3) } \otimes V_{ [n]}^{ (S_n)}   ) \cr 
&& = \sC ( R , R , R )^2 
\eea  
Similarly, the reader can easily convince herself that 
\bea
\sum_{ Y } \dim ( \cK^{ ( R , R , S ) }_{  Y } ( n ) ) = 3 ( \sC ( R , R , S ) )^2
\eea
The $3$ comes from the fact that when the ordered list of Young 
diagrams  $ [ R_1 , R_2 , R_3 ] $ takes values from the set $ \{ R , R , S \} $, there are three possibilities. 

An interesting consequence of the multiplication rule given 
in (\ref{multselrules}) is that $ \cK_{ Y_0} (n)\oplus \cK_{ Y_2}(n) $ is a closed associative algebra. It will inherit a non-degenerate bilinear form from 
the $ \mC ( S_n ) \otimes \mC ( S_n ) \otimes \mC ( S_n )$ (or from 
$ \mC ( S_n ) \otimes \mC ( S_n ) $)  if we are working with the gauge-fixed formulation.  
So we expect that its dimension will be a sum of squares. The counting in terms of 
representations above automatically leads to such a sum of squares. 
For the $ ( R , S , T )$ subspace 
\bea 
\dim ( \cK_{Y_0}^{ ( R , S , T ) } ) + \dim ( \cK_{ Y_1}^{ ( R , S , T ) }  ) 
 =  2 ( \sC ( R , S , T ) )^2 
\eea
For the $ ( R , R , S ) $ subspace, 
\bea 
&& \dim \cK^{ ( R , R , S ) }_{ Y_0} + \dim \cK^{ ( R , R , S ) }_{ Y_2 } \cr 
&& = ( \Mult ( \Sym^2 ( V_R ) , V_S  ) )^2  + ( \Mult ( \Lambda^2 ( V_R ) , V_S ) )^2\crcr
&&  + 2  \Mult ( \Sym^2 ( V_R ) , V_S  )\Mult ( \Lambda^2 ( V_R ) , V_S )\cr
&& = ( \Mult ( \Sym^2 ( V_R ) , V_S  ) + \Mult ( \Lambda^2 ( V_R ) , V_S ) )^2 \cr 
&& = ( \sC ( R , R , S ) )^2 
\eea
For the $ ( R , R , R ) $ subspace 
     \bea 
 && \left( \Mult ( V_R^{ \otimes 3 } , V^{ (S_3)}_{ [3] } \otimes V^{ (S_n)}_{ [n]} ) \right)^2 
 + \left( \Mult ( V_R^{ \otimes 3 } , V^{ (S_3)}_{ [1^3] } \otimes V^{ (S_n)}_{ [n]} ) \right)^2 \cr
 &&
  +  4 \left( \Mult ( V_R^{ \otimes 3 } , V^{ (S_3)}_{ [2,1] } \otimes V^{ (S_n)}_{ [n]} ) \right)^2 \crcr
  && 
  + 2  \left( \Mult ( V_R^{ \otimes 3 } , V^{ (S_3)}_{ [3] } \otimes V^{ (S_n)}_{ [n]} ) \right) 
  \left( \Mult ( V_R^{ \otimes 3 } , V^{ (S_3)}_{ [1^3] } \otimes V^{ (S_n)}_{ [n]} ) \right) \cr 
&&  =  \left ( \Mult ( V_R^{ \otimes 3 } , V^{ (S_3)}_{ [3] } \otimes V^{ (S_n)}_{ [n]} ) \right)^2   
      + \left ( \Mult ( V_R^{ \otimes 3 } , V^{ (S_3)}_{ [1^3] } \otimes V^{ (S_n)}_{ [n]} ) \right )^2 
     \cr
     &&  +  4 \left ( \Mult ( V_R^{ \otimes 3 } , V^{ (S_3)}_{ [2,1] } \otimes V^{ (S_n)}_{ [n]} ) \right )^2 \cr 
       && 
       \eea

These  counting formulae for $ \dim \cK_{ Y } ( n ) $ in terms of representation theory data sets the stage for developing  representation theoretic bases. 
Using the basic technique of using permutations to construct observables (\ref{mapPermsObs}), 
these elements of $ \cK_{ Y  } ( n )$ will be expected to  give a refined orthonormal basis 
for the gauge-invariant observables, with good quantum numbers for the  $S_3$ color-exchange, as we described earlier for $Y=Y_0$.  This will be an interesting refinement of the results on orthogonal bases given earlier in section  \ref{sect:CCorr} and in  \cite{DR1706,DGT1707}.

\subsection{Counting color-symmetrised tensor invariants for general $d$ }
\label{sec:gensolsymd} 
 
 This way of approaching the calculation using Fourier transforms,
 presented for $d=3$ at the start of this section,  allows  us to generalize to any $ d$.  
 We get 
 \bea\label{generalformula} 
&&  \hbox { Number of color symmetrised tensor invariants of rank $d$ } = \cr 
&& { 1 \over d!  (n!)^2 }   \sum_{ \alpha \in S_d } \sum_{ R_1 , \cdots , R_d }  \sum_{ \sigma_1 \in S_n  } \sum_{ \sigma_2 \in S_n }  \prod_a \chi^{ R^{(a)}  } ( \sigma_1^{ l_a   } )
    \chi^{ R^{(a)}  } ( \sigma_2^{ l_a } )  \delta ( R_{ \alpha_a^{ 1} } , R_{ \alpha_a^{2}} , \cdots , R_{ \alpha_a^{ l_a} }   , R^{ (a) } )
      \cr 
     && 
 \eea
 Here the index $a$ runs over the cycles of the permutation $ \alpha $; $l_{a} $ is the length of the cycle. 
 Each such cycle is of the form $ ( \alpha_a^1 , \alpha_a^2 , \cdots ,  \alpha_a^{ l_a}  ) $ where the entries in  the cycle 
 are integers chosen from $ \{ 1 , \cdots d  \} $. Such a cycle leads to delta functions 
 enforcing $ R_{ \alpha_a^1} = R_{ \alpha_a^2 } \cdots  = R_{ \alpha_a^{ l_a}  } $, which leads to the definition 
 \be
 R^{(a)} =R_{ \alpha_a^1} = R_{ \alpha_a^2 } \cdots  = R_{ \alpha_a^{ l_a}  }
 \ee

We can re-write the counting as 
\bea 
&&  \hbox { Number of color symmetrised tensor invariants of rank $d$ } = \cr 
&& { 1 \over  (n!)^2 } \sum_{ p \vdash d }  \sum_{ \substack { R_{ i , j } \vdash d \\  \{ 1 \le i \le d , 1 \le j \le p_i  \} }}                 
\sum_{ \sigma_1 \in S_n  } \sum_{ \sigma_2 \in S_n } { 1 \over  | \Sym ( p ) | } \prod_{ i =1}^d \prod_{ j =1}^{ p_i } \chi^{ R_{ i , j } } ( \sigma_1^i ) 
 \chi^{ R_{ i , j } } ( \sigma_2^i )   
\eea
We have collected from (\ref{generalformula}) 
all the terms coming from a fixed conjugacy class  of $ \alpha $, which corresponds to a partition $p$ of $d$, specified 
by multiplicities $p_i$ of cycles lengths $i$ in the permutation $ \alpha $, i.e. $ \sum i p_i = d $.  
Given the delta functions on the representations, for a given $p$, the number of distinct representations being summed 
after using these delta functions is $ \sum_{ i=1}^d p_i $. We denote these representation labels $ R_{ i , j }$, where 
$i$ runs over the possible cycle lengths and $ j $ runs over the distinct cycles of the same length $i$.  

This is also the dimension  of the projection of $ \cK ( n , d ) $ to the subspace belonging to the one-dimensional irrep of $S_d$. 
We have in general 
\be
\cK ( n , d ) = \bigoplus_{ Y \vdash d  }(  \cK ( n , d ))_{Y }   
\ee
The above gives the projection to the $ S_d$ invariant subspace.  
\bea 
&& \dim (  \cK ( n , d ))_{Y  = [ n] } )  =   \hbox { Number of color symmetrised tensor invariants of rank $d$ } \cr 
&& { 1 \over  (n!)^2 }  \sum_{ p \vdash d }  \sum_{ \substack { R_{ i , j } \vdash d \\  \{ 1 \le i \le d , 1 \le j \le p_i  \} }}                 
 \sum_{ \sigma_1 \in S_n  } \sum_{ \sigma_2 \in S_n } { 1 \over  | \Sym ( p ) | } \prod_{ i =1}^d \prod_{ j =1}^{ p_i } \chi^{ R_{ i , j } } ( \sigma_1^i ) 
 \chi^{ R_{ i , j } } ( \sigma_2^i )  
\eea
For general representations $Y$, we have 
\bea 
&& \dim ( (  \cK ( n , d ))_{Y }  ) = \\
&&
 { d(Y) \over  (n!)^2 } \sum_{ p \vdash d }  \chi^Y ( \sigma_p )  \sum_{ \substack { R_{ i , j } \vdash d \\  \{ 1 \le i \le d , 1 \le j \le p_i  \} }}       \sum_{ \sigma_1 \in S_n  } \sum_{ \sigma_2 \in S_n }          
 { 1 \over  | \Sym ( p ) | } \prod_{ i =1}^d \prod_{ j =1}^{ p_i } \chi^{ R_{ i , j } } ( \sigma_1^i ) 
 \chi^{ R_{ i , j } } ( \sigma_2^i )  \nonumber
\eea
For practical computations, expressions for the  $ S^{ (d)}_{p} (n)  $  were also given 
in \cite{Sanjo}, for  $ d \le 4 $,   in terms of sums over partitions of $n$, with weights obtained by applying appropriate substitutions to the generating function of cycle indices of $S_n$ (equations (63) and (77) of arXiv version). 
  The generalization of these expressions in terms of partition sums to general $d$ is left as an interesting exercise for the future.

\subsection{$\cK^{ \infty} $ and color symmetry } 

We explained in section \ref{KinfCorr}  that the infinite direct sum $ \cK^{ \infty}$
\bea 
\cK^{ \infty} = \bigoplus_{n=0}^{ \infty} \cK ( n ) 
\eea
 has 
two products, which both play a role in correlators. If we restrict to the color-symmetrised 
subspace  
\bea 
\cK^{ \infty}_{ Y_0} = \bigoplus_{ n=0}^{ \infty} \cK_{Y_0}   ( n ) 
\eea
we again have a vector space with two products. We have already seen that the 
product at fixed $n$ of two elements in $ \cK_{ Y_0}  ( n )$ is in $ \cK_{ Y_0} ( n ) $. 
Likewise the outer product of two color-symmetrized elements in $ \cK_{ Y_0} ( n_1)  $ 
and $ \cK_{ Y_0} ( n_2 ) $ is a color-symmetrized element in $ \cK_{ Y_0 } ( n_1 + n_2)$. 
An easy way to see this is to think about the multiplication of color-symmetrized observables. 
Systematic investigations of color-symmetrized correlators is left for the future. We expect that 
the $ \cK^{ \infty} $ will prove to be a useful tool in these investigations.

\section{Summary and Discussion }
\label{concl}

\subsection{Summary } 

We have developed the description of the counting and correlators of 
general gauge invariant observables in  a class of  
tensor models started in \cite{Sanjo}. We focus on bosonic tensor models 
with a complex scalar field having $d$ indices.  We have showed that the permutation centralizer algebras introduced in \cite{PCA1601} provide a powerful framework for 
elucidating many aspects of correlators in the Gaussian model. The vector space of gauge-invariant observables in the rank-3 tensor model is isomorphic to the vector space of the algebra $ \cK ( n ) $.
This algebra  is spanned by elements in $ \mC ( S_n ) \otimes \mC ( S_n )$ which commute with the diagonally embedded $ \mC ( S_n )$. 
There is a also an equivalent description in terms of the subspace of 
$ \mC ( S_n ) \otimes \mC ( S_n ) \otimes \mC ( S_n )$ which is invariant under left and right action of the diagonal $ \mC ( S_n)$. 

$ \cK ( n )$ is a semi-simple associative algebra, i.e. an associative algebra with non-degenerate bilinear pairing. As a result, by the Wedderburn-Artin theorem, it is isomorphic to a direct sum of matrix algebras. The number of blocks in $ \cK ( n )$ is 
the number of ordered triples $ [ R_1 ,R_2 , R_3   ]$ of Young diagrams with $n$ boxes
which have a non-vanishing Kronecker coefficient. The sizes of the blocks are the Kronecker coefficients $ \sC ( R_1 , R_2 , R_3 )$. The basis elements corresponding to the matrix decomposition are constructed using Clebsch-Gordan coefficients for the invariant in 
$ R_1 \otimes R_2 \otimes R_3$.  These basis elements of $ \cK ( n )$ correspond to gauge invariant observables which diagonalize the 2-point function of normal ordered observables. 
A subspace of observables corresponds to the centre of $ \cK ( n )$. These observables can be constructed without the detailed knowledge of Clebsch-Gordan coefficients. They only require characters of $ S_n$. A basis for the centre is given by triples of Young diagrams 
$ R_1, R_2 , R_3 $ which have non-vanishing $ \sC ( R_1 , R_2 , R_3 )$. One point functions of central observables are proportional to $ \sC ( R_1 , R_2 , R_3 )$. 

The above results are based on a few key ingredients: the parameterization of gauge invariant observables using equivalence classes of permutations, the use of representation theory to give a Fourier transformed description of observables in terms of Young diagrams, and Clebsch-Gordan coefficients. These methods have found extensive use in multi-matrix models over the recent years (an overview is in \cite{PermsGI}). 

The algebra $ \cK ( n)$ allows a systematic study of the implications of 
the color-exchange symmetry in  tensor models. In \cite{Sanjo} we had described 
counting formulae for color-symmetrized observables, which correspond to color-symmetrized graphs.  Here we give the complete decomposition of $ \cK ( n)$ in terms of a direct sum, labelled by Young diagrams of the $S_3$ color-exchange symmetry. 
\bea 
\cK ( n ) = \bigoplus_{ Y \vdash 3 }  \cK_Y  ( n )
\eea
The color-symmetrized subspace $ \cK _{ Y_0 } ( n)$ is a closed sub-algebra of $ \cK ( n)$. 
Again as a result of the WA theorem, we immediately expect that it should be a direct sum of 
matrix algebras. The corresponding counting formula as a sum of squares is given in (\ref{WAK0}), and the corresponding refined Clebsch-Gordan coefficients are described. 
Similar group theoretic decompositions for $ \cK_Y  ( n )$ are given in section \ref{sect:colsym}.

 The counting formulae for $ \cK_{Y_0}  ( n)$ can be expressed in terms of sums over characters of $S_n$ parametrized by partitions of $S_d$ (with $d=3$). Such formulae were derived using the Burnside lemma in   \cite{Sanjo} for $ d=3,4$ and used 
  to get explicit number sequences for dimensions of 
 the space of color-symmetrized observables (color-symmetrized graphs). We have generalized these character formulae (section (\ref{sec:gensolsymd})) 
  to general $d$, by exploiting an alternative derivation which makes 
  use  the representation theoretic Fourier basis for $( \mC ( S_n ))^{ \otimes d }$. 
 The group theoretic results we have developed for color-exchange symmetry 
   will be useful for the study of correlators in Gaussian models, as well as interacting models which are perturbations of Gaussians by color-symmetrized observables. This is an interesting direction for future investigations.

A number of other future research directions are suggested by the results of this paper.
We outline some of them below.

\subsection{Towards Young diagram statistical models and field theory} 
\label{sect:YD-QFT}

We have found above that interesting classes of observables in 
tensor models, related to the centres of permutation algebras built from 
equivalence classes of permutations describing general observables, are parametrized by 
sets of Young diagrams.  Their correlators are directly related to fundamental representation theoretic quantities, e.g.  \eqref{corrFourier}. Similar observations 
in the context of multi-matrix models are developed in \cite{PCA1601}. This  leads us to a natural question:  Is there a statistical model of Young diagrams (YD) for which 
the functions \eqref{corrFourier} are the correlators? 
Section \ref{sec:tft2}  describes a 
 mapping of  these correlators observables in a topological field theory on
2-complexes.  Here we explore  a different perspective, 
and  provide a partial answer of the above 
question of what this statistical/field theory model could be. 

Fix $n\in \mathbb{N}$, and consider $R\vdash n$ a Young diagram.
A real field over YDs is a function $Y: p(n)\to \mathbb{R}$. We define an action of a $Y^k-$ Young diagram  
model (YDM) by: 
\be
S_{k-YDM}[Y] = \sum_{R,R' \vdash n} Y_{R}\,K(R,R')\,Y_{R'} 
+ g\sum_{R_l\vdash n} I(\{R_l\})
\prod_{l=1}^kY_{R_l} 
\label{sydft}
\ee
where $K(R,R')$ and $I(\{R_l\})$ are  kernels, 
 $g$ a coupling constant. From now, let us restrict to 
a cubic action determined by  $k=3$, 
$K(R,R')=\delta_{R,R'}$ and $I(\{R_l\})=\sC(R_1,R_2,R_3)$,
the Kronecker coefficient. 
The use of complex fields, 
the choice of $I(\{R_l\})$ as 
the Littlewood-Richardson coefficient,
or more generic  $Y^k$-models
of the form $I(\{R_l\})=\sC_{k}(R_1,R_2,\dots, R_k)$,
see \eqref{corrFourier}, might be also interesting choices
let for future investigations. 

Being interested in perturbation theory, 
the positivity of the above action will not be addressed here.
The partition function of the model \eqref{sydft} under the above
restrictions is of the form
\be
Z_{3-YDM}[g, J] 
 = \int \prod_{R\vdash n} dY_R \,
e^{-S_{3-YDM}[Y]  - \sum_{R \vdash n} J_{R} Y_{R} }
\ee
where $J_R$ is a source term. 
Correlators are computed perturbatively using the Gaussian 
measure $d\mu(Y)=\prod_{R\vdash n} dY_Re^{- \sum_R |Y_R|^2}$ 
and we find 
\bea
&& 
\big\langle Y_{S_1}Y_{S_2}\dots Y_{S_k} \big\rangle_{3-YDM}
 = \crcr
&& 
\sum_{n=0}^{\infty} \frac{(-g)^n}{n!}\int 
d\mu(Y) \Big(Y_{S_1}Y_{S_2}\dots Y_{S_k}\Big) 
\Big( \sum_{R_l\vdash n} \sC(R_1,R_2,R_3)
Y_{R_1}Y_{R_2}Y_{R_3} \Big)^{n}.
\eea
The free propagator in this theory is naturally defined by 
$\langle Y_R Y_{R'} \rangle_{\rm free} = G_0(R,R') = \delta_{RR'} $. 
Via the Wick theorem,
$N$-point correlators expand in terms of Feynman graphs, as
$\langle Y_{S_1} Y_{S_2} \dots Y_{S_N} \rangle_{{3-YDM};\, {\rm pert.} }
 = \sum_{\cG} K_{\cG} A_{\cG}$, 
with $A_{\cG}(S_1.S_2,\dots ,S_N)$ a graph amplitude, 
and $K_{\cG}$ a combinatorial factor. 

We evaluate a connected 3-point function a first order of
perturbation:  
\bea
&& 
\langle Y_{S_1} Y_{S_2}Y_{S_3}  \rangle_{{3-YDM};\, {\rm pert.; \,connected }}
 = - g \sum_{R_l} \sC(R_1,R_2,R_3) \prod_{\s \in S_3}\prod_{i} G_0(S_i,R_{\s(i)}) \cr\cr
&& 
 = -3!\, g\, \sC(R_1,R_2,R_3)\, . 
\eea
There are 3! trees contributing to the correlator and each of them has 
the same weight $-g\sC(R_1,R_2,R_3)$. 
Thus at this first order of perturbation $\langle Y_{S_1} Y_{S_2}Y_{S_3}  \rangle_{{3-YDM};\, {\rm pert. ;\,connected}}$ is proportional to $\langle \cO_{S_1,S_2,S_3}\rangle $ up to the factor $\frac{1}{(-3! g)(n!)^2}  [\prod_{i=1}^3 f_N(S_l) ]$. 

Computing a connected 4-point function $\langle Y_{S_1} Y_{S_2}Y_{S_3}Y_{S_4} \rangle_{{3-YDM};\, {\rm pert.; \,connected} }$ at second order of perturbation, 
we have a sum of Feynman amplitudes. 
Consider the tree graph which appears in that expansion,
$
 \begin{tikzpicture}
\draw (0,0)  -- (1,0) -- (1.5,0.5) ; 
\draw(-0.5,-0.5) -- (0,0) -- (-0.5,0.5) ; 
\draw (1.5,-0.5) -- (1,0) ;
\draw[color=black] (0.5,0.2) node {$R$};
\draw[color=black] (-0.75,-0.6) node {$S_1$};
\draw[color=black] (-0.8,0.6) node {$S_2$};
\draw[color=black] (1.75,0.6) node {$S_3$};
\draw[color=black] (1.75,-0.6) node {$S_4$};
 \fill[fill=black] (0,0) circle (0.1);
\fill[fill=black] (1,0) circle (0.1);
\end{tikzpicture} 
$,
the amplitude of which  is given by
\be
A_{t}(S_1.S_2,S_3,S_4)  
= \frac{g^2}{2}
\sum_{R\vdash n} \sC(S_1,S_2,R)\sC(R,S_3,S_4)
\ee
which is proportional to $\cO_{S_1,S_2,S_3,S_4}$ \eqref{corrFourier}. 
We conjecture that at any order $m-2> 0$ of perturbation theory, 
$\langle  \cO_{S_1,S_2,\dots,S_m} \rangle $ corresponds to a tree (hence connected) Feynman graph  of the correlator $\langle  Y_{S_1} Y_{S_2} \dots Y_{S_m} \rangle_{{3-YDM};\, {\rm pert.; connected}} $ up to a constant. 

Dealing with an action $S_{d-YDM}$, it is direct to get
at first order of perturbation: 
\be
\langle  Y_{S_1} Y_{S_2} \dots Y_{S_d} \rangle_{{d-YDM};\, {\rm pert.;\,connected}} 
\propto   \langle  \cO_{S_1,S_2,\dots,S_d} \rangle 
\ee
Beyond tree level, generic amplitudes should involve 
free sums over Young diagrams associated with loops 
in the graphs,  hence factors of the number of partitions $p(n)$. 
In the limit $n\to \infty$,  where we have an  infinite number of  degrees of freedom, 
one should expect that a YDM will have divergent amplitudes. 
It will be interesting to investigate the application of 
quantum field theoretic renormalization techniques to make sense of 
this limit.

\subsection{The space of holographic duals of tensor models }

In the early applications of matrix models as holographic duals to 
 quantum gravity in low dimensions of the nineties \cite{GM9304,Di Francesco:1993nw}, a detailed map was achieved 
 where the holographic duals included minimal model CFTs with $ c <1 $ coupled to Liouville theory and  the standard string theory $ b,c$ ghost system, as well as a $ c=-2$ model coupled to Liouville and ghosts as a dual for the Gaussian model \cite{Distler90}. A dual to the Gaussian model in terms of Belyi maps 
  and topological strings  on $\mC \mathbb{P}^1$ has also been investigated \cite{Gal,Gopak1104,GP1303,DN1411}, which should be related to the earlier $ c=-2$ proposal.
  
  The AdS2 dual for the double-scaled limit  with a quartic interaction (Gurau-Witten model)  is currently of active interest, with motivations from black hole physics  \cite{wit}. The rich mathematical structure of the Gaussian model raises the very interesting question of what is the precise dual of this model. The description of  
 permutation TFT2 constructions for the correlators of the Gaussian model 
 given in section \ref{sec:tft2} is a good starting point for investigations along the lines of \cite{Gal,Gopak1104,GP1303,DN1411}.  The rich mathematics involving permutations, Fourier transforms of group algebras, the structure of associative algebras,  the role of color-exchange symmetry underlying the space of tensor model observables suggests that a complete description of holography for the space of tensor models will be a  fascinating challenge. 

\subsection{Computational complexity of central correlators in matrix versus tensor models } 

One of the interesting results is that the one-point function
in the representation basis,   for the  $ d=3$ complex tensor model,  is equal to the Kronecker coefficient (\ref{O1sss}), a number of 
fundamental importance in Computational Complexity Theory. 

Compare this with the extremal 3-point correlator for the half-BPS sector
which is directly proportional to the Littlewood-Richardson  (LR) coefficient $g ( R , S , T ) $ \cite{cjr}
\bea\label{corrgRST}  
\langle \chi^R ( Z ) \chi^S ( Z ) \chi^{ T } ( Z^{ \dagger} ) \rangle 
= g ( R , S , T ) { n! \Dim_N( T) \over d(T) } 
\eea
This correlator has been interpreted in terms of topology change \cite{BDRT06}. 
The half-BPS sector and its connections to topology change has also been investigated recently in \cite{BM1702,HZ1705}. In the AdS-CFT correspondence, Young diagrams can be 
used to parametrize space-times of different topologies \cite{cjr,Ber04,LLM}. 

The LR coefficient has been of interest in the context of Computational Complexity Theory.  
It has been shown that the determination of the vanishing or otherwise of 
the LR coefficient can be done in polynonial time \cite{MS05}. The actual evaluation of 
the LR coefficient for general Young diagrams is \# P-hard \cite{Narayanan06} (\# P is the analog of NP when we go from decision problems to counting problems).  
Recently it was found that deciding the 
vanishing of Kronecker coefficients is NP-hard \cite{IMW1507}. This is an interesting contrast between central correlators in the 1-matrix problem (\ref{corrgRST}) and the one-point function of central observables 
in the tensor model (\ref{O1sss}).  Characterizing
 the complexity of determining the vanishing of 
extremal correlators of central observables in the 2-matrix case studied in \cite{PCA1601}
would be a useful problem to solve in getting a more complete picture of 
the relative complexities in matrix and tensor models. 
 
It would be interesting to explore the implications of the above results 
in the context of physical applications of matrix/tensor models,  e.g. in  
black hole physics or early cosmology. Different  physical roles for computational complexity  in these contexts have been proposed \cite{Sussk1403}\cite{DDGZ1706}. 
Finite $N$ effects, of interest in the physics of the stringy exclusion priinciple and giant gravitons, turn out to have drastic effects on the complexity questions \cite{CDW12}, tending to allow polynomial  time algorithms.  

\vskip2cm 
\begin{center} 
{ \bf Acknowledgements} 
\end{center} 
SR's research is supported by the STFC consolidated grant ST/L000415/1 “String Theory, Gauge Theory \& Duality” at Queen Mary University of London and  a Visiting Professorship at the 
Mandelstam Institute for Theoretical Physics,  University of the Witwatersrand, funded by a Simons Foundation  grant. JBG thanks the Centre for Research in String Theory at QMUL for hospitality, while this work was in progress.

\newpage 

\section*{ Appendix}

\appendix

\renewcommand{\theequation}{\Alph{section}.\arabic{equation}}
\setcounter{equation}{0}

\section{Symmetric group, representation
theory and group algebra}
\label{app:symrep}

\subsection{Symmetric group and representation }
\label{sapp:syrep}

We collect some  basic facts in the representation theory of symmetric groups. 
A useful textbook discussion is in \cite{Hammermesh}.

Irreducible representations of symmetric group $S_n$ are labelled by Young diagrams or partitions $R$ of $n$, that we denote $ R \vdash n$. 
In the following, we interchangeably use and assimilate an irrep with $R$.  
As a consequence of the Schur-Weyl duality, one associates also a Young diagram $R$ with an irreducible representation of the unitary group $U(N)$, 
when 
the length $l(R)$ of the first column of $R$ is bounded by $N$, i.e $l(R)\leq N$.  

At fixed $n$,  denote $d(R)$ the dimension of the representation of $S_n$ and 
$\Dim_N(R)$ the dimension of representation of $U(N)$, we write 
\be\label{dims}
d(R)   =  n! / h(R)  \,,   \qquad  
\Dim_N(R) = f_N(R)/ h(R) 
\ee 
where $h(R)$ is the product of the so-called hook lengths, i.e. 
$h(R) = \prod_{i,j}(c_j-j +r_i-i+1)$ and $f_N(R)$ is the products of  box weights
\bea\label{fN} 
f_N(R)= \prod_{i,j}(N-i+j)
\eea
where the pairs $(i,j)$ label the boxes of the Young diagram: $ i$ is the row label and $j$ is the column label. $r_i$ is the row length of the $i$'th row. $c_j$ is the column length of the $ j$'th column.

The matrices $D^R_{ij}(\s)$ of the representation $R$ of a permutation $\s\in S_n$  are $d(R)\times d(R)$ 
and  satisfy the following basic properties 
\be
\sum_{i} D^R_{ a i} ( \s) 
 D^R_{ ib  } ( \s' ) = D^R_{  ab  } ( \s \s') \,, 
 \qquad 
D^R_{ab}(e) = \delta_{ab}  
\label{ddinv}
\ee
and are also orthogonal 
\be\label{ortho}
\sum_{\s \in S_n} D^R_{ij}(\s)  D^S_{kl}(\s)
 = \frac{n!}{d(R)} \,\delta_{RS}\,\delta_{ik}\delta_{jl}
\ee
This follows from Schur's lemma. Note that 
we choose to work with orthogonal (and so real) 
matrices obeying  
\be
D^{R}_{ij}(\s^{-1}) = D^{R}_{ji}(\s)\,
\ee
such that \eqref{ortho} is again $\sum_{\s} D^{R}_{ij}(\s) D^{S} _{lk}(\s^{-1})$. 

Another important object in representation theory is of course the character of a given 
representation. The character of the irrep $R$ is simply the trace of $D^R(\s)$, $\chi^R(\s)=\Tr (D^R(\s))= \sum_iD^R_{ii}(\s)$. 
It is immediate that 
\be
\chi^R(\s) = \chi^R(\s^{-1}) 
\ee
The Kronecker delta of the symmetric group (defined to be equal to 1 when the argument is the
identity and 0 otherwise) decomposes as
\be
\delta(\s) = \sum_{R \vdash n} \frac{d(R)}{n!} \, \chi^R(\s) 
\ee
The summation $ R \vdash n $ is  a sum over partitions of $R$ of $n$, equivalently 
over Young diagrams with $n$ boxes.  
We have also 
\bea\label{delgsg0}
&& 
\sum_{\g \in S_n} \delta( \g \s \g^{-1} \tau^{-1}) = 
\sum_{\g \in S_n} \sum_{R \vdash n} \frac{d(R)}{n!} \, \sum_{i,a,b,c} D^R_{ia}(\g) D^R_{ab}(\s) D^R_{cb}(\g) D^R_{ic}(\tau) 
 \crcr
&&
\overset{ \eqref{ortho} }{=} \sum_{R \vdash n} \, \sum_{i,a,b,c}    D^R_{ab}(\s) D^R_{ic}(\tau)   \,\delta_{ic}\delta_{ab}
  = 
\sum_{R \vdash n} \chi^R(\s)\chi^R(\tau) 
\eea
If $B$ is a central element, then  
\bea
&&
\sum_{\g\in S_n} \chi^R(A\g B\g^{-1} ) = n! \, \chi^R(AB)
 = \sum_{a,b,c,d} \sum_{\g\in S_n} D^R_{ab}(A)D^R_{bc}(\g) D^R_{cd}(B)
D^R_{da}(\g^{-1})  \cr\cr
&&
\overset{ \eqref{ortho} }{=} \frac{n!}{d(R)} \sum_{a,b,c,d}  D^R_{ab}(A)D^R_{cd}(B) \delta_{ba}\delta_{cd}  
= \frac{n!}{d(R)} \chi^{R}(A)\chi^{R}(B) 
\eea
Hence 
 \be \label{chiAB0}
 \chi^R ( A B ) = { 1 \over d(R) } \chi^R ( A ) \chi^R ( B )  
 \ee
 Also useful to know that 
 \be
{ 1 \over n! } \sum_{ \alpha }   \chi^R ( \alpha ) N^{ \cy( \alpha ) } = \Dim_N (R) 
 \ee
 where $ \Dim_N( R )$ is the  dimension of the $U(N)$ representation $R$ \eqref{dims}.

The same type of the above calculation which involves \eqref{ortho} leads to other formulae  given by
\bea
\sum_{ \s \in S_n}  \chi^R(\s \tau_1) \chi^S(\s \tau_2) = \frac{n!}{d(R)} \, \delta_{RS}\, \chi^R(\tau_1 \tau_2^{-1})  
\quad
\overset{\tau_{1}=\tau_2=\textrm{id}}{\Rightarrow} \quad 
\sum_{ \s \in S_n}  \chi^R(\s) \chi^S(\s) = n! \, \delta_{RS} 
\eea
Finally, concerning identities involving the dimension of irrep $R$ in $U(N)$, one has
\be
\sum_{\s \in S_n} D^R_{ij}(\s) N^{\cy(\s)} = \delta_{ij} f_N(R) 
\quad
\overset{\eqref{ortho}}{\Rightarrow} \quad 
\sum_{\s \in S_n}  \chi^R(\s) N^{\cy(\s)} = d(R)f(R) =  n!\, \Dim_N(R) 
\ee
where $\cy(\s)$ is the number of cycles of $\s$.

The following table lists the above formulas:
\bea
&&
D^{R}_{ij}(\s^{-1}) = D^{R}_{ji}(\s)\\
&&
\chi^R(\s) = \chi^R(\s^{-1})  = \chi^{R}(\g \s \g^{-1})\,, \qquad \forall \g \in S_n
\\
&& 
\delta(\s) = \sum_{R \vdash n} \frac{d(R)}{n!} \, \chi^R(\s) \label{deltasym}\\
&&
\sum_{\s \in S_n} D^R_{ij}(\s)  D^S_{kl}(\s)
 = \frac{n!}{d(R)} \,\delta_{RS}\,\delta_{ik}\delta_{jl}
\label{ortho2}\\
&&
\sum_{ \s \in S_n}  \chi^R(\s \tau_1) \chi^S(\s \tau_2) = \frac{n!}{d(R)} \, \delta_{RS}\, \chi(\tau_1 \tau_2^{-1})  
\label{cscs}\\
&&
\sum_{ \s \in S_n}  \chi^R(\s) \chi^S(\s) = n! \, \delta_{RS}  
\label{cscs2}\\
&&
\forall B \in \cZ(S_n)\,,\quad \chi^R ( A B ) = { 1 \over {d(R)} } \chi^R ( A ) \chi^R ( B ) 
 \label{chiAB} \\
&&
\label{delgsg}
\sum_{\g \in S_n} \delta( \g \s \g^{-1} \tau^{-1}) = 
\sum_{R \vdash n} \chi^R(\s)\chi^R(\tau) \\
&&
\sum_{\s \in S_n} D^R_{ij}(\s) N^{\cy(\s)} = \delta_{ij} f_N(R) 
\\
&&
\sum_{\s \in S_n}  \chi^R(\s) N^{\cy(\s)} = d(R)f(R) =  n!\, \Dim_N(R) 
\label{chiN}
\eea
Defining the central element $ \Omega \in \mC ( S_n)$,
\be
\Omega = \sum_{ \sigma \in S_n } N^{n - \cy( \sigma ) } \sigma 
\ee
equation \eqref{chiN}, can be also written as 
\be
{ N^n\over n! }  \chi^R ( \Omega ) = \Dim_N ( R ) 
\label{chiN2}
\ee

\subsection{Clebsch-Gordan coefficients }
\label{sapp:cgc}

Consider two irreps $V_{R_1} , V_{R_2}  $ of $S_n$ corresponding to Young diagrams $R_1 , R_2 $. 
We assume  that we have picked an orthogonal basis of states for the irreps e.g. 
 $ | R , i \ra $ obeying 
 \bea 
 \la R , j | R , i \ra = \delta_{ij} 
 \eea
A representation $\varrho_{R}: S_n \to {\rm End}(V_{R})$
is given by a matrix $D^{R}$ with entries determined by
$\varrho_{S}(\s) |R, i \rangle  =\sum_{l=1}^{d(R)} D^{R}_{li}(\s)|R, l \rangle$ with $\s\in S_n$. We write in short $\varrho_S (\s) = \s$
and then $\langle R,j |\s|R, i \rangle = D^{R}_{ji}(\s)$. 
The tensor product 
representation  $ V_{R_1}  \otimes V_{R_2}  $ can be decomposed into a direct sum of 
irreps $V_{R_3} $ with multiplicities
\bea 
V_{ R_1 } \otimes V_{R_2}  = \bigoplus_{ R_3 \vdash n } V_{R_3} \otimes V_{R_3}^{ \rm m } 
\eea
One set of basis vectors  in the tensor product space  is $| R_1, i_1\ra\otimes |R_2,i_2\ra = :|R_1,i_1;R_2,i_2 \ra  $ while the r.h.s
corresponds to a basis set  $ | R_3 , i_3 , \tau_{R_3} \ra $. The label $i_3$ runs over states in the irrep $R_3$, while $ \tau_{R_3}$  runs over an orthogonal basis in the multiplicity space $V_{R_3}^{\rm m}$. 
Clebsch-Gordan coefficients (CG's) are transition coefficients between the two types of bases
\be 
C^{R_1,R_2;\, R_3 ,\,\tau_{R_3} }_{\, i_1,i_2;\, i_3} :=    \la R_1,i_1; R_2,i_2 | R_3, \tau_{R_3}, i_3 \ra  = \la  R_3 , \tau_{R_3}, i_3 | R_1,i_1; R_2, i_2 \ra 
\ee
The last relation is obtained from the reality property of the CG's.  We will then use
$C^{R_1,R_2;\, R_3,\,\tau_{R_3} }_{\, i_1,i_2; \, i_3} = C^{\tau_{R_3}, {R_3}  ;\, R_1,R_2}_{\, i_3; \, i_1,i_2}$. 
A detailed discussion of the CG's for symmetric groups is 
in \cite{Hammermesh}.

Linear operators for $ \sigma \otimes \sigma $ in $ V_{ S } \otimes V_{R} $ have matrix elements 
\be
D^{R_1}_{i_1 j_1}(\s)D^{R_2}_{ i_2 j_2 }(\s) = \la R_1 ,i_1| \s | R_1,j_1\ra\, 
 \la R_2,i_2 | \s | R_2, j_2\ra  
  =: \la R_1,i_1; R_2,i_2 |\, \s \, | R_1,j_1;\, R_2, j_2 \ra 
\ee
Inserting a complete set of states
resolving the identity we get
\begin{align}\label{dd}
& D^{R_1}_{i_1j_1}(\s)D^{R_2}_{i_2 j_2}(\s)  =  
\sum_{ R_3 ,R_3', \tau_{ R_3} ,\tau_{ R_3}'}\sum_{i_3,j_3}
\la R_1,i_1; R_2,i_2 | R_3, \tau_{R_3}' , i_3\ra \, 
\la R_3, \tau_{R_3}, i_3 | \, \s \, |R_3', \tau'_{R_3'}, j_3 \ra \cr
&\qquad \qquad  \times 
\la R_3', \tau'_{R_3'}, j_3| R_1,j_1;\, R_2,j_2\ra \crcr
&= 
\sum_{R_3,R_3', \tau_{R_3} ,\tau'_{ R_3} }\sum_{i_3,j_3}
\la R_1,i_1; R_2, i_2| R_3, \tau_{R_3}, i_3 \ra \, 
\delta_{R_3R_3'}\delta_{\tau_{R_3}\tau'_{R_3'}} D^{R_3}_{i_3j_3}(\s)
\la R_3', \tau'_{R_3'}, j_3| R_1, i_2 ;\, R_2, j_2 \ra \crcr
&= 
\sum_{R_3,\tau_{ R_3} }\sum_{i_3 , j_3 }
\la R_1,i_1; R_2 , i_2 | R_3, \tau_{R_3} , i_3 \ra \, 
D^{R_3}_{ i_3 j_3 }(\s)
\la R_3 , \tau_{R_3} , j_3 | R_1, j_1;\, R_2,j_2 \ra 
\end{align}
Using the definition of the CG's \eqref{dd}, can be also written  as
\be\label{dd=cdc}
D^{R_1}_{i_1j_1}(\s)D^{R_2}_{i_2 j_2 }(\s)
 = \sum_{R_3,\tau }\sum_{i_3 , j_3 } C^{R_1 , R_2 ; R_3 ,\tau_{R_3} }_{i_1,i_2 ; i_3 }  D^{R_3}_{i_3 j_3 }(\s) C^{ R_1, R_2 ; R_3 ,\tau_{R_3} }_{j_1, j_2 ; j_3 }  
\ee 
Because there is no possible confusion, $\tau_{R_3} $ is sometimes  denoted $\tau$ in the text. 

The following identities hold
\bea
&&
\sum_{j_1,j_2}
D^{R_1}_{i_1 j_1}(\g)D^{R_2}_{i_2 j_2}(\g)C^{R_1,R_2; \, R_3,\,\tau}_{\, j_1,j_2; \, j_3}
 =\sum_{i_3 }  C^{R_1,R_2;\, R_3,\,\tau}_{\, i_1,i_2; \, i_3} \, D^{R}_{i_3 j_3}(\g)  
\label{ddc=cd} \\
&&
\sum_{i_1,i_2} 
C^{R_1,R_2; \, R_3 ,\,\tau}_{\, i_1,i_2; \,i_3} C^{R_1,R_2; \, R_3',\,\tau'}_{\, i_1,i_2; \,j_3}  
 = \delta_{R_3 R'_3 }\, \delta_{\tau\tau'} \, \delta_{i_3j_3} 
\label{cc=del} \\
&&
\sum_{R_3 ,i_3 , \tau } 
C^{R_1,R_2; \, R_3 ,\,\tau}_{\, i_1,i_2; \, i_3 } C^{R_1,R_2; \, R_3 ,\,\tau}_{\, j_1,j_2; \,i_3 }  
 = \delta_{i_1j_1}\, \delta_{i_2j_2} 
\label{cc=del2}  \\
&&
\sum_{R_3 ,\tau;\, i_3,j_3} 
C^{R_1,R_2; \, R_3,\,\tau}_{\, i_1,i_2; \,i_3} D^{R_3}_{i_3j_3}(\g) C^{R_1,R_2; \, R_3,\,\tau}_{\, j_1,j_2; \,j_3} 
=  D^{R_1}_{i_1j_1}(\g)D^{R_2}_{i_2j_2}(\g)
\label{dsd4}
\eea
Note that \eqref{dd=cdc} is \eqref{dsd4}. 

Furthermore, we have by applying twice \eqref{ddc=cd}: 
\bea
&& 
\sum_{j_1,j_2,j_3}
D^{R_1}_{i_1 j_1}(\g)D^{R_2}_{i_2 j_2}(\g) D^{R_3}_{i_3 j_3}(\g) C^{R_1,R_2; \, R_3,\,\tau}_{\, j_1,j_2; \,j_3}
 =\sum_{j_3} \sum_{l}  C^{R_1,R_2;\, R_3,\,\tau}_{\, i_1,i_2; \, l} \, D^{R_3}_{lj_3}(\g) 
D^{R_3}_{i_3 j_3}(\g)  
 \cr\cr
&& 
 =\sum_{l}  C^{R_1,R_2;\, R_3,\,\tau}_{\, i_1,i_2; \, l} \,
\sum_{j_3}  D^{R_3}_{lj_3}(\g) D^{R_3}_{ j_3 i_3}(\g^{-1}) 
 =   C^{R_1,R_2;\, R_3,\,\tau}_{\, i_1,i_2; i_3} 
\label{dddc=c}
\eea

These equations can be put in diagrammatrics  which 
 lighten the proofs. We now recall them. 
The diagrammatic notation for the CG coefficient will be a three valent black node:  
\begin{equation}
\label{eq:CG_diag}
C^{R_2 , R_2 ;R_3 , \tau}_{i_1,i_2; i_3 }
\quad = \quad
\mytikz{	
	\node (t) at (0,0) [circle,fill,inner sep=0.5mm,label=below:$\tau$] {};	
	\node (i1) at (-1.5,0.7) {$i_1$};
	\node (i2) at (-1.5,-0.8) {$i_2$};
	\node (m) at (1.5,0) {$ i_3 $};
	\draw [postaction={decorate}] (t) to node[above]{$R_3$} (m);		
	\draw [postaction={decorate}] (i1) to node[above]{$R_1$} (t);
	\draw [postaction={decorate}] (i2) to node[below]{$R_2$} (t);
}
\end{equation}
A representation matrix $D^R_{ij}(\s)$ is drawn like 
 $\mytikz{
		\node (g) at (0,0) [rectangle,draw] {$\s$};
		\draw (-0.5,0) -- (-0.26,0) ; 
		\draw (0.26,0) -- (0.5,0) ;  
	}$, 
the rest of the indices will be explicit when the matrix will 
be composed with others coefficients. 
Then the above identities can be translated as 
\begin{align}
\eqref{ddc=cd} \qquad  \qquad &	\mytikz{
		\node (m) at (0,0) [circle,fill,inner sep=0.5mm,label=below:$\tau$] {};	
		\node (i1) at (-2.3,0.7) {$i_1$};
		\node (i2) at (-2.3,-0.8) {$i_2$};
		\node (k) at (1.2,0) {$i_3$};
		\node (g1) at (-1,0.7) [rectangle,draw] {$\gamma$};
		\node (g2) at (-1,-0.7) [rectangle,draw] {$\gamma$};
		\draw [postaction={decorate}] (m) to node[above]{$R_3$} +(1,0);	
		\draw [postaction={decorate}] (g1) to node[above]{$R_1$} (m);
		\draw [postaction={decorate}] (g2) to node[below]{$R_2$} (m);		
		\draw [postaction={decorate}] ($(g1)+(-1.0,0)$) to (g1);
		\draw [postaction={decorate}] ($(g2)+(-1.0,0)$) to (g2);
	} 
\;= \;
	\mytikz{
            \node (i1) at (-2.8,0.7) {$i_1$};
		\node (i2) at (-2.8,-0.8) {$i_2$};
		\node (k) at (0.7,0) {$i_3$};
		\node (s) at (-0.2,0) [rectangle,draw] {$\gamma$};		
		\node (m) at (-1.5,0) [circle,fill,inner sep=0.5mm,label=below:$\tau$] {};	
		\draw [postaction={decorate}] (s) to +(0.7,0);	
		\draw [postaction={decorate}] (m) to node[above]{$R_3$} (s);		
		\draw [postaction={decorate}] ($(m)+(-1,0.7)$) to node[above]{$R_1$} (m);
		\draw [postaction={decorate}] ($(m)+(-1,-0.7)$) to node[below]{$R_2$} (m);
	}
\\		
\eqref{cc=del}  \qquad  \qquad	&\mytikz{				
		\node (n) at (-0.7,0) [circle,fill,inner sep=0.5mm,label=below:$\tau$] {};	
		\node (tn) at (0.7,0) [circle,fill,inner sep=0.5mm,label=below:$ \tau'$] {};	
		\node (i) at (-2,0) {$i_3$};				
		\node (j) at (2,0) {$j_3$};							
		\draw [postaction={decorate}] (n) to [bend left=45] node[above]{$R_1$} (tn);
		\draw [postaction={decorate}] (n) to [bend right=45] node[below]{$R_2$} (tn);		
		\draw [postaction={decorate}] (i) to node[above]{$R_3$} (n);
		\draw [postaction={decorate}] (tn) to node[above]{$R_3'$} (j);		
	}
\;  = \;	
	\mytikz{
		\node (i) at (-1,0) {$i_3$};				
		\node (j) at (1,0) {$j_3$};	
		\draw [postaction={decorate}] (i) to node[above]{$R_3$} (j);
	}
	\times 
	\delta_{R_3 R_3'} \delta_{\tau  \tau'}\delta_{i_3 j_3}
\\	
\eqref{cc=del2}  \qquad  \qquad &	\sum_{R_3 ,\tau} 
	\mytikz{				
		\node (n) at (-0.7,0) [circle,fill,inner sep=0.5mm,label=below:$\tau$] {};	
		\node (tn) at (0.7,0) [circle,fill,inner sep=0.5mm,label=below:$\tau$] {};	
		\node (i1) at (-2,0.7) {$i_1$};		
		\node (i2) at (-2,-0.7) {$i_2$};
		\node (j1) at (2,0.7) {$j_1$};		
		\node (j2) at (2,-0.7) {$j_2$};				
		\draw [postaction={decorate}] (i1) to node[above]{$R_1$} (n);
		\draw [postaction={decorate}] (i2) to node[below]{$R_2$} (n);
		\draw [postaction={decorate}] (n) to node[above]{$R_3$} (tn);		
		\draw [postaction={decorate}] (tn) to node[above]{$R_1$} (j1);
		\draw [postaction={decorate}] (tn) to node[below]{$R_2$} (j2);
	}
\; = \;	
	\mytikz{
		\node (i1) at (-1,0.5) {$i_1$};		
		\node (i2) at (-1,-0.5) {$i_2$};
		\node (j1) at (1,0.5) {$j_1$};		
		\node (j2) at (1,-0.5) {$j_2$};
		\draw [postaction={decorate}] (i1) to node[above]{$R_1$} (j1);
		\draw [postaction={decorate}] (i2) to node[below]{$R_2$} (j2);
	}		
	\times 
	\delta_{i_1j_1} \delta_{i_2j_2}
\\	
\eqref{dsd4}	\qquad \qquad  &
	\sum_{R_3 ,\tau}
	\mytikz{				
		\node (n) at (-1,0) [circle,fill,inner sep=0.5mm,label=below:$\tau$] {};
		\node (s) at (0,0) [rectangle,draw] {$\gamma$};		
		\node (tn) at (1,0) [circle,fill,inner sep=0.5mm,label=below:$\tau$] {};	
		\node (i1) at (-2,0.7) {$i_1$};		
		\node (i2) at (-2,-0.7) {$i_2$};
		\node (j1) at (2,0.7) {$j_1$};		
		\node (j2) at (2,-0.7) {$j_2$};				
		\draw [postaction={decorate}] (i1) to node[above]{$R_1$} (n);
		\draw [postaction={decorate}] (i2) to node[below]{$R_2$} (n);
		\draw [postaction={decorate}] (n) to node[above]{$R_3$} (s);		
		\draw [postaction={decorate}] (s) to (tn);		
		\draw [postaction={decorate}] (tn) to node[above]{$R_1$} (j1);
		\draw [postaction={decorate}] (tn) to node[below]{$R_2$} (j2);
	}
\;  = \;	
	\mytikz{
		\node (i1) at (-1.2,0.5) {$i_1$};		
		\node (i2) at (-1.2,-0.5) {$i_2$};
		\node (j1) at (1.2,0.5) {$j_1$};		
		\node (j2) at (1.2,-0.5) {$j_2$};
		\node (s1) at (0,0.5) [rectangle,draw] {$\gamma$};
		\node (s2) at (0,-0.5) [rectangle,draw] {$\gamma$};
		\draw [postaction={decorate}] (i1) to node[above]{$R_1$} (s1);
		\draw [postaction={decorate}] (s1) to (j1);
		\draw [postaction={decorate}] (i2) to node[below]{$R_2$} (s2);
		\draw [postaction={decorate}] (s2) to (j2);
	}			
\end{align}

The following lemma is useful in the text. 
\begin{lemma}\label{lem:2C3D}
The following relation holds: 
\bea
&&
 \sum_{i_l,j_l}
C^{R_1,R_2;R_3,\tau_1}_{i_1,i_2;i_3}
C^{R_1,R_2;R_3,\tau_2}_{j_1,j_2;j_3}
D^{R_1}_{i_1j_1}(\g_1\s_1\g_2)D^{R_2}_{i_2j_2}(\g_1\s_{2}\g_2)D^{R_3}_{i_3j_3}(\g_1\s_{3}\g_2)
 = \cr\cr
&&
\sum_{i_l,j_l}
C^{R_1,R_2;R_3,\tau_1}_{i_1,i_2;i_3}
C^{R_1,R_2;R_3,\tau_2}_{j_1,j_2;j_3}
 D^{R_1}_{i_1j_1}(\s_{1}) 
 D^{R_2}_{i_2j_2}(\s_{2})
 D^{R_3}_{i_3j_3}(\s_{3}) 
\eea
\end{lemma}
\proof  We have 
\bea
&&
 \sum_{i_l,j_l}
C^{R_1,R_2;R_3,\tau_1}_{i_1,i_2;i_3}
C^{R_1,R_2;R_3,\tau_2}_{j_1,j_2;j_3}
D^{R_1}_{i_1j_1}(\g_1\s_1\g_2)D^{R_2}_{i_2j_2}(\g_1\s_{2}\g_2)D^{R_3}_{i_3j_3}(\g_1\s_{3}\g_2) \cr\cr
&&
 = 
\sum_{a_l,b_l}
 D^{R_1}_{a_1b_1}(\s_{1}) 
 D^{R_2}_{a_2b_2}(\s_{2})
 D^{R_3}_{a_3b_3}(\s_{3}) 
\sum_{i_l}
C^{R_1,R_2;R_3,\tau_1}_{i_1,i_2;i_3} D^{R_1}_{i_1a_1}(\g_1)D^{R_2}_{i_2a_2}(\g_1)D^{R_3}_{i_3a_3}(\g_1)\cr\cr
&& \times
\sum_{j_l}
C^{R_1,R_2;R_3,\tau_2}_{j_1,j_2;j_3}
D^{R_1}_{b_1j_1}(\g_2)
 D^{R_2}_{b_2j_2}(\g_2)
 D^{R_3}_{b_3j_3}(\g_2) \crcr
&&
 = 
\sum_{a_l,b_l}
 D^{R_1}_{a_1b_1}(\s_{1}) 
 D^{R_2}_{a_2b_2}(\s_{2})
 D^{R_3}_{a_3b_3}(\s_{3}) 
C^{R_1,R_2;R_3,\tau_1}_{a_1,a_2;a_3}
C^{R_1,R_2;R_3,\tau_2}_{b_1,b_2;b_3} 
\eea
where we used \eqref{dddc=c}.  
 
\qed 

Integrating three representation matrices, the following relation is useful:  
\begin{lemma}\label{lem2:DDD=CC}
We have
\bea
\sum_{\s\in S_n} D^{R_1}_{i_1j_1}(\s)D^{R_2}_{i_2j_2}(\s)D^{R_3}_{i_3j_3}(\s)
 = \frac{n!}{d(R_3)}\sum_{\tau}
C^{R_1,R_2;R_3,\tau}_{i_1,i_2;i_3}
C^{R_1,R_2;R_3,\tau}_{j_1,j_2;j_3} 
\eea
\end{lemma}
\proof We use, successively,  \eqref{dd=cdc} and \eqref{ortho2} to get: 
\bea
&&
\sum_{\s\in S_n} D^{R_1}_{i_1j_1}(\s)D^{R_2}_{i_2j_2}(\s)D^{R_3}_{i_3j_3}(\s)
 = \sum_{\s} 
\Big[\sum_{R,\tau}\sum_{a,b} 
C^{R_1,R_2;R,\tau}_{i_1,i_2;a}
C^{R_1,R_2;R,\tau}_{j_1,j_2;b} D^{R}_{ab}(\s)\Big] 
D^{R_3}_{i_3j_3}(\s) \cr\cr
&&
 =\frac{n!}{d(R_3)}\sum_{R,\tau}\sum_{a,b} 
C^{R_1,R_2;R,\tau}_{i_1,i_2;a}
C^{R_1,R_2;R,\tau}_{j_1,j_2;b}
\delta_{RR_3} \delta_{ai_3}\delta_{bj_3} 
\eea
summing over $R,a$ and $b$ achieves the result.  
\qed

We can illustrate Lemma \ref{lem2:DDD=CC} in 
the following  way: 
\bea
&&
\mytikz{
		\node (g) at (0,0) [rectangle,draw] {$\s$};
		\node (g2) at (-1,0.2) []{$i_1$}; 
		\node (g3) at (1,0.2) []{$j_1$} ;
		\draw (-0.9,0) -- (-0.26,0) ; 
 		\draw (0.26,0) -- (0.9,0) ;  
		\node (g1) at (0,1) [rectangle,draw] {$\s$};
		\node (g21) at (-1,1.2) []{$i_2$}; 
		\node (g31) at (1,1.2) []{$j_2$} ;
		\draw (-0.9,1) -- (-0.26,1) ; 
		\draw (0.26,1) -- (0.9,1) ;  
		\node (g22) at (0,2) [rectangle,draw] {$\s$};
		\node (g221) at (-1,2.2) []{$i_3$}; 
		\node (g321) at (1,2.2) []{$j_3$} ;
		\draw (-0.9,2) -- (-0.26,2) ; 
 		\draw (0.26,2) -- (0.9,2) ;  
\node (r0) at (1.8,1) {$=$};
\node (r1) at (4,1.1)  {$\tau$}; 
\node (r2) at (4.5,1.1)  {$\tau$}; 
\fill[fill=black] (3.6,1.05) circle (0.07);
\fill[fill=black] (4.9,1.05) circle (0.07);
\draw  (2.5,1.05) -- (3.6,1.05) ; 
\draw (6,1.05) -- (4.9,1.05);
\draw (2.5,2) -- (3,2);  
\draw (2.5, 0.05) -- (3, 0.05); 
\draw  (5.5,2) -- (6,2) ; 
\draw (5.5,0.05) -- (6,0.05); 
 \draw[shorten <= 0.1cm, shorten >= 0.1cm] (2.9, 2) to[out=0, in=90]  (3.6,1);
\draw[shorten <= 0.1cm, shorten >= 0.1cm] (2.9, 0.05)  to[out=0, in=-90] (3.6,1.1);
 \draw[shorten <= 0.1cm, shorten >= 0.1cm] (5.6, 2) to[out=-180, in=90]  (4.9,1);
\draw[shorten <= 0.1cm, shorten >= 0.1cm] (5.6, 0.05)  to[out=-180, in=-90] (4.9,1.1);
}
\eea

Note that we have defined the Clebsch-Gordan coefficients in terms of inner products 
between states in $ V_{ R_1 } \otimes V_{ R_2} $ transforming as $V_{R_3} $ under the diagonal action of $S_n$. We could also have defined them in terms of the states 
in $ V_{ R_1 } \otimes V_{ R_2} \otimes V_{ R_3} $ which transform as the trivial representation  $ [n] $ of the diagonal $S_3$. If we use the latter formulation the Clebsch's 
\bea 
{ \tilde C }^{ R_1 , R_2 , R_3 ; \tau }_{ i_1 , i_2 , i_3 } = \langle R_1 , i_1 , R_2 , i_2 , R_3 , i_3 | [n] , \tau \rangle 
\eea
will appear on the r.h.s of Lemma \ref{lem2:DDD=CC} without the $ 1/ d (R_3) $ factor, so that 
\bea 
{ \tilde C }^{ R_1 , R_2 , R_3 ; \tau }_{ i_1 , i_2 , i_3 } 
= { 1 \over \sqrt { d (R_3) } } C^{R_1,R_2;R_3,\tau}_{i_1,i_2;i_3}
\eea

\subsection{Projectors for irreducible representations }
\label{app:proj}

We  discuss some properties of projectors in the group algebra of $ \mC ( S_n)$, which we use in the text. For every irreducible representation $R$ of $\mC (  S_n) $ we have 
the characters $ \chi^R ( \sigma) $  which are used to define projectors $P_R$
\bea 
P_{ R } = { d ( R)  \over n! } \sum_{ \sigma \in S_n } \chi^R ( \sigma ) \sigma 
\eea
Using character orthogonality we verify that 
\bea 
P_R P_S  = \delta_{ R , S  } P_R 
\eea
Consider a general representation $ W$ of $ S_n $ which has a decomposition into irreducible 
representations 
\bea 
W =  \bigoplus_{R \vdash n} V_R \otimes V_{R}^{\rm m}
\eea 
$V_R$ are irreducible representation spaces and $V_{R}^{\rm m}$ are multiplicity spaces of dimension $m_{W}^R$. 
Taking the trace 
\bea 
\tr_{ W } ( P_S ) && = \sum_{ R \vdash n } \tr_{ V_{ R } \otimes V^{ \rm m }_{ R } } ( P_S \otimes 1 ) \cr 
&& = \sum_{ R } m_{W}^R  \tr_{ V_R } ( P_S ) = \sum_{ R } d (R)  m_{W}^R \delta_{RS} \cr 
&& = m_W^S d(S) 
\eea
We used  
\bea 
\tr_{ V_R} ( P_S ) = \sum_{ \sigma \in S_n  } { d (S)  \over n! } \chi^S ( \sigma ) \chi^R ( \sigma )   = d(S)  \delta_{ R , S } 
\eea
which is an application of the orthogonality of characters. 
For the case where $R$ is the trivial representation $R_0 = [n] $, $ \chi^{ R_0} ( \sigma ) = 1 $  and  
\bea 
P_{ R_0} = { 1 \over n! } \sum_{ \sigma \in S_n } \sigma 
\eea
is the projection on the invariant space of $W$.

\section{ The permutation centralizer algebra $ \cK ( n )$ }
\label{app:algebra}

\subsection{Basis of $\cK(n)$}
\label{app:baskn}

The semi-simple algebra $ \cK ( n ) $ is defined in terms of 
permutation equivalences in $ \mC ( S_n ) \otimes \mC ( S_n )  \otimes \mC ( S_n )  $ (double coset description) 
or in  $ \mC ( S_n ) \otimes \mC ( S_n )$ (centralizer description). 
By Fourier transforming from the permutation basis of  $\mC ( S_n )$ 
to the representation basis $Q^{R}_{ij}$, we have Fourier bases for $\cK ( n ) $ 
in either formulation.  In this appendix, we prove  that the $Q$-basis
elements in each formalism are indeed invariant under the appropriate equivalence relations, that they multiply like matrices (thus giving the WA decomposition) and that they are orthogonal with respect to the non-degenerate bilinear pairing.  

\

\noindent{\bf $Q$-Basis of $\cK(n)$ -}
We start by checking the invariance of the basis $Q^{R,S,T}_{\tau_1,\tau_2}$ \eqref{qbasis}  under the diagonal action:  
\bea
&&
(\gamma\otimes \gamma)\cdot Q^{R,S,T}_{\tau_1,\tau_2} \cdot
(\gamma^{-1}\otimes \gamma^{-1}) = \crcr
&&\kappa_{R,S}
\sum_{\s_1, \s_2 \in S_n}
\sum_{i_1,i_2,i_3, j_1,j_2}
C^{R , S ; T , \tau_1  }_{ i_1 , i_2 ; i_3 } C^{R , S ; T , \tau_2  }_{ j_1 , j_2 ; i_3 } 
 D^{ R }_{ i_1 j_1} ( \sigma_1  ) D^S_{ i_2 j_2 } ( \sigma_2 ) \,  \gamma \sigma_1
\gamma^{-1} \otimes \gamma\sigma_2 \gamma^{-1} \crcr
&&
=  \kappa_{R,S}
\sum_{\s_1, \s_2 \in S_n}
\sum_{i_1,i_2,i_3, j_1,j_2}
C^{R , S ; T , \tau_1  }_{ i_1 , i_2 ; i_3 } C^{R , S ; T , \tau_2  }_{ j_1 , j_2 ; i_3 } 
 D^{ R }_{ i_1 j_1} (\gamma^{-1}  \sigma_1 \gamma ) D^S_{ i_2 j_2 } ( \gamma^{-1}  \sigma_2\gamma ) \,   \sigma_1
\otimes \sigma_2  \cr\cr
&&
 =  \kappa_{R,S}
\sum_{\s_1, \s_2 \in S_n}
\sum_{i_1,i_2,i_3, j_1,j_2}
\sum_{a_l,b_l}
C^{R , S ; T , \tau_1  }_{ i_1 , i_2 ; i_3 } C^{R , S ; T , \tau_2  }_{ j_1 , j_2 ; i_3 } 
 D^{ R }_{ i_1 a_1} (\gamma^{-1})  
D^{ R }_{ a_1 b_1} ( \sigma_1  )
D^{ R }_{ b_1 j_1} (\gamma)
\cr\cr
&& \times 
 D^S_{ i_2 a_2 } ( \gamma^{-1} ) 
 D^S_{ a_2 b_2 } ( \sigma_2 ) 
 D^S_{ b_2 j_2 } ( \gamma ) \,
   \sigma_1
\otimes \sigma_2   
\label{inq1}
\eea
Using \eqref{ddc=cd} of appendix \ref{sapp:cgc}, we  get
\bea
&&
(\gamma\otimes \gamma)\cdot Q^{R,S,T}_{\tau_1,\tau_2} \cdot
(\gamma^{-1}\otimes \gamma^{-1}) = \kappa_{R,S} \times \crcr
&&
\sum_{\s_1, \s_2 \in S_n}
\sum_{a_l,b_l,i_3}
\Big[
\sum_{i_1,i_2 }
D^{ R }_{ a_1i_1 } (\gamma)  
D^S_{ a_2 i_2 } ( \gamma) 
C^{R , S ; T , \tau_1  }_{ i_1 , i_2 ; i_3 }  
\Big] 
\Big[\sum_{j_1,j_2}
D^{ R }_{ b_1 j_1} (\gamma)
 D^S_{ b_2 j_2 } ( \gamma )
 C^{R , S ; T , \tau_2  }_{ j_1 , j_2 ; i_3 } 
 \Big] 
\crcr
&& \times 
 D^{ R }_{ a_1 b_1} ( \sigma_1  )
 D^S_{ a_2 b_2 } ( \sigma_2 ) 
 \,  \sigma_1
\otimes \sigma_2  \cr\cr
&&
 = \kappa_{R,S}
\sum_{\s_1, \s_2 \in S_n}
\sum_{a_l,b_l,i_3}
\Big[
\sum_{l_1}
C^{R , S ; T , \tau_1  }_{ a_1 , a_2 ; l_1 }  
D^T_{ l_1 i_3 } ( \gamma) 
\Big] 
\Big[\sum_{l_2}
 C^{R , S ; T , \tau_2  }_{ b_1 , b_2 ;l_2 } 
 D^T_{ l_2 i_3 } ( \gamma )
 \Big] 
\crcr
&& \times 
 D^{ R }_{ a_1 b_1} ( \sigma_1  )
 D^S_{ a_2 b_2 } ( \sigma_2 ) 
 \,  \sigma_1
\otimes \sigma_2  \cr\cr
&&
 = \kappa_{R,S}
\sum_{\s_1, \s_2 \in S_n}
\sum_{a_1,a_2,b_1,b_2,l}
C^{R , S ; T , \tau_1  }_{ a_1 , a_2 ; l }  
 C^{R , S ; T , \tau_2  }_{ b_1 , b_2 ;l } 
 D^{ R }_{ a_1 b_1} ( \sigma_1  )
 D^S_{ a_2 b_2 } ( \sigma_2 ) 
 \,  \sigma_1
\otimes \sigma_2  \cr\cr
&& = Q^{R,S,T}_{\tau_1,\tau_2} 
\label{inq2}
\eea
where the factor $\sum_{i_3}D^T_{ l_1 i_3 } ( \gamma) D^T_{ l_2 i_3 } ( \gamma ) $ evaluates using \eqref{ddinv}. 

We prove now that $Q^{R,S,T}_{\tau_1,\tau_2}$'s multiply like matrices:  
\bea
&&
Q^{R,S,T}_{\tau_1,\tau_2}Q^{R',S',T'}_{\tau_2',\tau_3} 
= \kappa_{R,S}\kappa_{R',S'}
\sum_{\s_i, \s_i' \in S_n}
\sum_{i_l, j_m, i'_l, j'_m}
C^{R , S ; T , \tau_1  }_{ i_1 , i_2 ; i_3 } C^{R , S ; T , \tau_2  }_{ j_1 , j_2 ; i_3 } 
 D^{ R }_{ i_1 j_1} ( \sigma_1  ) D^S_{ i_2 j_2 } ( \sigma_2 )  \cr\cr
&&\times
C^{R' , S' ; T' , \tau_2'  }_{ i'_1 , i'_2 ; i'_3 } C^{R' , S' ; T' , \tau_3  }_{ j'_1 , j'_2 ; i'_3 } 
 D^{ R' }_{ i'_1 j'_1} ( \sigma_1'  ) D^{S'}_{ i'_2 j'_2 } ( \sigma_2' )
\;     \sigma_1  \sigma_1' \otimes \sigma_2   \sigma_2'  \cr\cr
&& 
 =\kappa_{R,S}\kappa_{R',S'}
\sum_{\s_i, \s_i' \in S_n}
\sum_{i_l, j_m, i'_l, j'_m}
C^{R , S ; T , \tau_1  }_{ i_1 , i_2 ; i_3 } C^{R , S ; T , \tau_2  }_{ j_1 , j_2 ; i_3 } 
 D^{ R }_{ i_1 j_1} ( \sigma_1 \s_1'^{-1} ) D^S_{ i_2 j_2 } ( \sigma_2 \s_2'^{-1})  \cr\cr
&& \times
C^{R' , S' ; T' , \tau_2'  }_{ i'_1 , i'_2 ; i'_3 } C^{R' , S' ; T' , \tau_3  }_{ j'_1 , j'_2 ; i'_3 } 
 D^{ R' }_{ i'_1 j'_1} ( \sigma_1'  ) D^{S'}_{ i'_2 j'_2 } ( \sigma_2' )
\;     \sigma_1  \otimes \sigma_2    \cr\cr
&& 
=\kappa_{R,S}^2\frac{(n!)^2}{d(R)d(S)} \delta_{RR'} \delta_{SS'}   \sum_{\s_i \in S_n}
\sum_{i_1,i_2,i_3,i_3',a,b }
C^{R , S ; T , \tau_1  }_{ i_1 , i_2 ; i_3 }
C^{R , S ; T' , \tau_3  }_{a ,b ; i'_3 } 
 D^{ R }_{ i_1 a} ( \sigma_1) 
D^S_{ i_2b} ( \sigma_2)   \sigma_1  \otimes \sigma_2 \cr\cr
&& \times
\Big[ \sum_{j_1,j_2} 
C^{R , S ; T' , \tau_2'  }_{ j_1 , j_2 ; i'_3 } 
 C^{R , S ; T , \tau_2  }_{ j_1 , j_2 ; i_3 }  \Big]    \cr\cr
&& 
=\kappa_{R,S} \delta_{RR'} \delta_{SS'}  \delta_{TT'} \delta_{\tau_2 \tau_2'} \crcr
&&
\times 
\sum_{\tau_2}\sum_{\s_i \in S_n}
\sum_{i_1,i_2,i_3,i_3',a,b }\delta_{i_3i_3'}
C^{R , S ; T , \tau_1  }_{ i_1 , i_2 ; i_3 }
C^{R , S ; T' , \tau_3  }_{a ,b ; i'_3 } 
 D^{ R }_{ i_1 a} ( \sigma_1) 
D^S_{ i_2b} ( \sigma_2)   \sigma_1  \otimes \sigma_2 \cr\cr
&& 
 = \delta_{RR'} \delta_{SS'}  \delta_{TT'}\delta_{\tau_2 \tau_2'}  Q^{R,S,T}_{\tau_1,\tau_3}  
\label{qqmatrix}
\eea
where we used the orthogonality relations \eqref{ortho2} and 
\eqref{cc=del}. 

Next, we evaluate the pairing  between 
two basis elements $Q$'s and check that
there are orthogonal: 
\bea
&&
\bdel_2 (Q_{\tau_1,\tau'_1}^{R,S,T}; Q_{\tau_2,\tau'_2}^{R',S',T'})
 =\kappa_{R,S}^2
\sum_{\s_i,\g_i, \g'_i \in S_n}
\sum_{i_l, j_l,i'_l, j'_l}
 C^{R , S ; T , \tau_1  }_{ i_1 , i_2 ; i_3 } C^{R , S ; T , \tau'_1  }_{ j_1 , j_2 ; i_3 } 
C^{R' , S' ; T' , \tau_2  }_{ i'_1 , i'_2 ; i'_3 } C^{R' , S' ; T' , \tau'_2  }_{ j'_1 , j'_2 ; i'_3 } \cr\cr
&& \times 
 D^{ R }_{ i_1 j_1} ( \sigma_1  ) D^S_{ i_2 j_2 } ( \sigma_2 )  
 D^{ R' }_{ i'_1 j'_1} ( \sigma_1  ) D^{S'}_{ i'_2 j'_2 } ( \sigma_2 )  
 \cr\cr
&&
 = \kappa_{R,S}^2
\sum_{i_l, j_l,i'_l, j'_l}
 C^{R , S ; T , \tau_1  }_{ i_1 , i_2 ; i_3 } C^{R , S ; T , \tau'_1  }_{ j_1 , j_2 ; i_3 } 
C^{R' , S' ; T' , \tau_2  }_{ i'_1 , i'_2 ; i'_3 } C^{R' , S' ; T' , \tau'_2  }_{ j'_1 , j'_2 ; i'_3 } \cr\cr
&& \times 
 \frac{(n!)^2}{d(R)d(S)} \delta_{RR'}\delta_{SS'}\delta_{i_1i_1'}\delta_{i_2i_2'}
\delta_{j_1j_1'}\delta_{j_2j_2'} \cr\cr
&&
 = \kappa_{R,S}
\delta_{RR'}\delta_{SS'}
\sum_{i_l, j_l, i'_3, j'_l}
 C^{R , S ; T , \tau_1  }_{ i_1 , i_2 ; i_3 } 
C^{R , S; T' , \tau_2  }_{ i_1 , i_2 ; i'_3 }
C^{R , S ; T , \tau'_1  }_{ j_1 , j_2 ; i_3 } 
 C^{R , S; T' , \tau'_2  }_{ j_1 , j_2 ; i'_3 } \cr\cr
&& 
 =  \kappa_{R,S} d(T)\,  \delta_{RR'} \delta_{SS'}
\delta_{TT'} \delta_{\tau_1\tau_2} \delta_{\tau'_1\tau'_2}  
\label{orhoqq}
\eea
which shows that according to the normalization that we are
using the matrices $Q$ are orthogonal but not normalized. 

\
 
\noindent{\bf $Q_{\ung}$-basis of $\cK_{\ung}(n)$ -}
We now give few properties of the  $Q_{\ung}$-basis \eqref{qunbasis}.
We will skip steps since the derivations are similar to the
above case. 

The  $Q_{\ung}$-basis is stable under left and right actions 
of the diagonal $\diag(\mC(S_n))$: 
\bea
&&
(\gamma_1^{\otimes 3} )\cdot Q^{R,S,T}_{\ung; \tau_1,\tau_2} \cdot
(\gamma_2^{\otimes 3})= 
\label{inqUN2}\\
&&
 =  \kappa_{R,S,T}
\sum_{\s_i \in S_n}
\sum_{i_l, j_l}
\sum_{a_l,b_l}
C^{R , S ; T , \tau_1  }_{ i_1 , i_2 ; i_3 } C^{R , S ; T , \tau_2  }_{ j_1 , j_2 ; i_3 } 
 D^{ R }_{ i_1 a_1} (\gamma_1^{-1})  
D^{ R }_{ a_1 b_1} ( \sigma_1  )
D^{ R }_{ b_1 j_1} (\gamma_2^{-1})
\cr\cr
&& \times 
 D^S_{ i_2 a_2 } ( \gamma_1^{-1} ) 
 D^S_{ a_2 b_2 } ( \sigma_2 ) 
 D^S_{ b_2 j_2 } ( \gamma_2^{-1} ) \,
 D^T_{ i_3 a_3 } ( \gamma_1^{-1} ) 
 D^T_{ a_3 b_3 } ( \sigma_3 ) 
 D^T_{ b_3 j_3 } ( \gamma_2^{-1} ) \,
  \s_1 \otimes \s_2 \otimes \s_3    
\nonumber
\eea
We use \eqref{ddc=cd} and \eqref{ddinv} (of appendix \ref{sapp:cgc}) to  reduce the above to $Q^{R,S,T}_{\ung;\tau_1,\tau_2}$. 

Taking a product of $Q_{\ung}$-elements, 
we change variables $\s_i \to \s_i {\s'}_i^{-1} $, use \eqref{ortho2} 
to get 
\bea
&& Q^{R,S,T}_{\ung;\tau_1,\tau_2} Q^{R',S',T'}_{\ung;\tau_2',\tau_3}
= \cr\cr
&& 
\kappa_{R,S,T} \kappa_{R',S',T'} 
\sum_{\s_l,\s_l'}\sum_{i_l,i_l',j_l,j_l'}
C^{R , S ; T , \tau_1  }_{ i_1 , i_2 ; i_3 } C^{R , S ; T , \tau_2  }_{ j_1 , j_2 ; j_3 } C^{R' , S' ; T' , \tau'_2  }_{ i'_1 , i'_2 ; i'_3 } C^{R' , S' ; T' , \tau_3  }_{ j'_1 , j'_2 ; j'_3 } \cr\cr
&&\times 
D^{R}_{ i_1 , j_1 } ( \s_1 {\s'}_1^{-1}  ) D^{S}_{ i_2, j_2 } ( \s_2{\s'}_2^{-1}) D^{T}_{ i_3 , j_3 } (\s_3 {\s'}_3^{-1})
D^{R'}_{ i'_1 , j'_1 } ( \sigma'_1 ) D^{S'}_{ i'_2, j'_2 } ( \sigma'_2 ) D^{T'}_{ i'_3 , j'_3 } ( \sigma'_3 )
\sigma_1 \otimes \sigma_2 \otimes \sigma_3 \cr\cr
&&
 = \delta_{RR'}\delta_{SS'}\delta_{TT'}
\delta_{\tau_2\tau'_2}Q^{R,S,T}_{\ung;\tau_1,\tau_3}
\label{qqUNmatrix} 
\eea
which shows that $Q_{\ung}$ multiply like matrices.  

Computing the pairing of two elements of the
$Q_{\ung}$-basis,  we obtain
\bea
&&
\bdel_3(Q^{R,S,T}_{\ung;\tau_1,\tau_2} Q^{R',S',T'}_{\ung;\tau_2',\tau_3})
 = 
\kappa_{R,S,T} 
\delta_{RR'}\delta_{SS'}\delta_{TT'}
\sum_{i_l,j_l}
C^{R , S ; T , \tau_1  }_{ i_1 , i_2 ; i_3 } 
 C^{R, S ; T , \tau'_2  }_{ i_1 , i_2 ; i_3 }
C^{R , S ; T , \tau_2  }_{ j_1 , j_2 ; j_3 }
 C^{R , S ; T , \tau_3  }_{ j_1 , j_2 ; j_3 } 
\cr\cr
&& 
= \kappa_{R,S,T} d(T)^2
\delta_{RR'}\delta_{SS'}\delta_{TT'}
\delta_{\tau_1\tau'_2} \delta_{\tau_2\tau_3}  
\label{qqUNin}
\eea

\

\noindent{\bf $Q_{\ung}$-basis from a tensor product 
basis-}
There is a nice way to arrive at the $Q_{\ung}$-basis for $ \cK ( n ) $ 
by starting from the representation theoretic Fourier basis for 
the tensor product $ \mC ( S_n ) \otimes \mC ( S_n) \otimes \mC ( S_n)$.

Consider in the group algebra  $ \mC ( S_n )$, the elements  
\be
Q^R_{ ij}   =   { \kappa_R \over n! } \sum_{ \sigma \in S_n } D^R_{ ij} ( \sigma ) \sigma  
\ee
with $\kappa_R$ is a normalization factor that we will fix
after introducing the pairing.  $R$ is any irreducible representation of $ S_n$, parametrised by partitions of $n$ or Young diagrams with$n$ boxes. The indices  
$i,j$ run over an orthonormal basis set for the representation. The number of these $Q^R_{ij}$ is equal to $n!$ thanks to a standard group theory identity
\bea 
\sum_{ R \vdash n } (d(R))^2 = n! 
\eea
These $Q^R_{ij}$ form a representation theoretic Fourier basis for $ \mC ( S_n)$. 

We have the following important properties: 
\be \label{tauq}
\tau \,  Q^R_{ ij} = 
\sum_{l} D^R_{ li} ( \tau ) \,  Q^R_{ l j} \,, \qquad  
 Q^R_{ ij} \, \tau  =    \sum_{l} Q^R_{ i  l} ~ D^R_{ j l } ( \tau )  
\ee
There is a pairing on $ \mC ( S_n ) ^{d}$, 
such that 
$$\delta(\s_1 \otimes \dots \otimes \s_d; 
\s_1' \otimes \dots \otimes \s_d') =\delta(\s_1\s_1'^{-1})\dots
\delta(\s_d^{-1}\s_d'^{-1})$$
 and such that 
\bea
\delta(Q^{ R}_{ i j };Q^{ R'}_{ i' j' }) 
 = \frac{\kappa_{R}\kappa_{R'}}{(n!)^2}
\sum_{\s} D^R_{ij}(\s)D^{R'}_{i'j'}(\s)
 =\frac{\kappa_R^2}{n!d(R)}\delta_{RR'}\delta_{ii'}\delta_{jj'} 
= \delta_{RR'}\delta_{ii'}\delta_{jj'}   
\eea
with $\kappa_R^2 = n!d(R)$. Then 
\be
\delta(
Q^{ R_1}_{ i_1 j_1 } \otimes \dots \otimes Q^{ R_d}_{ i_d  j_d }\,;\,
Q^{ R'_1}_{ i'_1 j'_1 } \otimes \dots \otimes Q^{ R'_d}_{ i'_d  j'_d })
 = 
\delta_{R_1R_1'}\delta_{i_1i_1'}\delta_{j_1j_1'}
\dots
\delta_{R_dR'_d}\delta_{i_di'_d}\delta_{j_dj'_d} 
\label{pairing}
\ee
Hence, the basis $\{Q^{ R_1}_{ i_1 j_1 } \otimes \dots \otimes Q^{ R_d}_{ i_d  j_d }\}$ is an orthonormal (Fourier-like) basis 
for $ \mC ( S_n )^d$. 

Let us restrict now to rank $d=3$ (the following extends to any $d$
easily). Consider 
$\rho_{L}(\tau_1)$ and $\rho_{R}(\tau_2)$ the left and right diagonal 
action on the tensor product $ \mC ( S_n )^3$. Then 
we write: 
\bea
&&
\sum_{\s_1,\s_2}
\rho_{L}(\s_1)\rho_{R}(\s_2)\, Q^{ R_1}_{ i_1 j_1 }\otimes 
Q^{ R_2}_{ i_2 j_2 } \otimes Q^{ R_3}_{ i_3 j_3 }
  = 
  \sum_{\s_1,\s_2}
\s_1\, Q^{ R_1}_{ i_1 j_1 } \s_2 \otimes 
\s_1 Q^{ R_2}_{ i_2 j_2 }\s_2 \otimes 
\s_1 Q^{ R_3}_{ i_3 j_3 }\s_2 \cr\cr
&& 
 = 
\sum_{\s_1,\s_2}
\sum_{p_l,q_l}
D^{R_1}_{p_1 i_1}(\s_1)Q^{ R_1}_{ p_1 q_1 } D^{R_1}_{j_1 q_1}(\s_2)
\otimes 
D^{R_2}_{p_2 i_2}(\s_1)Q^{ R_2}_{ p_2 q_2 } D^{R_2}_{j_2 q_2}(\s_2)
\otimes 
D^{R_3}_{p_3 i_3}(\s_1)Q^{ R_3}_{ p_3 q_3 } D^{R_3}_{j_3 q_3}(\s_2)\cr\cr
&& 
 = 
\frac{(n!)^2}{d(R_3)^2}
\sum_{p_l,q_l}
\sum_{\tau,\tau'}
C^{R_1,R_2;R_3,\tau}_{p_1,p_2;p_3}
C^{R_1,R_2;R_3,\tau}_{i_1, i_2; i_3}
C^{R_1,R_2;R_3,\tau'}_{j_1 ,j_2 ;j_3}
C^{R_1,R_2;R_3, \tau'}_{ q_1, q_2;q_3}\, 
Q^{ R_1}_{ p_1 q_1 }\otimes 
Q^{ R_2}_{ p_2 q_2 } \otimes 
Q^{ R_3}_{ p_3 q_3 } \cr\cr
&& 
\eea
where we used \eqref{tauq} and Lemma \ref{lem2:DDD=CC} 
for summing $\s_1,\s_2$, 
(in appendix \ref{sapp:cgc}). Then we couple the last result with
two CG's, and use \eqref{cc=del} , in such a way to have: 
\bea
&&
\sum_{i_l,j_l} 
C^{R_1,R_2;R_3,\tau}_{i_1, i_2; i_3}
C^{R_1,R_2;R_3,\tau'}_{j_1 ,j_2 ;j_3}
\sum_{\s_1,\s_2}
\rho_{L}(\s_1)\rho_{R}(\s_2)\, Q^{ R_1}_{ i_1 j_1 }\otimes 
Q^{ R_2}_{ i_2 j_2 } \otimes Q^{ R_3}_{ i_3 j_3 } \cr\cr
&& 
= 
\frac{(n!)^2}{d(R_3)^2}
\sum_{p_l,q_l}
\sum_{\varrho,\varrho'}
\sum_{i_l,j_l} 
C^{R_1,R_2;R_3,\tau}_{i_1, i_2; i_3}
C^{R_1,R_2;R_3,\varrho}_{i_1, i_2; i_3}
C^{R_1,R_2;R_3,\tau'}_{j_1 ,j_2 ;j_3}
C^{R_1,R_2;R_3,\varrho'}_{j_1 ,j_2 ;j_3}
\cr\cr
&& \times 
C^{R_1,R_2;R_3,\varrho}_{p_1,p_2;p_3}
C^{R_1,R_2;R_3, \varrho'}_{ q_1, q_2;q_3}\, 
Q^{ R_1}_{ p_1 q_1 }\otimes 
Q^{ R_2}_{ p_2 q_2 } \otimes 
Q^{ R_3}_{ p_3 q_3 } \cr\cr
&&
 = (n!)^2
\sum_{p_l,q_l}
C^{R_1,R_2;R_3,\tau}_{p_1,p_2;p_3}
C^{R_1,R_2;R_3, \tau'}_{ q_1, q_2;q_3}\, 
Q^{ R_1}_{ p_1 q_1 }\otimes 
Q^{ R_2}_{ p_2 q_2 } \otimes 
Q^{ R_3}_{ p_3 q_3 }  
\eea
The last expression matches $Q_{\ung;\tau,\tau'}^{R_1,R_2,R_3}$
up to a normalization \eqref{qunbasis}. 

From 
\bea
\sum_{i_l,j_l} 
C^{R_1,R_2;R_3,\tau}_{i_1, i_2; i_3}
C^{R_1,R_2;R_3,\tau'}_{j_1 ,j_2 ;j_3}
\sum_{\s_1,\s_2}
\rho_{L}(\s_1)\rho_{R}(\s_2)\, Q^{ R_1}_{ i_1 j_1 }\otimes 
Q^{ R_2}_{ i_2 j_2 } \otimes Q^{ R_3}_{ i_3 j_3 } 
\eea
we can infer invariance under left and right diagonal action of
$Q_{\ung;\tau,\tau'}^{R_1,R_2,R_3}$:  fix $\g_1, \g_2$, 
$$\rho_{L}(\g_1)\rho_{R}(\g_2)[\sum_{\s_1,\s_2}\rho_{L}(\s_1)\rho_{R}(\s_2)]
= \sum_{\s_1,\s_2}\rho_{L}(\g_1\s_1)\rho_{R}(\g_2\s_1)
 = \sum_{\s_1,\s_2} \rho_{L}(\s_1)\rho_{R}(\s_2).$$ 
Finally, similar derivations
allows one to get $Q_{\tau,\tau'}^{R_1,R_2,R_3}$ \eqref{qbasis}
in terms of diagonal adjoint action coupled with 
CG's. 

\subsection{Basis of $\cZ(\cK(n))$}
\label{app:Zbaskn}

\noindent{\bf Overcompleteness of  the $z_{R_1,R_2;R_3}$-basis in $\cZ(\cK(n))$ -}
Consider the elements
\bea
z_{R_1,R_2;R_3}
 = (z_{R_1}\otimes z_{R_2})\cdot z_{R_3} \,, \quad
z_{R_{1,2} } = \sum_{\s} \chi^{R_{1,2}}(\s) \s\,,
\quad z_{R_3} =\sum_{\s} \chi^{R_3}(\s) \s\otimes \s    
\eea
which are elements of $\cZ(\cK(n))$. 
We evaluate now the overlapping between the basis $P^{R_1,R_2,R_3}$
and the elements $z_{R'_1,R'_2;R'_3}$:
\bea
&&
\bdel_2(P^{R_1,R_2,R_3};z_{R'_1,R'_2;R'_3})  
 = \sum_{\tau}  \sum_{\s_i \in S_n} \chi^{R'_1}(\s_1) \chi^{R'_2}(\s_2)
 \chi^{R'_3}(\s_3)
\bdel_2(Q^{R_1,R_2,R_3}_{\tau,\tau} ;\s_1\s_3 \otimes  \s_2\s_3)  \cr\cr
&&
= \kappa_{R_1,R_2}
 \sum_{\tau}  
\sum_{\s_l' ,\s_l\in S_n}
\sum_{i_l, j_l}
C^{R_1,R_2;R_3 , \tau  }_{ i_1 , i_2 ; i_3 } C^{R_1,R_2;R_3 , \tau  }_{ j_1 , j_2 ; i_3 } \cr\cr
&&\times
 D^{ R_1}_{ i_1 j_1} ( \sigma'_1  ) D^{R_2}_{ i_2 j_2 } ( \sigma'_2 ) 
\chi^{R'_1}(\s_1) \chi^{R'_2}(\s_2)
 \chi^{R'_3}(\s_3) \, 
 \bdel_2(\sigma'_1 \otimes \sigma'_2; \s_1\s_3 \otimes  \s_2\s_3)
\cr\cr
&&
= \kappa_{R_1,R_2}
 \sum_{\tau}  
\sum_{\s_l\in S_n}
\sum_{i_l, j_l}
C^{R_1,R_2;R_3 , \tau  }_{ i_1 , i_2 ; i_3 } C^{R_1,R_2;R_3 , \tau  }_{ j_1 , j_2 ; i_3 } \cr\cr
&&\times
\sum_{a_l,b_l}
 D^{ R_1 }_{ i_1 a_1}( \s_1)  D^{ R_1 }_{a_1 j_1} (\s_3 ) 
 D^{ R_2 }_{ i_2 a_2}( \s_2)  D^{ R_2 }_{a_2 j_2} (\s_3 )
D^{R'_1}_{b_1b_1}(\s_1) D^{R'_2}_{b_2b_2}(\s_2)
 D^{R'_3}_{b_3b_3}(\s_3) 
\cr\cr
&&
= \kappa_{R_1,R_2}
\frac{(n!)^2}{d(R_1)d(R_2)} \delta_{R_1R_1'}\delta_{R_2R_2'}
 \sum_{\tau}  
\sum_{i_l, j_l}\sum_{\s}
C^{R_1,R_2;R_3 , \tau  }_{ i_1 , i_2 ; i_3 } C^{R_1,R_2;R_3 , \tau  }_{ j_1 , j_2 ; i_3 } \cr\cr
&&\times
\sum_{a_l,b_l} \delta_{i_1b_1}\delta_{a_1b_1}
\delta_{i_2b_2}\delta_{a_2b_2}
 D^{ R_1 }_{a_1 j_1} (\s ) 
 D^{ R_2 }_{a_2 j_2} (\s )
 D^{R'_3}_{b_3b_3}(\s)  
\cr\cr
&&
= \kappa_{R_1,R_2}
\frac{(n!)^2}{d(R_1)d(R_2)} \delta_{R_1R_1'}\delta_{R_2R_2'}
 \sum_{\tau}  
\sum_{b_3}
\sum_{i_l, j_l}\sum_{\s}
C^{R_1,R_2;R_3 , \tau  }_{ i_1 , i_2 ; i_3 } C^{R_1,R_2;R_3 , \tau  }_{ j_1 , j_2 ; i_3 } \cr\cr
&&\times
 D^{ R_1 }_{i_1 j_1} (\s) 
 D^{ R_2 }_{i_2 j_2} (\s )
 D^{R'_3}_{b_3b_3}(\s)  
\cr\cr
&&
= \delta_{R_1R_1'}\delta_{R_2R_2'}
 \sum_{\tau}  
\sum_{b_3,l}
\sum_{i_l}\sum_{\s}
C^{R_1,R_2;R_3 , \tau  }_{ i_1 , i_2 ; i_3 }
C^{R_1,R_2;R_3 , \tau  }_{ i_1 , i_2 ;l} 
 D^{R'_3}_{b_3b_3}(\s) 
D^{ R_3 }_{l  i_3 } (\s )
\cr\cr
&& 
 = 
\delta_{R_1R_1'}\delta_{R_2R_2'}
\sC(R_1,R_2,R_3)
\sum_{b,l} \sum_{\s}
D^{R'_3}_{bb}(\s) 
D^{ R_3 }_{l  l } (\s )\cr\cr
&& 
 = n!\, \delta_{R_1R_1'}\delta_{R_2R_2'}\delta_{R_3R_3'}
\sC(R_1,R_2,R_3)  
\label{eq:Pz}
\eea

\

\noindent{\bf Overcompleteness of  the $z_{\ung}^{R_1,R_2,R_3}$-basis in $\cZ(\cK_{\ung}(n))$ -}
The elements of interest are of the form 
\bea
z^{R_1,R_2,R_3}_{\ung}
 = z_{R_1}\otimes z_{R_2}\otimes z_{R_3} \,, \qquad
z_{R_i } = \sum_{\s} \chi^{R_i}(\s) \s 
\eea
which are elements of $\cZ(\cK(n))$. Note that they are more
symmetric in the three indices than the previous central element. 
We compute the overlap: 
\bea
&&
\bdel_{3} (z_{\ung}^{R_1,R_2,R_3};P_{\ung}^{R'_1,R'_2,R_3'})   =
\sum_{\tau}
\sum_{\s_i}\chi^{R_1}(\s_1)\chi^{R_2}(\s_2)\chi^{R_3}(\s_3) 
\bdel_3( \s_1 \otimes \s_2 \otimes \s_3; Q^{R'_1,R'_2,R_3'}_{\ung; \tau,\tau}) \cr\cr
&&
 = \kappa_{R'_1,R'_2,R_3'}\sum_{\tau}\sum_{\s_l,\s'_l \in S_n}  \sum_{i_l,j_l} 
C^{R'_1,R'_2;R_3' , \tau  }_{ i_1 , i_2 ; i_3 } C^{R'_1,R'_2;R_3' , \tau  }_{ j_1 , j_2 ; j_3 } \cr\cr
&& \times 
\chi^{R_1}(\s_1)\chi^{R_2}(\s_2)\chi^{R_3}(\s_3)  
D^{R_1'}_{ i_1 , j_1 } ( \sigma'_1 ) D^{R_2'}_{ i_2, j_2 } ( \sigma'_2 ) D^{R_3'}_{ i_3 , j_3 } ( \sigma'_3 )
\bdel_3( \s_1 \otimes \s_2 \otimes \s_3; \s'_1 \otimes \s'_2 \otimes \s'_3) 
\cr\cr
&&
 = \kappa_{R'_1,R'_2,R_3'}\sum_{\tau}\sum_{\s_l\in S_n}  \sum_{i_l,j_l} 
C^{R'_1,R'_2;R_3' , \tau  }_{ i_1 , i_2 ; i_3 } C^{R'_1,R'_2;R_3' , \tau  }_{ j_1 , j_2 ; j_3 } 
\cr\cr
&&
 \times 
\chi^{R_1}(\s_1)\chi^{R_2}(\s_2)\chi^{R_3}(\s_3)  
D^{R_1'}_{ i_1 , j_1 } ( \sigma_1 ) D^{R_2'}_{ i_2, j_2 } ( \sigma_2 ) D^{R_3'}_{ i_3 , j_3 } ( \sigma_3 )
\cr\cr
&& 
 = 
\delta_{R_1R'_1}\delta_{R_2R'_2}\delta_{R_3R_3'}
\sum_{\tau}
\sum_{i_l} 
C^{R_1,R_2;R_3  , \tau  }_{ i_1 , i_2 ; i_3 } C^{R_1,R_2;R_3, \tau  }_{i_1 , i_2 ; i_3}  
 \cr\cr 
&& 
 =
d(R_3)\sC(R_1,R_2,R_3)\delta_{R_1R'_1}\delta_{R_2R'_2}\delta_{R_3R_3'}
\label{eq:PUNz}
\eea
That coefficient is not vanishing in general hence 
$P^{R_1,R_2,R_3}$ admits an expansion in the $z_{R_1,R_2,R_3}$-basis.  

\section{Multiplication table of rank $d=3,$ $n=3$ colored tensor graphs}
\label{app:11m} 

We list, in this appendix, the multiplication table of the $10$ invariants
(recall that there are 11 invariants but the row for the identity element $E= \idtr$  
is trivial) of rank $d=3$ at $2n=6$ number of vertices. 
The index $c=1,2,3$ is a color label of a graph.  
Define $\check{c}=1,2,3$ such that $\check{c}\ne c$, and
$\check{\check{c}}=1,2,3$ which can be neither $\check{c}$
nor $c$, we obtain Table \ref{table} giving the multiplication
between the 10 possible elements. 

The table teaches us that the basic products are all commutatives
and so is the algebra $\cK_{\ung}(3)$. Furthermore, some 
elements admit a or multiple factorizations: 
\bea
&& 
\scalebox{1.5}{ \dtwo{$c$} } = \scalebox{1.5}{ \dtwo{$c$} \cc{$c$} } \\
&& 
\scalebox{1.5}{  \ac{$c$} }  =   \scalebox{1.5}{ \dtwo{$c$} \cc{$\check{c}$} }  = 
  \scalebox{1.5}{ \cc{$\check{c}$} \ac{$c$} }  =  
 \scalebox{1.5}{ \dtwo{$c$} \ktr  }
\nonumber
\eea

\begin{center}
\begin{sidewaystable}
\centering
\begin{tabular}{|c||c|c|c|c|c|c|c|}
\hline 
$(\downarrow) \times (\to) $ & \dtwo{$c$}  & \dtwo{$\check{c}$}  &\cc{$c$}  & \cc{$\check{c}$}& \ac{$c$} &\ac{$\check{c}$}  & \ktr\\
\hline 
\hline 
\dtwo{$c$}&  $ \frac{2}{3}\cc{$c$}  + \frac13\idtr $ &  
$\frac23\ac{ $\check{\check{c}}$ }+\frac13\dtwo{$\check{\check{c}}$} $ &
\dtwo{$c$} &
\ac{$c$} &
$\frac13 \Big[ \cc{$\check{c}$} + \cc{$\check{\check{c}}$} $   &  $\frac23 \ac{$\check{\check{c}}$}+ \frac13\dtwo{$\check{\check{c}}$} $&
 \ac{$c$}\\ 
& &  & & &
$ + \ktr\Big] $   &  & \\ 
\hline 
\cc{$c$}& \dtwo{$c$}  & \ac{$\check{c}$} & $\frac12\Big[ \cc{$c$} + \idtr \, \Big]$   & 
$\frac12\Big[ \cc{$\check{\check{c}}$} + \ktr \Big]$  & 
\ac{$c$} & $\frac12\Big[ \ac{$\check{c}$} + \dtwo{$\check{c}$} \Big]$  &
$\frac12\Big[ \cc{$\check{c}$} + \cc{$\check{\check{c}}$} \Big]$\\ 
\hline 
\ac{$c$}&  $\frac{1}{3}\Big[ \cc{$\check{c}$} + \cc{$\check{\check{c}}$} $ &
$\frac23 \ac{$\check{\check{c}}$} + \frac13 \dtwo{$\check{\check{c}}$} $ & \ac{$c$} & 
$\frac12 \Big[  \ac{$c$}  + \dtwo{$c$} \Big]$ & $\frac16\Big[ 2\cc{$c$}+\cc{$\check{c}$}$&  
$\frac23 \ac{$\check{\check{c}}$} + \frac13 \dtwo{$\check{\check{c}}$}$ &
$\frac12\Big[ \ac{$c$} + \dtwo{$c$}   \Big]$ \\ 
&  $ + \ktr   \Big]$ &  & & & $+\cc{$\check{\check{c}}$} + \ktr + \idtr \,\Big]$ &     &\\ 
\hline 
\ktr &\ac{$c$} &\ac{$\check{c}$} & 
 $\frac12\Big[ \cc{$\check{c}$}+ \cc{$\check{\check{c}}$}  \Big]$ &
 $\frac12\Big[ \cc{$c$}+ \cc{$\check{\check{c}}$}  \Big]$  
&  $\frac12\Big[ \ac{$c$}+ \dtwo{$c$} \Big]$ 
& $\frac12\Big[ \ac{$\check{c}$}+ \dtwo{$\check{c}$} \Big]$  &
 $\frac12\Big[ \ktr + \idtr \,\Big]$  
\\ 
\hline 
\end{tabular}
\caption{Multiplication table of rank 3 invariants up to order $n=3$.}
\label{table}
\end{sidewaystable}
\end{center}

\section{Correlators of tensor observables}
\label{app:correlators} 

Correlators involves insertions and evaluation of 
general observables in the Gaussian path integral.
We will restrict attention to $d=3$ and 
consider the Gaussian model 
\be
\cZ = \int d \Phi d \bar \Phi \; e^{ - { 1 \over 2 } \sum_{i_l} \Phi_{i_1 i_2 i_3 } \bar \Phi_{i_1 i_2 i_3 } }  
\ee
The index $i_a$ takes values in $\{ 1 \cdots N_a \}$, for $ a \in \{ 1,2,3 \} $.  
The propagator or 2-point function is of the form
\be
\la \Phi_{ i_1 i_2 i_3    } \bar \Phi_{ j_1 j_2 j_3 } \ra = \delta_{ i_1 j_1 } \delta_{i_2 j_2 } \delta_{i_3 j_3 }  
\ee
The observables, invariant under $U(N) \times U(N) \times U(N)$, are labelled by permutations 
$ ( \s_1 , \s_2 , \s_3 ) $ subject to equivalence $ ( \s_1 , \s_2 , \s_3 ) \sim ( \gamma_1 \s_1 \gamma_2 , \gamma_1 \s_2 \gamma_2 , \gamma_1 \s_3 \gamma_2 ) $. We will write these observables subjected
to the equivalence as $ \cO_{ \sigma_1 , \sigma_2 , \sigma_3} =\cO_{ \gamma_1 \s_1 \gamma_2 , \gamma_1 \s_2 \gamma_2 , \gamma_1 \s_3 \gamma_2 } $. 
We recall 
\bea
&&
\cO_{ \sigma_1 , \sigma_2 , \sigma_3}  = 
\sum_{i_l,j_l,k_l}
\Phi_{i_1j_1k_1} \Phi_{i_2j_2k_2} \dots \Phi_{i_nj_nk_n} \cr\cr
&& \times 
\, 
\bar\Phi_{i_{\s_1(1)}j_{\s_2(1)}k_{\s_3(1)}}\bar \Phi_{i_{\s_1(2)}j_{\s_2(2)}k_{\s_3(2)}} \dots \bar\Phi_{i_{\s_1(n)}j_{\s_2(n)}k_{\s_3(n)}}  
\eea
The integral the such operators is  given by the Wick theorem 
\bea 
&&
\la \cO_{ \s_1 , \s_2 , \s_3 }  \ra
=
{ 1 \over \cZ }  \int d \Phi d \bar \Phi \; e^{ - { 1 \over 2 } \sum_{i,j,k}   \Phi_{ijk} \bar \Phi_{ijk} }  
 \cO_{ \s_1 , \s_2 , \s_3 }  \crcr
&& =  \sum_{i_l,j_l,k_l}
\sum_{\mu \in S_n} 
\delta_{i_1 i_{\mu(\s_1(1))}} \delta_{i_2 i_{\mu(\s_1(2))}}  \dots \delta_{i_n i_{\mu(\s_1(n))}} \cr\cr
&& \times 
\delta_{j_1 j_{\mu(\s_2(1))}} \delta_{j_2 j_{\mu(\s_2(2))}}  \dots \delta_{j_n j_{\mu(\s_2(n))}} \;
\delta_{k_1 k_{\mu(\s_3(1))}} \delta_{k_2 k_{\mu(\s_3(2))}}  \dots \delta_{k_n k_{\mu(\s_3(n))}} \cr\cr
&& 
 = \sum_{\mu \in S_n} N^{\cy(\mu \s_1)+\cy(\mu \s_2)+\cy(\mu \s_3)} 
\eea
where $\cy(\alpha)$ is the number of cycles of $\alpha\in S_n$. 
 We have used the fact that, given $\s\in S_n$, $i_l \in [\![1, N ]\!]$, $l=1,\dots,n$, 
\bea
\sum_{i_l} \delta^{i_1}_{i_{\s(1)}}\dots \delta^{i_n}_{i_{\s(n)}} = N^{\cy(\s)} 
\eea
This allows us to recover \eqref{corrd}. 

Two point-functions $\la \cO_{ \s_1 , \s_2 , \s_3 } \bar{\cO}_{ \tau_1 , \tau_2 , \tau_3 } \ra $ can be also computed in a similar way.
We give a summary of appendix C in \cite{Sanjo} (note that we correct a few mistakes
appearing this appendix below). 
We are taking the observables to be ``normal ordered''
so we only allow contractions to take place between the $\Phi$'s from the first observable to the $\bar \Phi$'s 
from the second (parametrized by $\mu_a$) and  between the $\bar \Phi$'s from the first observable to the $ \Phi$'s 
from the second (parametrized by $\nu_a$): 
\be
\la \cO_{ \s_1 , \s_2 , \s_3 } \bar{\cO}_{ \tau_1 , \tau_2 , \tau_3 } \ra 
= \sum_{\mu\in S_n  }  \sum_{\nu \in S_n  }
\tr_{ V_1^{ \otimes n } } ( \s_1 \mu  \tau^{-1}_1 \nu ) \tr_{ V_2^{ \otimes n } } ( \s_2 \mu  \tau^{-1}_2 \nu )
\tr_{ V_3^{ \otimes n } } ( \s_3 \mu \tau^{-1}_3   \nu ) 
\ee
that could be translated as
\bea 
&& \la \cO_{ \s_1 , \s_2 , \s_3 } \bar{\cO}_{ \tau_1 , \tau_2 , \tau_3 } \ra
= \sum_{\mu \in S_n  }  \sum_{\nu\in S_n  }  
 N_1^{ \cy( \s_1 \mu  \tau^{-1}_1 \nu) }  N_2^{ \cy( \s_2 \mu  \tau^{-1}_2 \nu) }  N_3^{\cy( \s_3 \mu  \tau^{-1}_3\nu) } \cr 
 &&  =  \sum_{\mu \in S_n  }  \sum_{\nu \in S_n  }  
\sum_{\alpha_i  \in S_n  }  
N_1^{ \cy(\alpha_1)  }N_2^{ \cy(\alpha_2)}   N_3^{ \cy(\alpha_3) } 
\delta (  \s_1 \mu  \tau^{-1}_1\nu \alpha_1  ) 
\delta (  \s_2 \mu \tau^{-1}_2 \nu \alpha_2  )  
\delta (  \s_3 \mu  \tau^{-1}_3  \nu \alpha_3 )  \cr 
&&   =  \sum_{\mu \in S_n  }  
\sum_{\nu \in S_n  }   
N_1^{ n }N_2^{ n } N_3^{ n   } 
\delta (  \s_1 \mu  \tau^{-1}_1 \nu \Omega_1  ) 
\delta(\s_2 \mu \tau^{-1}_2 \nu\Omega_2)
\delta(\s_3 \mu \tau^{-1}_3 \nu\Omega_3) \cr 
&& 
\eea
with $\cy(\alpha) = \cy(\alpha^{-1})$, and
 \be
N^{n} \Omega = N^{n}  \sum_{ \alpha  \in S_n } 
N^{ \cy(\alpha) - n}  \alpha\; .
\ee
Finally, setting $\gamma_1 = \mu$, and $\gamma_2 = \nu$, and keeping in mind that $\bar  \cO_{ \tau_1 , \tau_2 , \tau_3 } =  \cO_{ \tau_1 ^{-1}, \tau_2^{-1} , \tau_3^{-1} } $, 
\be
\langle \cO_{ \sigma_1 , \sigma_2 , \sigma_3 } \cO_{ \tau_1 , \tau_2 , \tau_3 } \rangle 
= \sum_{ \gamma_1 , \gamma_2 } N^{ 3 n } \bdel [(  \sigma_1 \otimes \sigma_2 \otimes \sigma_3 ) \gamma_1^{ \otimes 3 }   (  \tau_1 \otimes \tau_2 \otimes \tau_3 ) \gamma_2^{ \otimes 3 } 
(\Omega_1 \otimes \Omega_2 \otimes \Omega_3 ) ]
\ee
which is the starting point \eqref{ooc}.

\end{document}